\RequirePackage{lineno}

\documentclass[12pt]{revtex4}
\usepackage{setspace}
\usepackage{times}

\usepackage{booktabs}
\usepackage{amsmath}
\usepackage{amssymb}

\usepackage{graphicx}
\usepackage{subfig}

\usepackage{layout}

\usepackage{booktabs}
\usepackage{mathrsfs}

\usepackage{multirow}

\usepackage{caption} 
\newcommand{\be}{\begin{eqnarray}}
\newcommand{\ee}{\end{eqnarray}}
\newcommand{\ba}{\begin{array}}
\newcommand{\ea}{\end{array}}

\begin{document}

\verb| |\\   % Seems to fix some latex bug

{\large
\begin{center}
\protect{\vspace{-2cm}}
A Jefferson Lab PAC 48 Experiment Proposal
\end{center}
}
\title{Backward-angle Exclusive $\pi^0$ Production above the Resonance Region}

%GH: William and Mary authors, Bill and Justin first, others in alphabetical order
\author{Wenliang Li (Spokesperson and contact person)} \email[E-mail: ]{wenliang.billlee@gmail.com} 
\affiliation{College of William and Mary, Williamsburg, VA, USA}

\author{Justin Stevens (Spokesperson)}
\affiliation{College of William and Mary, Williamsburg, VA, USA}

\author{David Armstrong}
\affiliation{College of William and Mary, Williamsburg, VA, USA}

\author{Todd Averett}
\affiliation{College of William and Mary, Williamsburg, VA, USA}

\author{Andrew Hurley}
\affiliation{College of William and Mary, Williamsburg, VA, USA}

\author{Lydia Lorenti}
\affiliation{College of William and Mary, Williamsburg, VA, USA}

\author{Arkaitz Rodas}
\affiliation{College of William and Mary, Williamsburg, VA, USA}

\author{Amy Schertz}
\affiliation{College of William and Mary, Williamsburg, VA, USA}
%GH: Regina authors, Garth first, others in alphabetical order

\author{Garth Huber (Spokesperson)}
\affiliation{University of Regina, Regina, SK  Canada}

\author{Muhammad Junaid}
\affiliation{University of Regina, Regina, SK  Canada}

\author{Stephen Kay}
\affiliation{University of Regina, Regina, SK  Canada}

\author{Vijay Kumar}
\affiliation{University of Regina, Regina, SK  Canada}

\author{Zisis Papandreou}
\affiliation{University of Regina, Regina, SK  Canada}

\author{Dilli Paudyal}
\affiliation{University of Regina, Regina, SK  Canada}

\author{Ali Usman}
\affiliation{University of Regina, Regina, SK  Canada}
%GH: TDA authors

\author{Kirill Semenov-Tian-Shansky}
\affiliation{National Research Centre Kurchatov Institute: Petersburg Nuclear Physics Institute, RU-188300 Gatchina, Russia}
\affiliation{Saint Petersburg National Research Academic University of the Russian Academy of Sciences, RU-194021 St. Petersburg, Russia}

\author{Bernard Pire}
\affiliation{CPHT, CNRS, \'Ecole Polytechnique, IP Paris, 91128-Palaiseau, France}

\author{Lech Szymanowski}
\affiliation{National Centre for Nuclear Research (NCBJ),  02-093 Warsaw, Poland}

% GH: Please put multi-author per institution below.  Larger #/institution first, then in 
% alphabetical order by institution name.

\author{Alexandre Camsonne}
\affiliation{Jefferson Lab, Newport News, Virginia, USA}

\author{Jian-Ping Chen}
\affiliation{Jefferson Lab, Newport News, Virginia, USA}

\author{Silviu Covrig Dusa}
\affiliation{Jefferson Lab, Newport News, Virginia, USA}

\author{Filippo Delcarro}
\affiliation{Jefferson Lab, Newport News, Virginia, USA}

\author{Markus Diefenthaler}
\affiliation{Jefferson Lab, Newport News, Virginia, USA}

\author{Dave Gaskell}
\affiliation{Jefferson Lab, Newport News, Virginia, USA}

\author{Ole Hansen}
\affiliation{Jefferson Lab, Newport News, Virginia, USA}

\author{Doug Higinbotham}
\affiliation{Jefferson Lab, Newport News, Virginia, USA}

\author{Astrid Hiller Blin}
\affiliation{Jefferson Lab, Newport News, Virginia, USA}

\author{Mike McCaughan}
\affiliation{Jefferson Lab, Newport News, Virginia, USA}

\author{Brad Sawatzky}
\affiliation{Jefferson Lab, Newport News, Virginia, USA}

\author{Greg Smith}
\affiliation{Jefferson Lab, Newport News, Virginia, USA}

\author{Arthur Mkrtchyan}
\affiliation{A. Alikhanyan National Science Laboratory (Yerevan Physics Institute), Yereven, Armenia}

\author{Vardan Tadevosyan}
\affiliation{A. Alikhanyan National Science Laboratory (Yerevan Physics Institute), Yereven, Armenia}

\author{Hakob Voskanyan}
\affiliation{A. Alikhanyan National Science Laboratory (Yerevan Physics Institute), Yereven, Armenia}

\author{Hamlet Mkrtchyan}
\affiliation{A. Alikhanyan National Science Laboratory (Yerevan Physics Institute), Yereven, Armenia}

\author{Stefan Diehl}
\affiliation{University of Connecticut, Mansfield, Connecticut, USA}

\author{Eric Fuchey}
\affiliation{University of Connecticut, Mansfield, Connecticut, USA}

\author{Kyungseon Joo}
\affiliation{University of Connecticut, Mansfield, Connecticut, USA}

\author{Werner Boeglin}
\affiliation{Florida International University, Miami, Florida, USA}

\author{Mariana Khachatryan}
\affiliation{Florida International University, Miami, Florida, USA}

\author{Pete E. Markowitz}
\affiliation{Florida International University, Miami, Florida, USA}

\author{Carlos Yero}
\affiliation{Florida International University, Miami, Florida, USA}

\author{Moskov Amaryan}
\affiliation{Old Dominion University, Norfolk, VA, USA}

\author{Florian Hauenstein}
\affiliation{Old Dominion University, Norfolk, VA, USA}

\author{Charles Hyde}
\affiliation{Old Dominion University, Norfolk, VA, USA}

\author{Gabriel Niculescu}
\affiliation{James Madison University, Harrisonburg, Virginia, USA}

\author{Ioana Niculescu}
\affiliation{James Madison University, Harrisonburg, Virginia, USA}

\author{Paul King}
\affiliation{Ohio University, Athens, Ohio, USA}

\author{Julie Roche}
\affiliation{Ohio University, Athens, Ohio, USA}

% GH: Single author per institution below.
% GH: Please put in alphabetical order by last name.

\author{Darko Androi\'{c}}
\affiliation{University of Zagreb, Zagreb , Croatia}

\author{Konrad Aniol}
\affiliation{California State University, Los Angeles, California, USA}

\author{Marie Boer}
\affiliation{University of New Hampshire, Durham, New Hampshire, USA}
\affiliation{Virginia Polytechnic Institute and State University, Blacksburg, Virginia, USA}

\author{Wouter Deconinck}
\affiliation{University of Manitoba, Winnipeg, Manitoba, Canada}

\author{Maxime Defurne}
\affiliation{CEA, Universit\'{e} Paris-Saclay, Gif-sur-Yvette, France}

\author{Mostafa Elaasar}
\affiliation{Southern University at New Orleans, New Orleans, Louisiana, USA}

\author{Cristiano Fanelli} 
\affiliation{Massachusetts Institute of Technology, Cambridge, Massachusetts, USA}

\author{Stuart Fegan}
\affiliation{University of York, Heslington, York, UK }

\author{Carlos Ayerbe Gayoso}
\affiliation{Mississippi State University, Starkville, MS, USA}

\author{Narbe Kalantarians}
\affiliation{Virginia Union University, Richmond, VA, USA}

\author{Daniel Lersch}
\affiliation{Florida State University, Tallahassee, Florida, USA}

\author{Rafayel Paremuzyan}
\affiliation{University of New Hampshire, Durham, New Hampshire, USA}

\author{Kijun Park}
\affiliation{Hampton University Proton Therapy Institute, Hampton, Virginia, USA}

\author{Igor Strakovsky} 
\affiliation{The George Washington University, Washington, DC, USA}

\date{\today}

\begin{abstract}

The proposed measurement is a dedicated study of the exclusive electroproduction process, $^1$H$(e, e^{\prime}p)\pi^0$, in the backward-angle regime ($u$-channel process) above the resonance region. Here, the produced $\pi^0$ is emitted 180 degrees opposite to the virtual-photon momentum (at large momentum transfer). This study also aims to apply the well-known Rosenbluth separation technique that provides the model-independent (L/T) differential cross-section at the never explored $u$-channel kinematics region ($-t=-t_{max}$, $-u=-u_{min}$). 

Currently, the ``soft-hard transition'' in $u$-channel meson production remains an interesting and unexplored subject. The available theoretical frameworks offer competing interpretations for the observed backward-angle cross section peaks. In a "soft" hadronic Regge exchange description, the backward meson production comes from the interference between nucleon exchange and the meson produced via re-scattering within the nucleon. Whereas in the ``hard'' GPD-like backward collinear factorization regime, the scattering amplitude factorizes into a hard subprocess amplitude and baryon to meson transition distribution amplitudes (TDAs), otherwise known as super skewed parton distributions (SuperSPDs). Both TDAs and SPDs are universal non-perturbative objects of nucleon structure accessible only through backward-angle kinematics. 

The separated cross sections: $\sigma_T$, $\sigma_L$ and ($\sigma_T$/$\sigma_L$) ratio at $Q^2=$2-6 GeV$^2$, provide a direct test of two predictions from the TDA model: $\sigma_T \propto 1/Q^8$ and the $\sigma_T \gg \sigma_L$ in $u$-channel kinematics. The magnitude and $u$-dependence of the separated cross sections also provide a direct connection to the re-scattering Regge picture. The extracted interaction radius (from $u$-dependence) at different $Q^2$ can be used to study the soft-hard transition in the $u$-channel kinematics. The acquisition of these data will be an important step forward in validating the existence of a backward factorization scheme (TDA and SuperSPD) of the nucleon structure function and establishing its applicable kinematic range.

\end{abstract}

\maketitle

\tableofcontents

\section{Introduction}

\begin{figure}[hb]
\centering
\includegraphics[width=0.8\textwidth]{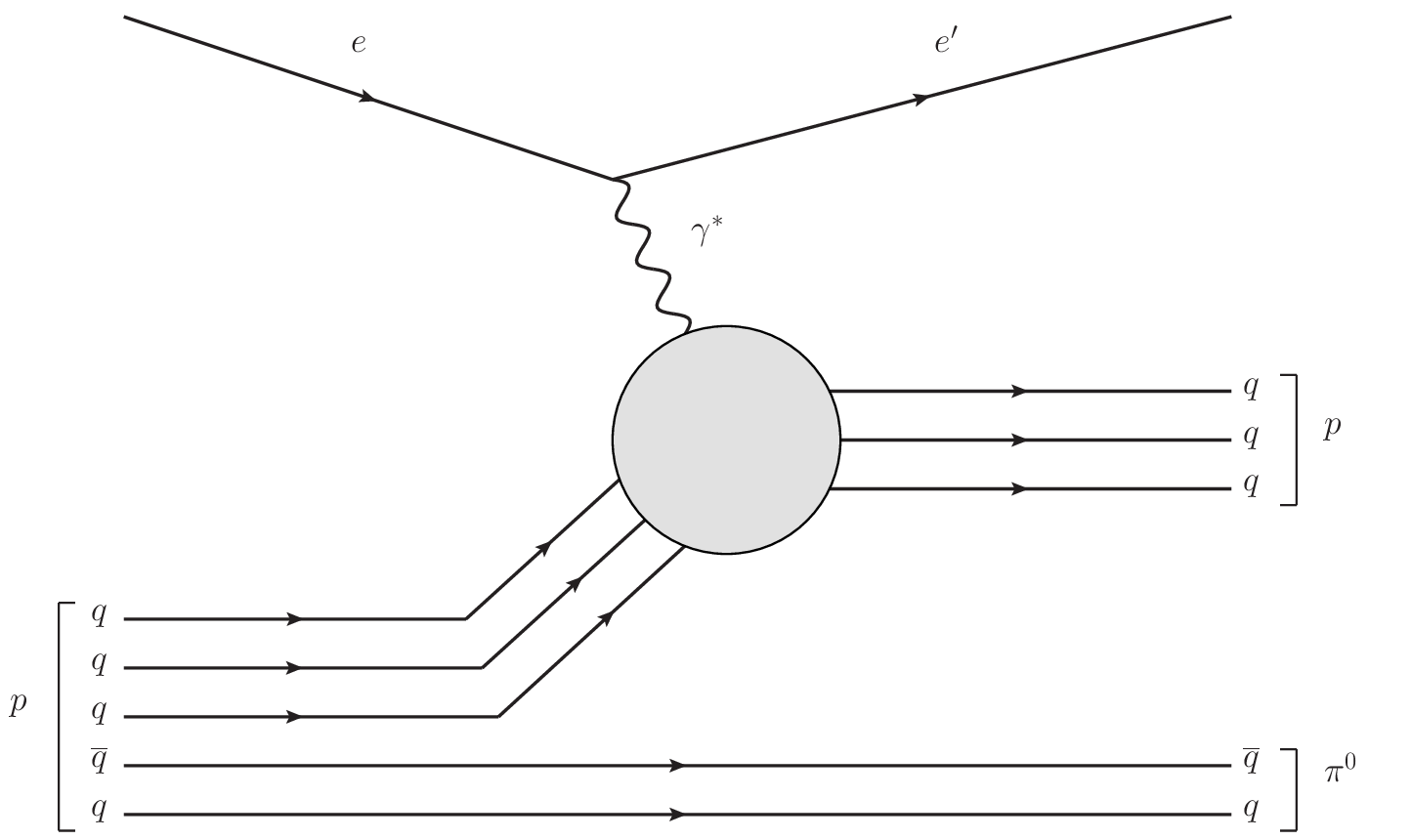}
\caption{Cartoon demonstration of a ``knocking a proton out of a proton process'' above the resonance region ($\sqrt{s}=W>2$~GeV)~\cite{weiss17}. In this case, a backward $\pi^0$ is produced nearly at rest.}
\label{fig:pi0_cartoon}
\end{figure}

In this proposal, we present a unique opportunity to access deep exclusive meson production (DEMP) in the backward-angle ($u$-channel kinematics) regime. The primary experimental observable involves exclusive $\pi^0$ electroproduction: $^1$H$(e, e^{\prime}p)\pi^{0}$, with a kinematic coverage of $2 < Q^2 < 6.25$~GeV$^2$ at fixed $x_{\rm B}=0.36$ and $W>2$~GeV. Since the $\pi^0$ is produced almost at 180$^\circ$ opposite to the direction of the virtual-photon momentum (corresponding to extreme backward angles), the Mandelstam variable for crossed four-momentum transfer squared is $u^{\prime} = u-u_{\textrm{min}} \approx 0$~GeV$^2$. At selected $Q^2$ settings, the full L/T/LT/TT cross section separation will be performed. Due to its unusual kinematics, the backward-angle meson production reaction is often referred to as a ``knocking a proton out of a proton process'', as shown in Fig.~\ref{fig:pi0_cartoon}.  

The proposed measurement uses the standard Hall~C equipment, polarized electron beam up to 70~$\mu$A (at standard accelerator gradient settings at the time of running) and standard unpolarized liquid hydrogen (LH$_2$) target. Since the produced $\pi^0$ are not directly detected, the missing mass reconstruction method will be applied. This technique permits access to a unique backward-angle kinematics region which was previously unexplored. The L/T separation technique used here is identical to the ones used successfully by many previous Hall A and C experiments during the 6~GeV era of CEBAF, an example being the pion form factor experiment~\cite{volmer01, blok08}.

The most important goals of the proposed measurement are to:
\begin{enumerate}
\item{Determine if exclusive $\pi^0$ electroproduction has a significant backward-angle peak, as it was demonstrated recently in exclusive $\omega$ electroproduction \cite{li19}, where backward angle data from Hall C were combined with forward-angle data from CLAS.  Here, we have chosen kinematics compatible with E12-13-010 \cite{E12-13-010} (forward-angle) and CLAS 12 (forward and wide angle) to facilitate a complete coverage in $-t$ for the $\pi^0$ production at certain $W$ and $Q^2$ settings. A complete $-t$ evolution would  reveal a forward-angle peak (at $t_{min}$), a wide angle plateau ($-t\approx{-u}$) and a backward-angle peak (at $t_{max}$). }

\item{A phenomenology study of extracting the $u$-dependence for the separated cross sections would be the good handle to determine transverse size of interaction, which can be used to study the transition from a ``soft'' Regge-exchange type picture (transverse size of interaction is of order of the hadronic size) to the ``hard'' QCD regime (transverse size of interaction $\ll$ hadronic size). See further detail in Sec.~\ref{sec:u_dep}.}

\item{Assuming the backward-angle peak is present, as expected, the next important objective of the proposed measurement is to demonstrate the (model independent) dominance of the transverse cross section ($\sigma_{T}$) over the longitudinal: $\sigma_{T} > \sigma_{L}$, at $2 < Q^2 < 6$~GeV$^2$ above the resonance region ($W>2$~GeV). } 

\item{The last objective is to measure the $Q^2$-dependence of the $\sigma_{T}$ cross section at fixed $x_{B}=0.36$.}
\end{enumerate}

The outcome of the measured result is a critical step towards finding the applicable factorization region in the backward-angle ($u$-channel) kinematic regime. These scientific motivations are further elaborated in Sec.~\ref{sec:TDA}.  Additional to the main objectives, there are two potential opportunistic studies which will come for free with the planned measurements:
\begin{enumerate}
    \item As part of the physics background to $\pi^0$, backward-angle Virtual Compton Scattering (VCS) above the resonance region has generated high community interest. An exploratory effort with proposed data into this challenging measurement will gain important experimental insights which may lead to a dedicated study of this interesting process. See more detail in Sec~\ref{sec:background}. 
    \item By default, CEBAF offers polarized electron beam, this provides an opportunity to study the Beam Spin Asymmetry (BSA) with the proposed data. See detail in Sec.~\ref{sec:pi0_BSA}. 
\end{enumerate}

It is also worth mentioning that the proposed $\pi^0$ measurement was submitted to PAC 46 as a letter of intent with reference number LOI-12-18-005. In the final PAC report, the committee members acknowledged the uniqueness of the proposed study and the fact that Jefferson Lab is the best venue to carry out such study. In addition, the feedback by experimental experts regarded the measurements as technically straightforward. It is also important to remember that the $u$-channel meson production reaction is not a new concept among past experiments at JLab, Sec.~\ref{sec:exp_summary} provides a brief summary of these experimental efforts. In addition, the difference between the proposed $\pi^0$ measurement and other approved Hall C $\pi^0$ experiments is addressed in Sec.~\ref{sec:other_mea}.

We also would like to emphasize that the proposed $\pi^0$ measurement is not an isolated measurement; it marks the beginning of a comprehensive plan to study the backward-angle ($u$-channel) factorization scheme of nucleon structure. This plan involves JLab 12 GeV measurements, collaborative efforts with the $\overline{\rm P}$ANDA experiment and the future EIC. See Sec.~\ref{sec:TDA_study_program} for full details.

\section{Summary of Backward-angle Physics from JLab 6~GeV}
\label{sec:exp_summary}

At Jefferson Lab, direct or indirect measurements of exclusive meson electroproduction at large scattering angles are not a new concept.  Here, indirect measurement implies the usage of the missing mass reconstruction technique. During the 6~GeV era, there have been a few examples of such studies. In this section, we present a short overview of some of the important pioneering studies of backward-angle physics.

\subsection{Backward VCS and $\pi^0$ Electroproduction at Hall A and C}

\begin{figure}[htb]
\centering
\includegraphics[width=0.90\textwidth]{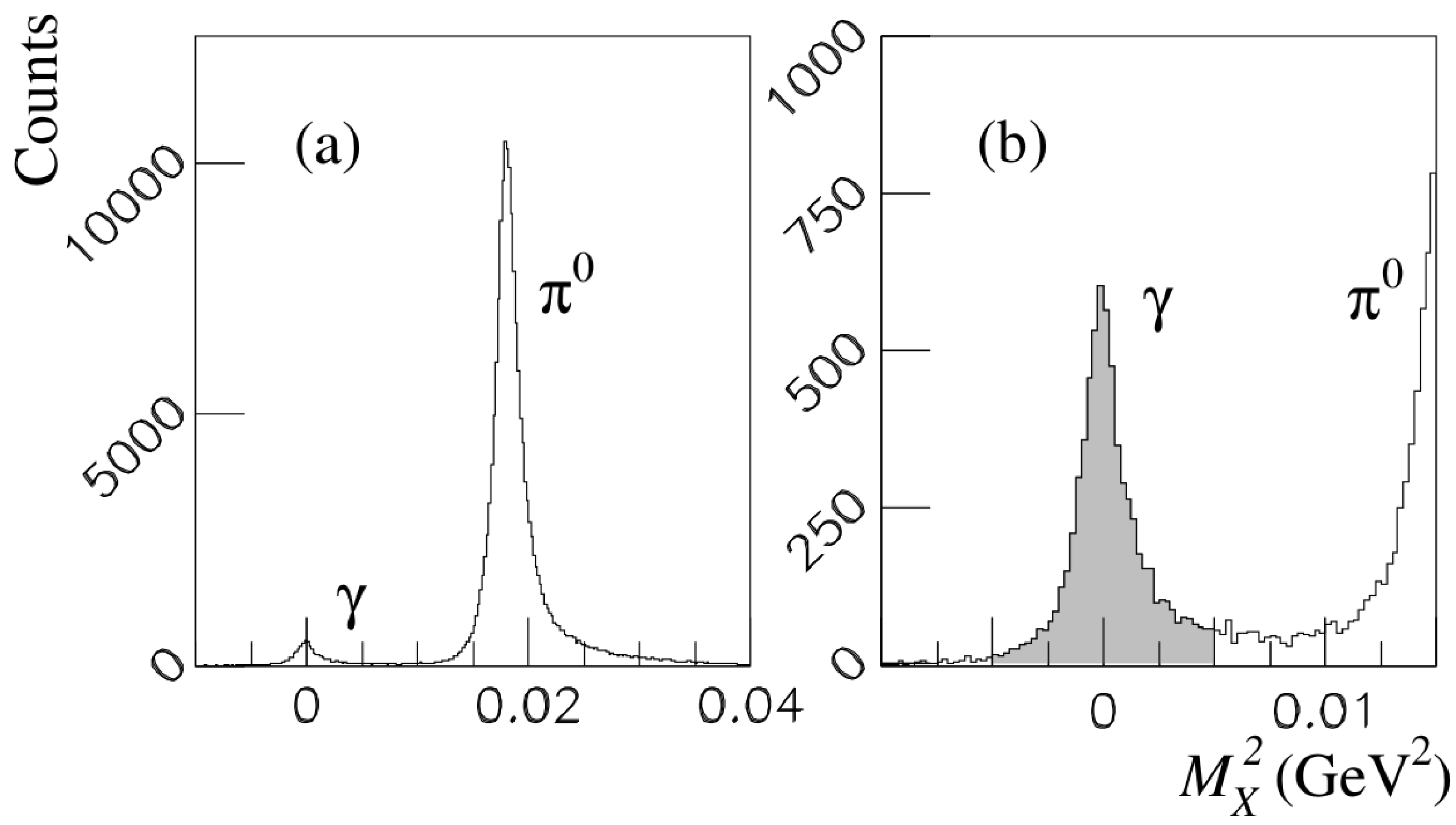}
\caption{Squared missing mass $M^2_X$ for an experimental setting $W=1.2$~GeV is shown in plot (a). The zoomed distribution around $\gamma$ peak is shown in (b). These plots were published in Ref.~\cite{laveissiere09}.}
\label{fig:mx2}
\end{figure}

Since the early stage of JLab (1993), backward angle $^1$H$(e, e^{\prime}p)\gamma$ and $^1$H$(e, e^{\prime}p)\pi^0$ measurements were attempted by a dedicated Hall A experiment E93-050~\cite{audit93, Fonvieille:2012cd, laveissiere04} in the nucleon resonance region. E93-050 used the 4~GeV electron beam colliding with a liquid hydrogen target, where a pair of High Resolution Spectrometers (HRSs) were used to detect the scattered electron and proton in coincidence.  The forward-going proton was detected in parallel kinematics and the `recoil' $\pi^0,~\gamma$ was emitted at backward angle at low momentum. The missing mass reconstruction technique was used to reconstruct the final state $\gamma$ as well as $\pi^0$ events.  An example of the reconstructed missing mass squared distribution from E93-050 is shown in Fig.~\ref{fig:mx2}.

The physics objective was to access the Compton photon scattered at backward angles in the nucleon resonance region ($S_{11}$ and $D_{13}$), whereas the $\pi^0$ was detected as the dominant background. Thanks to the good particle momentum resolution of the HRSs, separating the $\gamma$ and $\pi^0$ peaks was a relatively easy task (as shown in Fig,~\ref{fig:mx2}). 

E93-050 (and later E00-110) obtained a great deal of information about VCS, even when the BH amplitude was larger than VCS (DVCS-BH interfere is at the amplitude level). Also, $(e,e'p)\gamma$ and $(e,e'p)\pi^0$ cross sections were published in Ref.~\cite{Fonvieille:2012cd, laveissiere09}.

In 2008, Laveissiere, et al., published the first measurement of the backward-angle VCS cross section with the data from E93-050~\cite{laveissiere09}. This experiment was performed at $Q^2=1$ GeV$^2$ in the nucleon resonance region from threshold to $W = 1.9$ GeV. Despite the differences in physics motivations, Experiment E93-050 provides important insight to this proposal. The relative height and width of $\gamma$ and $\pi^0$ peaks from this measurement are useful benchmarks for estimating cross sections and determining the mass resolution requirements. 

In the 12 GeV era, the backward-angle VCS program is further explored by E12-15-001 at Hall C~\cite{camsonne12}. The measurement aims to extract the two scalar Generalized Polarizabilities of the proton in the range of $Q^2 = 0.3$ to $0.75$ GeV$^2$, near the $\Delta(1232)$ resonance region. Most interestingly, the equipment configuration (including 10 cm target cell) and missing mass reconstruction technique used by E12-15-001 are identical to the one used in this proposal. The partial completion of E12-15-001 (in 2019) is a great validation to the experimental methodology.

\subsection{High $-t$ charged $\pi$ Electroproduction at Hall B}
\label{sec:clas}

\begin{figure}[htb]
\centering
\includegraphics[width=0.7\textwidth]{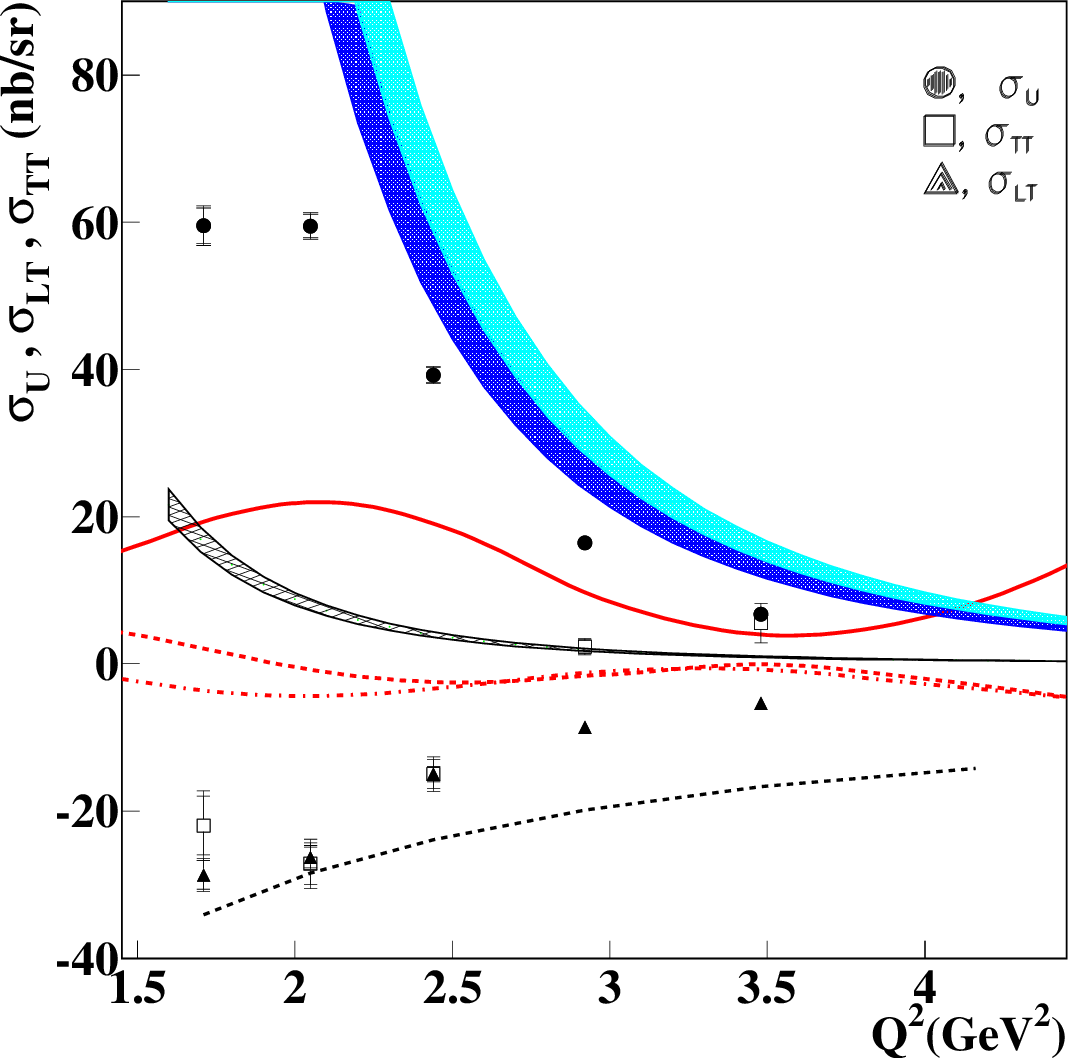}
\caption{The structure functions $\sigma_{u}$, $\sigma_{\rm TT}$ and $\sigma_{\rm LT}$ as a function of $Q^2$. The bands refer to model calculations of $\sigma_{u}$ in the TDA description with different nucleon DA models; dark blue band: COZ~\cite{chernyak89} $N$ DA model, light blue band: KS~\cite{king87}, black band: BLW NNLO~\cite{lenz09}. This plot was published in Ref.~\cite{park18}.} 
\label{fig:park18}
\end{figure}

The CLAS detector, in comparison to the Halls A and C spectrometers, offers the great advantage of a wide angular acceptance. Since the cross section for a given electroproduction reaction falls exponentially as a function of $-t$ (a larger $-t$ value corresponds to a wider scattering angle), it is difficult to determine the detector efficiency for wide scattering angles. After years of careful study, K. Park et al. published results for exclusive $\pi^+$ electroproduction, $^1$H$(e, e^\prime \pi^+)n$, near the backward angle above the resonance region~\cite{park18}. The $Q^2$ coverage is $1.5< Q^2 < 4.5$~GeV$^2$, at $W\sim$2.2 GeV, $-u =$0.5 GeV$^2$. 
The publication of this result was an important step for $u$-channel physics. Evidence of $Q^2$-scaling (particularly for $Q^2>2$ GeV$^2$) was observed, and it is consistent with the prediction of the GPD-like Transition Distribution Amplitude (TDA) factorization scheme at a much lower $Q^2$ range than originally expected.  This is demonstrated by the close agreement between the blue TDA band and the unseparated $\sigma_U$ in Fig.~\ref{fig:park18}.

\subsection{BSA on High-$t$ charged $\pi$ Electroproduction at Hall B}
\label{sec:BSA}

Recently \cite{BSApaper}, the CLAS collaboration reported the results of the analysis of hard exclusive single pion ($\pi^+$) electroproduction with CEBAF at $6$ GeV aiming on the study of the beam-spin asymmetry of the reaction above the resonance region. 

\begin{figure}[htb]
 \begin{center}
\includegraphics[width=.5\textwidth]{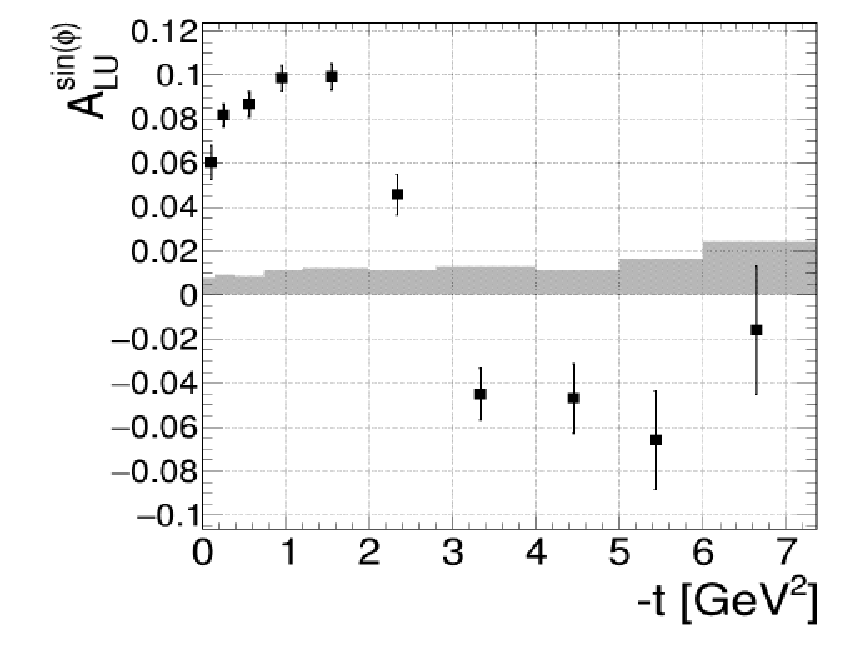}
\end{center}
\caption{$A^{\sin \phi}_{LU}$
as a function of $-t$.  The data are binned in $-t$ and integrated
over the complete $Q^2$ distribution ranging from $1$~GeV$^2$
to $4.5$~GeV$^2$ and $x_B$ ranging from $0.1$ to $0.6$.
The shaded area represents the systematic uncertainty (see detailed discussion in Ref.~\cite{BSApaper}).}
\label{Fig_BSA_t_dependence}
\end{figure}

The beam spin asymmetry for the reaction is defined as
\be
\quad B S A\left(t, \phi, x_{B}, Q^{2}\right)=\frac{d \sigma^{+}-d \sigma^{-}}{d \sigma^{+}+d \sigma^{-}}
=\frac{A_{L U}^{\sin \phi} \sin \phi}{1+A_{U U}^{\cos \phi} \cos \phi+A_{U U}^{\cos 2 \phi} \cos 2 \phi},
\label{Def_BSA}
\ee
where $\sigma^{\pm}$ is the differential cross section for each beam
helicity state ($\pm$). For the positive/negative helicity the
spin is parallel/anti-parallel to the beam direction. The
subscripts $ij$ represent the longitudinal ($L$) or unpolarized ($U$) state of the beam and the target, respectively.
$\phi$ is the azimuthal angle between the electron scattering plane and the hadronic reaction plane, on which the differential cross sections depend. The extraction of the beam spin asymmetry (\ref{Def_BSA}) provides access to the $A_{L U}^{\sin \phi}$ moment. It probes the interference between the amplitudes for longitudinal ($L$) and transverse ($T$) virtual-photon polarizations and is proportional to the polarized structure function $\sigma_{LT^{\prime}}$ (the symbol $^\prime$ signifies the structure function is the backward-angle):
\be
A_{L U}^{\sin \phi}=\frac{\sqrt{2 \varepsilon(1-\varepsilon)} \sigma_{L T ^{\prime}
}}{\sigma_{T}+\varepsilon \sigma_{L}},
\label{eqn:bsa}
\ee
where $\varepsilon$ is the polarization parameter of the virtual- photon. 
These measurements were performed with nearly full coverage
from forward to backward angles in the center-of-mass pion scattering angle. 
As shown in Fig.~\ref{Fig_BSA_t_dependence},  
the kinematic region for the extraction of $A_{L U}^{\sin \phi}$ 
was extended up to
$-t = 6.6\, {\rm GeV}^2$, which is close to the maximal 
accessible $-t$ value for given kinematical setup. 

The presented data provides important constraints for the development
of a factorized reaction mechanism describing the complete kinematic regime, including
the near-forward regime, with a possible collinear factorized description in terms of GPDs and pion DAs,
the intermediate kinematical regime and the near-backward kinematical regime,
with the eventual collinear factorized description in terms of  
$\pi N$ TDAs and nucleon DAs.   

In particular, the sign of  $A^{\sin \phi}_{LU}$ in near-forward kinematics (GPD region) is
clearly positive (Fig.~\ref{Fig_BSA_t_dependence}). However, a sign change of $A^{\sin \phi}_{LU}$ has been observed around
$\theta_{C M}=90^{\circ}$, bringing the BSA clearly negative in the backward hemisphere, and quite small in near-backward kinematics. This suggests a completely distinct reaction mechanism in the backward regime and hints at the leading twist dominance in the small $(-u)$ domain for $Q^2 < 4 \text{GeV}^2$, which is a central feature of the $u$-channel TDA factorization mechanism. 

\begin{figure}[htb]
 \begin{center}
\includegraphics[width=.8\textwidth]{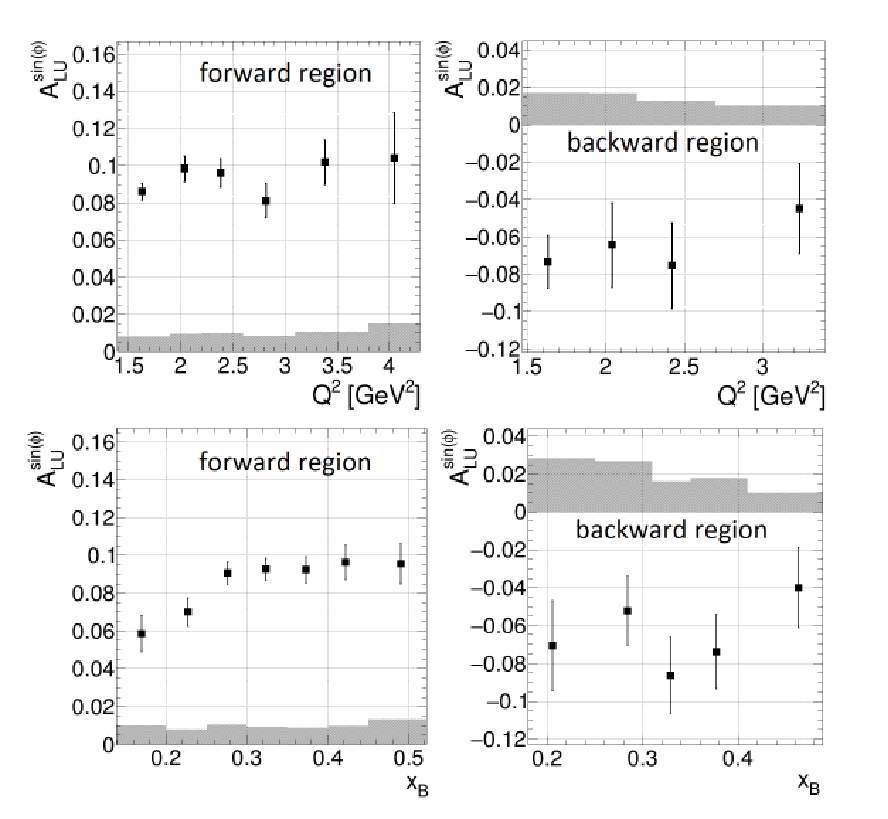}
\end{center}
\caption{$A^{\sin \phi}_{LU}$
 as function of $Q^2$ (top) and $x_B$ (bottom) for
pions going in the forward (left) and backward (right) regions.
The shaded area represents the systematic uncertainty (see details in Ref.~\cite{BSApaper}).}
\label{Fig_BSA}
\end{figure}

The data presented in Ref.~\cite{BSApaper} provide important constraints for the development of reaction mechanisms that describe the complete kinematic regime including GPDs and TDAs, as well as the intermediate regime.  Fig.~\ref{Fig_BSA} shows $A^{\sin \phi}_{LU}$ as function of $Q^2$ (top) and $x_B$ (bottom) for pions going in the near-forward (left) and near-backward (right) kinematics.

Here, we focus on the impact of these measurements in the near-backward kinematical regime, where a description in terms of $\pi N$ TDAs and nucleon DAs might be applied.
Assuming the collinear factorized description of the single pion electroproduction in the near-backward kinematics regime in terms of $\pi N$ TDAs and nucleon DAs, the cross section $\sigma_{LT}$ turns to be a subleading twist-$4$ effect.  Therefore, the expression for the BSA  involves the twist-$4$ nucleon DAs and nucleon-to-pion TDAs.
\begin{itemize}
\item 
For the leading twist transverse amplitude in terms of $\pi N$ TDAs,
$H_i^{{\rm tw}=3}$, and nucleon DAs, $\phi_i^{{\rm tw}=3}$,
we employ the notation
$\langle H_i^{{\rm tw}=3} \phi_j^{{\rm tw}=3} \rangle $.
\item To describe the next-to-leading twist longitudinal amplitude we need to introduce 
$\pi N$ TDAs, $H_i^{{\rm tw}=4}$, and nucleon DAs, $\phi_i^{{\rm tw}=4}$
\cite{Braun:1999te,Braun:2000kw}.
\end{itemize} 
Then, to twist-$4$ accuracy the appropriate amplitude can be written as
$ \langle H_i^{{\rm tw}=4} \phi_j^{{\rm tw}=3} \rangle + \langle H_i^{{\rm tw}=3} \phi_j^{tw=4} \rangle$.
Therefore, the cross section $\sigma_{LT'}$ within the TDA framework can be written as:
\begin{equation}
 \sigma_{LT}
\big|_{\rm Backward \atop regime} \approx {\rm Im} \left[\langle H_i^{{\rm tw}=3} \phi_j^{tw=3} \rangle \left( \langle  H_i^{tw=4} \phi_j^{tw=3} \rangle + \langle H_i^{tw=3} \phi_j^{tw=4} \rangle \right)^* \right].
\end{equation}

A complete theoretical study of this twist-$4$ longitudinal amplitude is not yet available, but is anticipated to be quite similar to the analysis done in Ref.~\cite{Belitsky:2002kj} for the calculation of the Pauli nucleon form factor. From Fig.~\ref{Fig_BSA} (top right), one could see the size and the sign flip in the $Q^2$ behavior of BSA fits (despite large error bars) in the backward angle, when compared to the forward-angle counterpart in Fig.~\ref{Fig_BSA} (top-left), this is similar to the prediction by the twist counting rules of collinear TDA/DA factorization mechanism in the near-backward regime. These findings are further elaborated in Ref.~\cite{BSApaper}.

A dedicated higher precision BSA measurement in a larger range of $Q^2$ will be enabled with the upgraded $12$~GeV CEBAF accelerator at JLab. This definitely would boost the theoretical studies, needed to provide the still lacking quantitative estimates of the effect.

\subsection{Backward $\omega$ Electroproduction at Hall C}

\label{sec:omega}

The recently published results from Hall C~\cite{li19, wenliang17} demonstrated that the missing mass reconstruction technique, in combination with operating the Hall C high precision spectrometers in coincidence mode, can be used to extract the backward-angle $\omega$ cross section reliably through the exclusive reaction $^1$H$(e, e^{\prime}p)\omega$, while performing a full L/T separation. The experiment has central $Q^2$ values of 1.60 and 2.45 GeV$^2$, at $W = 2.21$ GeV. There was significant coverage in $\phi$ and $\epsilon$, which allowed separation of $\sigma_{T,L,LT,TT}$. The data set has a unique $u$ coverage near $-u \sim 0$, which corresponds to $-t > 4$ GeV$^2$.

\begin{figure}[htb]
\centering
\includegraphics[width=0.8\textwidth]{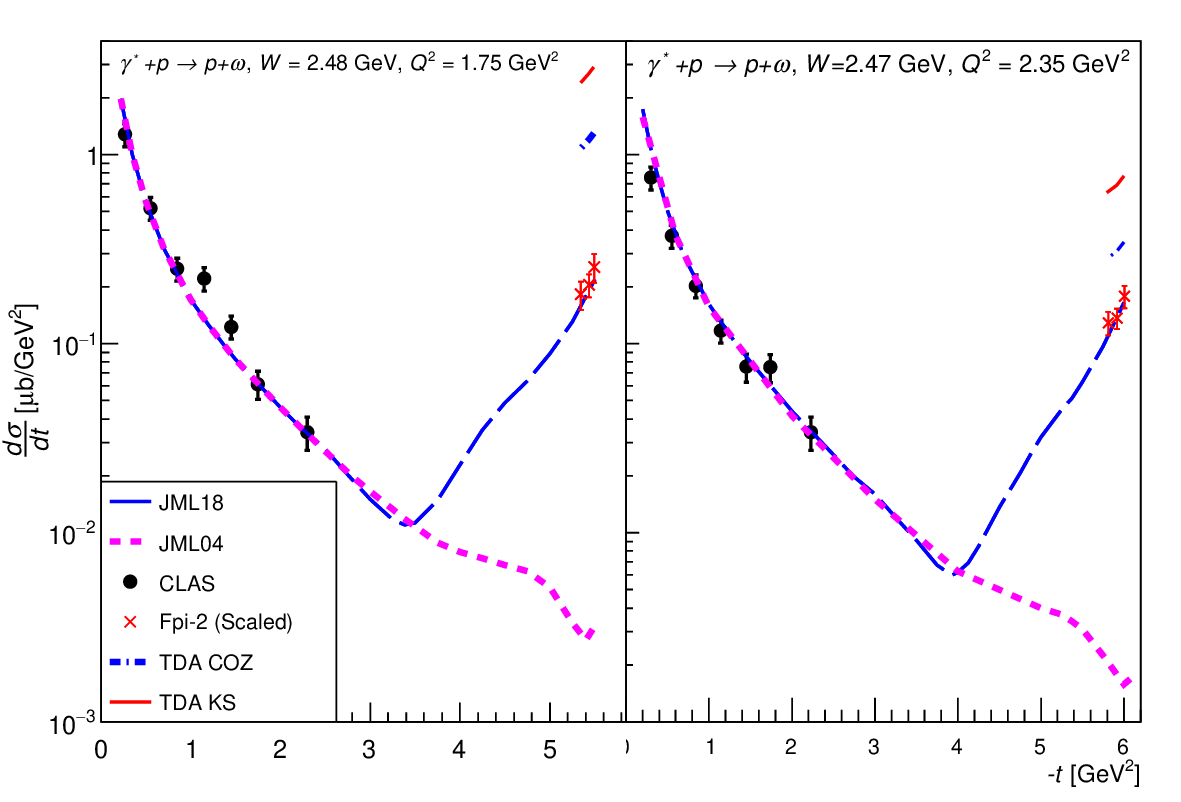}
\caption{Total differential cross section, $d\sigma_{u}/dt$ versus $-t$ for $W$=2.48~GeV, $Q^2$=1.75 (left) GeV and $W$=2.47~GeV, $Q^2$=2.35~GeV (right). The black dots are published CLAS results~\cite{morand05}. The red crosses are reconstructed $\sigma_u$ using $\sigma_{\rm T}$ and $\sigma_{\rm L}$ from Hall C (scaled to same kinematics)~\cite{li19, wenliang17}, the systematic error bands are shown in blue. The magenta and blue dashed lines represents the prediction of the hadronic Regge-based model, without ~\cite{laget04}, and with $\rho-N$ and $\rho-\Delta$ unitary rescattering (Regge) cuts \cite{laget18}. 
This plot was published in Ref.~\cite{li19}.} 
\label{fig:omega}
\end{figure}

The extracted cross sections (red crosses) show evidence of a backward-angle peak for $\omega$ exclusive electroproduction; angular distributions at $Q^2=1.75$ and 2.35 GeV$^2$ are shown in Fig.~\ref{fig:omega}. The forward-angle ($t$-channel) peak from the CLAS-6 data~\cite{morand05} is also shown. Previously, the the appearance of both forward and backward-angle peaks was only observed in meson photoproduction data~\cite{vgl96, guidal97}.
Furthermore, the Regge model description of Laget \cite{laget04}, involving re-scattering, provides a natural description of both the magnitude and slope of the observed backward-angle peak (discussed further in Sec.~\ref{regge}).
The investigation whether such a backward-angle peak also exists in $\pi^0$ electroproduction, and whether it persists over a wide $Q^2$ range, is the first goal of this proposal.

\begin{figure}[htb]
\centering
\includegraphics[width=0.8\textwidth]{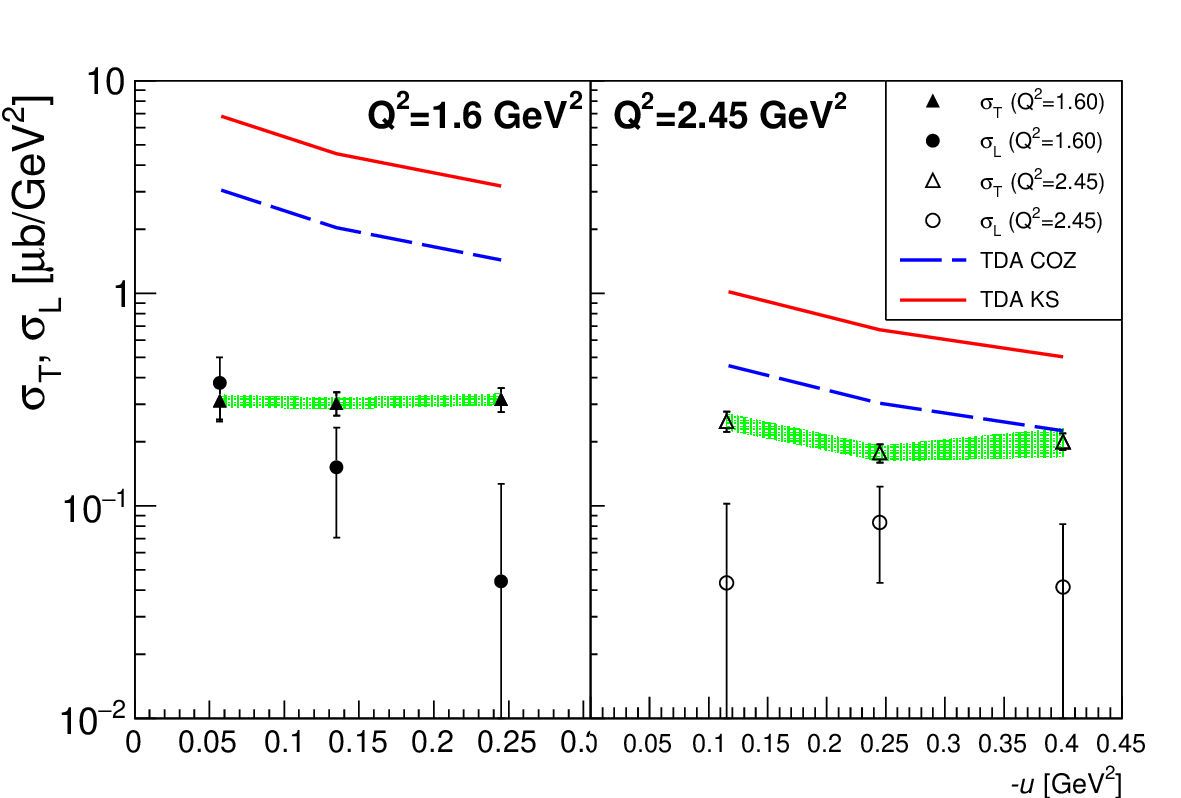}
\caption{Separated differential cross section $\sigma_{\textrm{T}}$ versus $-u$ for $Q^2=1.60$~GeV$^2$ (left) and $Q^2=2.45$~GeV$^2$ (right). The blue dashed and red solid and lines represent the TDA calculation \cite{pire15} using the COZ~\cite{chernyak89} and KS~\cite{king87} nucleon DA models, respectively.  The green bands indicate correlated systematic uncertainties for $\sigma_{\textrm{T}}$.  The separated cross sections shown in this figure are determined at the $Q^2$ and $W$ values at individual $-u$ bin, therefore cant not be used to determine the $u$ dependence directly. A scaling procedure is required when making the comparison at a nominal set of  $Q^2$ and $W$ values, such as in Fig.~\ref{fig:omega}. This plot was published in Ref.~\cite{wenliang17,li19}.} 
\label{fig:omega_sep}
\end{figure}

\begin{figure}[htb]
\includegraphics[width=0.9\textwidth]{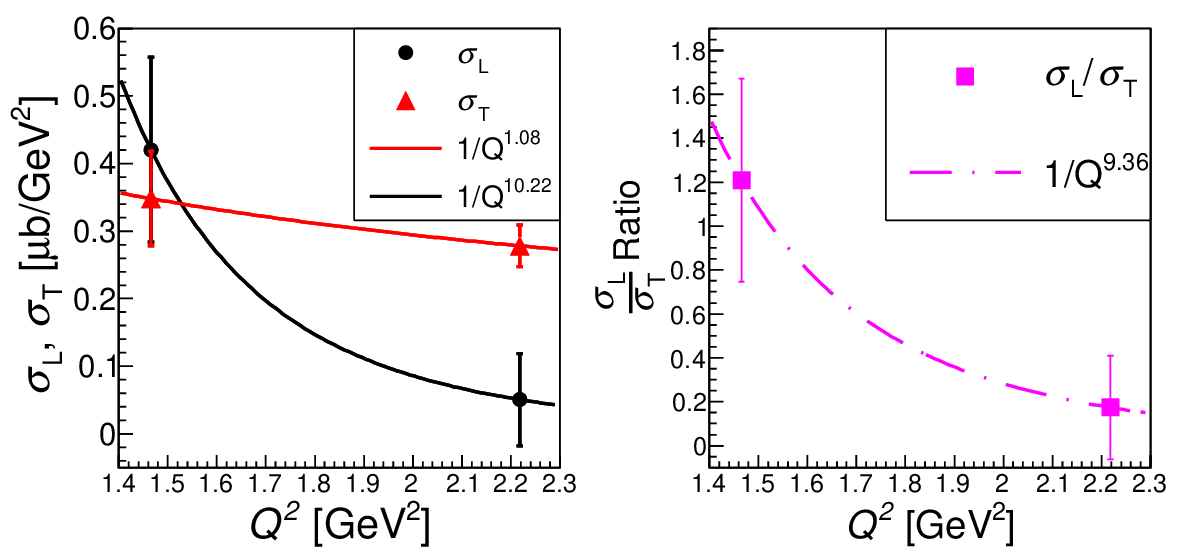}
\caption{(a) $\sigma_{T}$ and $\sigma_{L}$ as function of $Q^2$ at $u^{\prime}=0$ GeV$^2$; (b) $\sigma_L/\sigma_T$ ratio as function of $Q^2$.This plot was published in Ref.~\cite{li19, wenliang17}.} 
\label{fig:ratio}
\end{figure}

The extracted $d\sigma_{\rm L}/dt$ and $d\sigma_{\rm T}/dt$ from the Hall C $p(e,e^{\prime}p)\omega$ data are shown versus $-u$ in Fig.~\ref{fig:omega_sep}.  The data are compared to the TDA model prediction \cite{pire15}.  
At $Q^2=$2.45~GeV$^2$, the TDA predictions are within the same order of magnitude as the data, whereas at $Q^2=$1.6~GeV$^2$, the TDA model overpredicts the data by a factor of $\sim$10.  This is very similar to the behavior shown for the CLAS data in Fig.~\ref{fig:park18}.  Together, the datasets suggest that the backward-angle collinear factorization (TDA model) regime may begin to apply around $Q^2\approx$2~GeV$^2$.  The data proposed here would go a long way to confirm or reject whether this interpretation is correct.

The most important finding from the backward-angle $\omega$ analysis was the demonstration of $\sigma_{\rm T}$ dominance over $\sigma_{\rm L}$ at $Q^2 = 2.45$ GeV$^2$, see Fig.~\ref{fig:ratio}(b) for the $\sigma_L/\sigma_T$ cross section ratio as function of $Q^2$. Note that this was predicted by the TDA framework. As the JLab 12~GeV experiments can reach higher $Q^2$ values, the TDA formalism must be carefully studied and tested in more meson channels. An example of further study is this proposed $\pi^0$ meson measurement.

\section{Theoretical Context for Backward-angle $\pi^0$ Electroproduction}

In the framework of the $e$-$p$ scattering representation, the exclusive $\pi^0$ electroproduction $^1$H$(e, e^{\prime}p)\pi^{0}$ reaction can be written as
\begin{equation}
e(k) + p(p_1) \rightarrow e^{\prime}(k^\prime) + \pi(p_{\pi}) + p^\prime(p_2)\,.
\end{equation}
If the virtual-photon is considered as the projectile, then
\begin{equation}
\gamma^*(q) + p(p_1) \rightarrow \pi(p_{\pi}) + p^\prime(p_2)\,.
\label{eqn:reaction}
\end{equation}
Here, $p$ and $p^\prime$ are the proton before and after the interaction; $e$ and $e^\prime$ are the electron before and after the interaction; $\gamma^{*}$ is the space-like virtual-photon. The associated four-momentum for each particle is given inside the bracket. For this reaction, the Mandelstam variables are defined as
\begin{equation}
s=(p_1 + q)^2; ~~~~ u =(p_\pi - p_1)^2; ~~~~~ t= (p_2 -p_1)^2.
\end{equation}
In the case of the forward-angle ($t$-channel) meson production process, the $\pi^0$ is produced in the same direction as the virtual-photon momentum $q$ (known as the $q$-vector), and $-t\rightarrow t_{min}$ (i.e. parallel kinematics). Correspondingly, the backward-angle ($u$-channel) process produces $\pi^0$ in the opposite direction as the $q$-vector, and $-u \rightarrow u_{min}$ (anti-parallel kinematics). 

In the different kinematic regions, backward meson production can be explained using different nucleon structure models. When the process is within the resonance region ($W < 2$~GeV), the $u$-channel process can be described using the nucleon fragmentation model which has a mild $Q$ dependence~\cite{weiss17}; when above the resonance region ($W > 2$~GeV) a more complicated parton based model is required to describe the $Q^n$ dependence. The latter is the research interest of this proposal.

Within the 6 GeV JLab kinematics coverage: $ W>2$~GeV, $Q^2<3$~GeV$^2$, $x_{\rm B}=0.36$, there are two independent models capable of describing the existing backward angle data. The first is a QCD GPD-like model known as the TDA~\cite{pire05} (also Skewed Distribution Amplitude in the pioneering work of Ref.~\cite{FPPS}), which offers direct description of the individual partons within the nucleon;
the other model, a hadronic Regge-based model known as the JML model~\cite{laget04,laget18}, that explores meson-nucleon dynamics of hadron production reactions. In this section, we introduce how a backward-angle $\pi^0$ is produced according to both models and describe the benefits of studying them. L/T-separated cross sections can be calculated in both models and the leading twist TDAs predict $\sigma_{\textrm L}\sim0$~\cite{lansberg07}. 

\subsection{GPDs and Skewed Parton Distributions (SPDs)}

Generalized parton distributions (GPDs) are a modern description of the complex internal structure of the nucleon, which provides access to the correlations between the transverse position and longitudinal momentum distribution of the partons in the nucleon. In addition, GPDs give access to the orbital momentum contribution of partons to the spin of the nucleon~\cite{ji97, jo12}.

Currently, there is no known direct experimental access to the information encoded in GPDs~\cite{ji04}. The prime experimental channels for studying the GPDs are through the DVCS and DEMP processes~\cite{ji97}. Both processes rely on the collinear factorization (CF) scheme~\cite{collins97,radyushkin87}. An example DEMP reaction, $\gamma^*p\rightarrow p\pi^0$, is shown in Fig.~\ref{fig:GPD_TDA}(a). In order to access the forward-angle GPD collinear factorization regime ($\gamma^*p\rightarrow p\pi^0$ interaction), the kinematic variable requirements are as follows: sufficiently high $Q^2$, large $s$, fixed $x_{\rm B}$ and $t\sim0$~\cite{ji04, pire15}. Here, the definition of ``sufficiently high $Q^2$'' is process-dependent terminology. Based on the existing DIS data~\cite{girod08, Camacho:2006qlk, Defurne:2015kxq}, the GPD physics has shown that the range of ``sufficiently high $Q^2$'' lies between 1 and 5 GeV$^2$; this is sometimes referred to as ``early scaling''~\cite{pire18, voutier09}.

Under the collinear factorization regime, a parton is emitted from the nucleon GPDs ($N$ GPDs) and interacts with the incoming virtual-photon, then returns to the $N$ GPDs after the interaction~\cite{ji04}. Studies~\cite{kroll16, liuti10} have shown that perturbative calculation methods can be used to calculate the CF process (top oval in Fig.~\ref{fig:GPD_TDA} (a)) and extract GPDs through factorization, while preserving the universal description of the hadronic structure in terms of QCD principles.
One limitation in the GPD description of Fig.~\ref{fig:GPD_TDA} (a) requires $t\sim t_{min}$, namely, the process defaults a fast-meson and slow nucleon final state. Processes such as the one in this proposal could not be correctly accounted for by such a description.

In a 2002 paper~\cite{FPPS}, M. Strikman and others presented an innovative approach to resolve this issue. In it, they discuss the specific scenario when three valence quarks collapse to a small size color singlet configuration in a nucleon, or of valence quark and antiquark in a meson. As a result, a fast proton and a slow meson are created. Such a setup implies the manifestation of a ``cluster'' structure within the initial state that co-existed with three valence quarks. See a visualization of such a process in Fig.~\ref{fig:pi0_cartoon}. 

Application of this knowledge to the context of Reaction~\ref{eqn:reaction}: In the QCD description, the hard exclusive processes one needs to use generalized (skewed) parton distributions. In the case of describing the $N\rightarrow N$ transitions and non-diagonal transitions like $N \rightarrow \Lambda$, $\Delta$, the first type of distributions are known as generalized parton distributions (GPDs), while in the case of non-diagonal transitions (latter case) the used term is skewed Parton Distributions (SPDs). Under the case of extreme skewness, $\xi\rightarrow 0$, (for appropriate quantum numbers of the current) one would use super-SPDs to describe nucleon distribution amplitude~\cite{FPPS}. Note that this proposed $\pi^0$ measurement fulfills the super-SPD kinematics.

Although the above stated qualitative prediction was not made regarding $\pi^0$ electroproduction, one could still examine the predicted $1/(1-t/m^2)$ (where $m^2\sim 1$ GeV$^2$) cross section dependence using the proposed data.

\subsection{Meson-Nucleon Transition Distribution Amplitude}
\label{sec:TDA}

\begin{figure}[htb]
\centering
\includegraphics[width=0.49\textwidth]{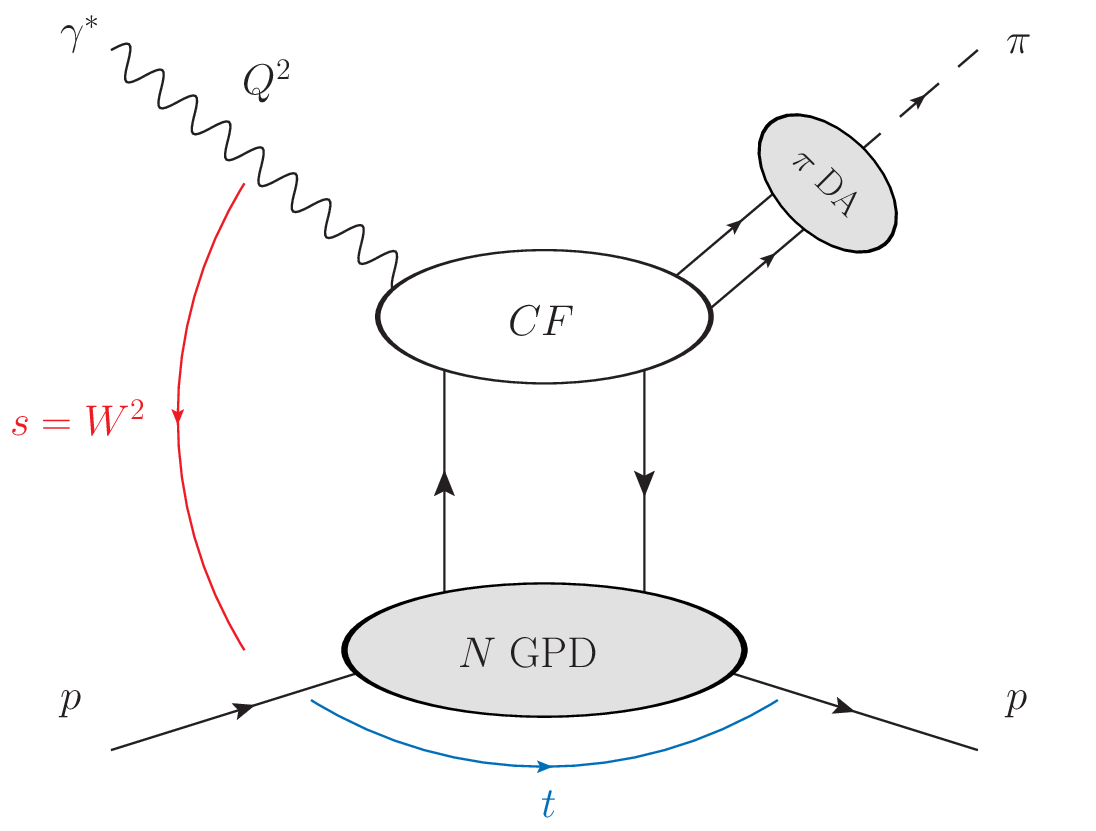}
\includegraphics[width=0.49\textwidth]{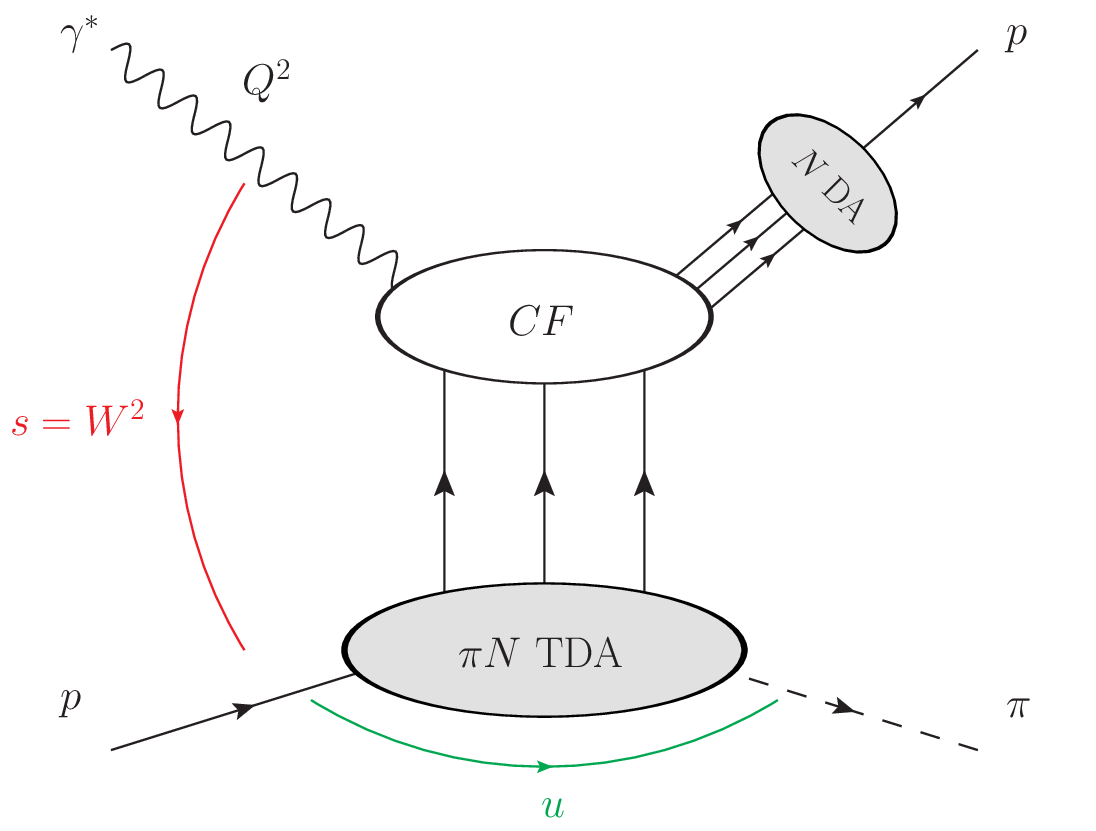}

\caption{(a) shows the $\pi^0$ electroproduction interaction ($\gamma^*p\rightarrow p\pi^0$) diagram under the (forward-angle) GPD collinear factorization regime (large $Q^2$, large $s$, fixed $x_{\rm B}$, fixed $t\sim0$). $N$ GPD is the quark nucleon GPD (note that there are also gluon GPD that is not shown). $\pi$ DA stands for the vector meson distribution amplitude. The CF corresponds to the calculable hard process amplitude.  (b) shows the (backward-angle) TDA collinear factorization regime (large $Q^2$, large $s$, fixed $x_{\rm B}$, $u\sim0$) for $\gamma^*p\rightarrow  p \pi^0$. The $\pi N$ TDA is the transition distribution amplitude from a nucleon to a vector meson. These plots were created based on the original ones published in Ref.~\cite{lansberg11}. }

\label{fig:GPD_TDA}
\end{figure}

A few years after this pioneering work, B. Pire, L. Szymanowski, J.P Lansberg and K. Semenov-Tian-
 Shansky rediscovered and developed the QCD formalism appropriate to describe the backward electroproduction of photons or mesons. In their transition distribution amplitude (TDA) formalism, they called  baryon-to-meson transition distribution amplitude ($\pi$N TDA) the backward analog of GPDs. TDAs describe the underlying physics mechanism of how the target proton transitions into a $\pi$ meson in the final state, shown as the gray oval in Fig.~\ref{fig:GPD_TDA}(b). One fundamental difference between GPDs and TDAs is that the TDAs require three parton exchanges between $\pi N$ TDA and CF.

Relevant to this discussion is the definition of skewness. For forward-angle kinematics, in the regime where the handbag mechanism and GPD description may apply, the skewness is defined in the usual manner,
\begin{equation}
\xi_t=\frac{p_1^+-p_2^+}{p_1^++p_2^+},
\label{eqn:xi_t}
\end{equation}
where $p_1^+$, $p_2^+$ refer to the light-cone plus components of the initial and final proton momenta in Eqn.~\ref{eqn:reaction}, calculated in the CM frame \cite{kroll04}.  The subscript $t$ has been added to indicate that this skewness definition is typically used for forward-angle kinematics, where $-t\rightarrow -t_{min}$.  In this regime, $\xi_t$ is related to Bjorken-$x$, and is approximated by $\xi_t=x/(2-x)$, up to corrections of order $t/Q^2 < 1$ \cite{favart15}.  This relation is an accurate estimate of $\xi_t$ to the few percent level for forward-angle electroproduction.  

In backward-angle kinematics, where $-t\rightarrow -t_{max}$ and $-u\rightarrow -u_{min}$, also $-t/Q^2>1$. The skewness is defined with respect to $u$-channel momentum transfer in the TDA (Transition Distribution Amplitude) formalism \cite{lansberg07},
\begin{equation}
\xi_u=\frac{p_1^+-p_{\pi}^+}{p_1^++p_{\pi}^+}.
\label{eqn:xi_u}
\end{equation}
    
The GPDs depend on $x$, $\xi_t$ and $t$, whereas the TDAs depend on $x$, $\xi_u$ and $u$. The $\pi^0$ production process through GPDs in the forward-angle ($t$-channel) and through TDAs in the backward-angle ($u$-channel) are schematically shown in Figs.~\ref{fig:GPD_TDA}(a) and (b), respectively. In terms of the formalism, TDAs are similar to the GPDs, except they depend on three quark momentum fractions $x_i$ (with $x_1+x_2 + x_3 = 2 \xi_u$.

The backward-angle TDA collinear factorization scheme has similar requirements: $x$ is fixed, the $u$-momentum transfer is required to be small compared to $Q^2$ and $s$; $u\equiv\Delta^2$, which implies that $Q^2$ and $s$ need to be sufficiently large. Recall an optimistic estimate of early scaling for GPD physics occurs between $2 < Q^2 < 5$~GeV$^2$ (although the $\sigma_{TT}$ for $\pi^0$ forward electroproduction is quite far from expectation). The case for the backward processes was open before the pioneering studies from JLab 6 GeV ~\cite{li19,park18, wenliang17}. The backward $\pi^+$ and $\omega$ production results have shown indications of TDA $Q^2$-scaling at $Q^2\ll 10$~GeV$^2$. Furthermore, the parameter $\Delta=p_{\pi}-p_1$ is considered to encode new valuable complementary information on the hadronic 3-dimensional wave functions, whose detailed physical meaning still awaits clarification~\cite{pire15}.

Beyond the JLab 12 GeV program, backward $\pi^0$ production will be studied by the $\overline{\rm P}$ANDA experiment at FAIR~\cite{panda15}. This experimental channel can be accessed through observables including $ p + \overline{p} \rightarrow  \gamma^* + \pi^0$ and $p + \overline{p} \rightarrow J/\psi + \pi^0$. Note that this backward $\pi^0$ production involves the same TDAs as in the electroproduction case. They will serve as very strong tests of the universality of TDAs in different processes~\cite{lansberg07}.

\subsubsection{Further Detail on the $\pi^0 N$ TDAs}

At leading twist-3, the parameterization of the Fourier transform of the $\pi N$ transition matrix element of the three-local light cone quark operator
$\widehat{O}_{\rho \tau \chi}(\lambda_1 n, \lambda_2 n,
\lambda_3n)$~\cite{radyushkin97} can be written as~\cite{pire11}
\begin{align}
& 4 \mathcal{F} \langle \pi_{\alpha}(p_\pi)| \widehat{O}_{\rho \tau \chi}(\lambda_1 n, \lambda_2 n, \lambda_3n)| N_{\iota} (p_1) \rangle  \nonumber \\[5mm]
& = 4(P\cdot n)^3 \int \left[\, \prod^3_{j=1} \frac{d \lambda_j}{2\pi} \, \right] e^{i\sum^{3}_{k=1} x_k\lambda_k (P \cdot n)} \langle \pi_{\alpha}(p_\pi)| \widehat{O}_{\rho \tau \chi}(\lambda_1 n, \lambda_2 n, \lambda_3n)| N_{\iota} (p_1) \rangle  \nonumber \\[3mm]  
& = \delta(x_1 + x_2 + x_3 - 2\xi_u) \sum_{s.f.}\,(f_a)^{\alpha\beta\gamma}_{\iota} \, s_{\rho \tau, \chi} \, H^{\pi N}_{s.f.}(x_1, x_2, x_3, \phi, \Delta^2; \mu^2_{F})
\label{eqn:parameterization}
\end{align}
where $\mathcal{F}$ represents the Fourier transform; $P=p_1+p_\pi$ is the average $u$-channel momentum, and $\Delta=p_{\pi}-p_1$ is the $u$-channel momentum transfer, recall $\Delta^2 \equiv u$.  The spin-flavor ($s.f.$) sum over all independent flavor structure $(f_a)^{\alpha\beta\gamma}_\iota$ and Dirac structure $s_{\rho\tau,\chi}$ relevant at the leading twist; $\iota(a)$ is the nucleon (pion) isotopic index. The invariant transition amplitudes, $H^{\pi N}_{s.f.}$, which are often referred to as the leading twist $\pi N$ TDAs, are functions of the light-cone momentum fraction $x_i(i=1,2,3)$, the skewness variable $\xi_u$, the $u$-channel momentum-transfer squared $\Delta^2$, and the factorization scale $\mu_F$~\cite{lansberg11}. The full extended expression of $H^{\pi N}_{s.f.}$ can be found in Ref.~\cite{pire11}. Note that the cross section depends upon the squared modulus of defined amplitude in Eqn.~\ref{eqn:parameterization}.

In a simplified notation, $H^{\pi N} (x, \xi_u, \Delta^2)$ can be written in terms of invariant amplitudes $V^{\pi N}_{1,2}$, $A^{\pi N}_{1,2}$, $T^{\pi N}_{1,2,3,4}$ ~\cite{lansberg11,pire11},
\begin{equation}
H^{\pi N}_{s.f.} = \{ V^{\pi N}_{1,2}, A^{\pi N}_{1,2}, T^{\pi N}_{1,2,3,4}\}\,.
\end{equation}
Each invariant amplitude $V^{\pi N}_{1,2}$, $A^{\pi N}_{1,2}$, $T^{\pi N}_{1,2,3,4}$ is also a function of $x_i$, $\xi_u$ and $\Delta^2$. It is important to note that not all of the $\pi N$ TDA invariant amplitudes are independent~\cite{lansberg11}, and their relations are documented in Ref.~\cite{lansberg11}. 

Similar to early attempts in the GPD case~\cite{ralston02}, the most straightforward solution to determine a reasonable $\Delta^2$ dependence is to perform a factorized form of $\Delta^2$ dependence for quadruple distributions. Thus, the $\pi N$ factorized form of $\Delta^2$ dependence can be written as~\cite{lansberg11}:
\begin{equation}
H^{\pi N} (x, \xi_u, \Delta^2) = H^{\pi N} (x_i, \xi_u) \times G(\Delta^2),
\label{eqn:G-Delta}
\end{equation}
where $G(\Delta^2)$ is the $\pi N$ transition form factor of the three local quarks. Note that the determination of the $\Delta^2$ dependence and extraction of the $G(\Delta^2)$ form factor will be a distant goal for backward-angle physics.

\begin{figure}[htb]
\centering
\includegraphics[width=0.75\textwidth]{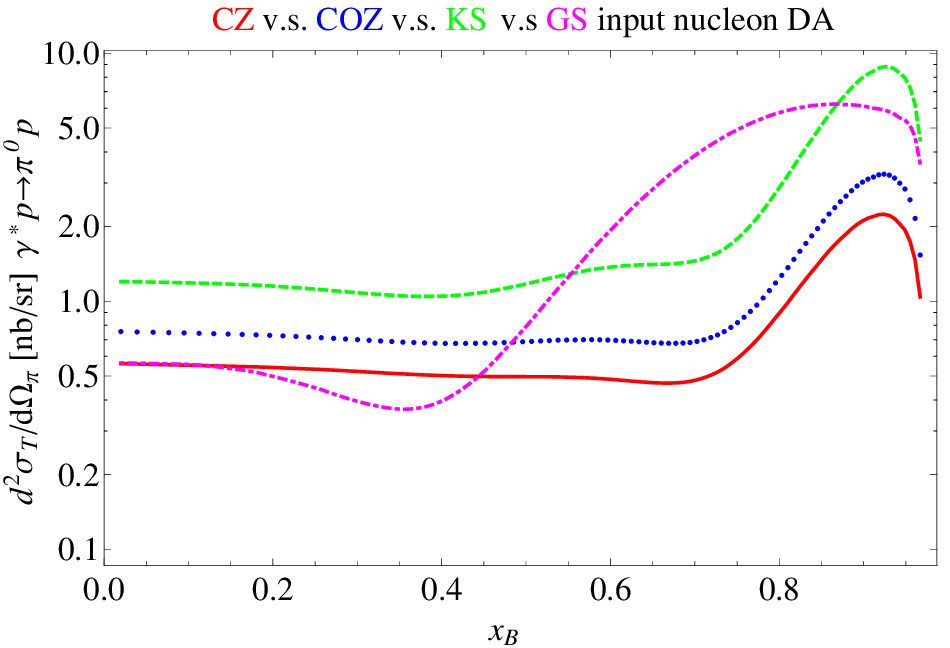} 
\caption{$d^2 \sigma_{T}/d\Omega_{\pi}$ for backward $\gamma^*p \rightarrow p\pi^0$ as a function of $x_{\textrm B}$ for $\pi N$ TDAs at $Q^2$ = 10~GeV$^2$, $u = -$0.5~GeV$^2$. CZ (solid line)~\cite{chernyak84}, COZ (dotted line) ~\cite{chernyak89}, KS (dashed line) ~\cite{king87} and GS (dash-dotted line) ~\cite{gari86} nucleon DAs were used as input. This plot was published in Ref.~\cite{lansberg11}}
\label{TDA_DA_cal}
\end{figure}

\begin{figure}[htb]
\centering
\includegraphics[width=0.32\textwidth]{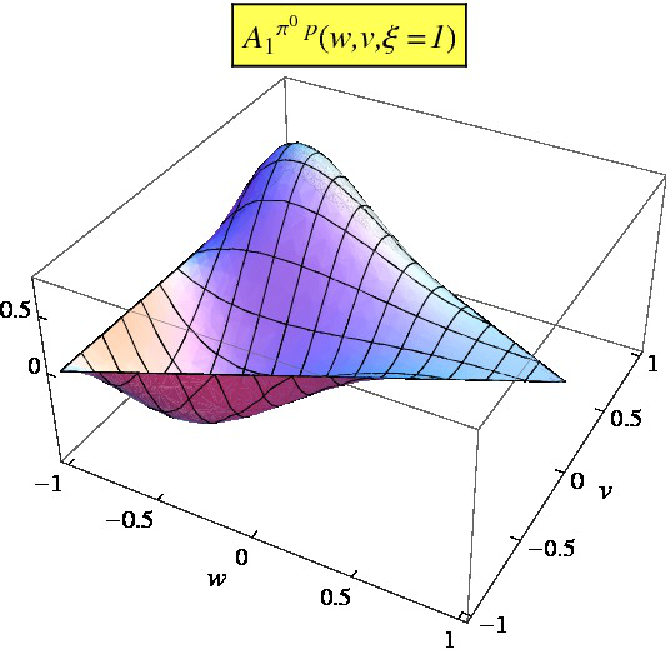}
\includegraphics[width=0.32\textwidth]{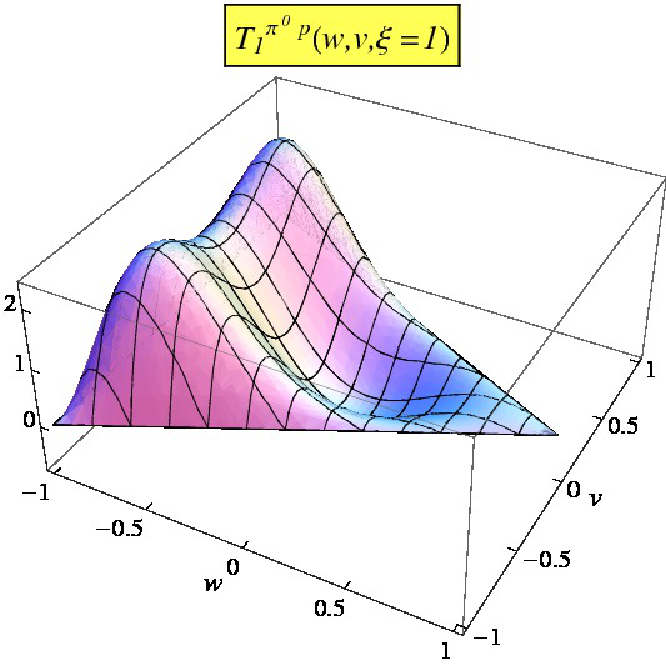}
\includegraphics[width=0.32\textwidth]{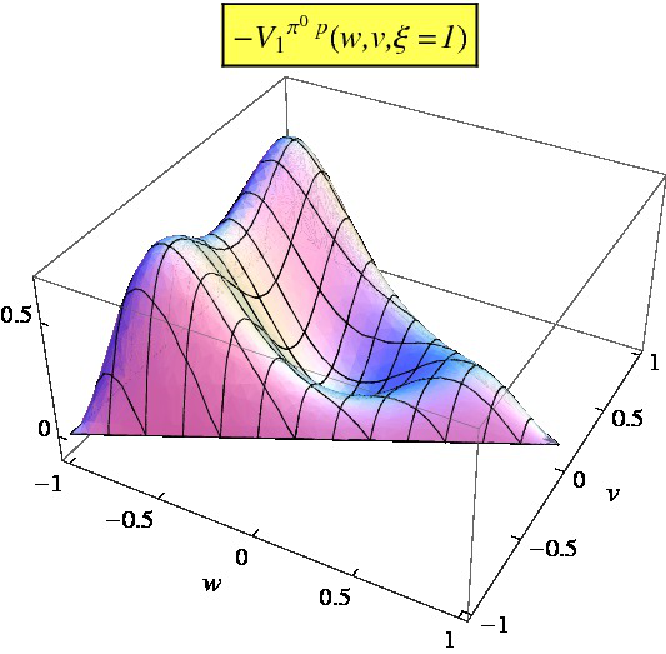}
\caption{$\pi^0p$ TDAs $V^{\pi^0 p}_1$, $A^{\pi^0 p}_1$ and $T^{\pi^0 p}_1$, computed as functions of quark-diquark coordinates, in the limit $\xi_u~\rightarrow 1$. CZ $N$ DAs are used as numerical input. These plots were published in Ref.~\cite{lansberg11}.}
\label{fig:TDA_shapes}
\end{figure}

In the $\xi_u=1$ limit, the $\pi^0 p$ TDAs: $V^{\pi^0p}_1$, $A^{\pi^0p}_1$, $T^{\pi^0p}_1$ can be simplified to the following combination of nucleon DAs~\cite{pire11}:
\begin{align}
V^{\pi^0p}_1 (x_1, x_2, x_3, \xi_u = 1) &= - \frac{1}{2} \times \frac{1}{4} \, V^P \left( \frac{x_1}{2}, \, \frac{x_2}{2}, \, \frac{x_3}{2} \right) \\[3mm]
A^{\pi^0p}_1 (x_1, x_2, x_3, \xi_u = 1) &= - \frac{1}{2} \times \frac{1}{4} \, A^P \left( \frac{x_1}{2}, \, \frac{x_2}{2}, \, \frac{x_3}{2} \right) \\[3mm]
T^{\pi^0p}_1 (x_1, x_2, x_3, \xi_u = 1) &= \frac{3}{2} \times \frac{1}{4}   \, T^P \left( \frac{x_1}{2}, \, \frac{x_2}{2}, \, \frac{x_3}{2} \right) .
\end{align}
A variety of nucleon ($N$) DAs, such as Chernyak-Zhitnitsky (CZ)~\cite{chernyak84}, Chernyak-Ogloblin-Zhitnitsky (COZ)~\cite{chernyak89}, King and Sachrajda (KS)~\cite{king87} and Gari and Stefanis (GS)~\cite{gari86} can be used as numerical input for $V^P$, $A^P$ and $T^P$, see their graphical representation in Fig.~\ref{fig:TDA_shapes}. A TDA calculation for $\pi^0$ production cross section versus $x_{\rm B}$ is shown in Fig.~\ref{TDA_DA_cal}, where all four $N$ DAs are used.

The $N$ DA model is an important part of the TDA model prediction, and depending on the choice of the $N$ DAs, the predicted experimental observables can change significantly. Therefore, improvements to the TDA parameterized formalism will rely on an accurate nucleon spectral distribution  by the $N$ DA models. In the same time, as more data are collected during JLab 12 GeV, a refined TDA model will help to discriminate between different $N$ DAs. This healthy iterative process can help improve our knowledge of proton structure~\cite{lansberg07}.

According to the TDA framework, the leading order (LO) backward angle $\gamma + p \rightarrow \pi^0 +p$ unpolarized cross section can be written as~\cite{lansberg07, lansberg11} 
\begin{equation}
\frac{d^2\sigma_{T}}{d \Omega_\pi} = |\mathcal{C}^2| \, \frac{1}{Q^6} \, \frac{\Lambda(s, m^2, M^2)}{128\,\pi^2 s (s-M^2)} \frac{1+\xi}{\xi} (|\mathcal{I}|^2 - \frac{\Delta^2_{T}}{M^2} |\mathcal{I}^\prime|^2).
\end{equation}
$\Lambda(s, m^2, M^2)$ is the Mandelstam function~\cite{lansberg11}, where $m$ corresponds to the meson mass and $M$ is the nucleon mass. In the backward-angle kinematics,
\begin{equation}
\Delta^2_T = \frac{(1-\xi) \left(\Delta^2 - 2\xi \left(\frac{M^2}{1+\xi} - \frac{m^2}{1-\xi}\right)\right) }{1+\xi}\,.
\end{equation}
The coefficients $\mathcal{I}$ and $\mathcal{I}^{\prime}$ are defined as~\cite{lansberg07}
\begin{equation}
\mathcal{I} = \int\left( 2\sum^7_{\alpha=1} T_{\alpha} + \sum^{14}_{\alpha=8}T_\alpha\right), ~~~ \mathcal{I}^{\prime} = \int\left( 2\sum^7_{\alpha=1} T^{\prime}_{\alpha} + \sum^{14}_{\alpha=8}T^{\prime}_\alpha\right) ,
\end{equation}
where the coefficients $T_{\alpha}$ and $T^{\prime}_\alpha (\alpha = 1, ..., 14)$ are functions of $x_i$, $y_j$, $\xi_u$ and $\Delta$. Here, $x_i$ and $y_j$ represent the momentum fractions for the initial and final state quarks. Also recall $\Delta^2 \equiv u$. Each of the components of $T_{\alpha}$ and $T^{\prime}$ represent one of the 21 diagrams contributing to the hard-scattering amplitudes (note that the last seven diagrams are the duplicates the of the first seven diagrams). 

Furthermore, $T_{\alpha}(\alpha = 1, ..., 14)$ can be written in terms of $V^{p\pi^0}_1$, $A^{p\pi^0}_1$, $T^{p\pi^0}_1$, $T^{p\pi^0}_4$ and $N$ DA ($V^p$, $A^p$, $T^p$);  $T^{\prime}_\alpha (\alpha = 1, ..., 14)$ can be written in terms of $V^{p\pi^0}_2$, $A^{p\pi^0}_2$, $T^{p\pi^0}_2$, $T^{p\pi^0}_3$ and $N$ DA~\cite{lansberg07}. This work has genuinely established the connection between the TDAs amplitudes to the cross section observables.

\subsubsection{Two Predictions from TDA Collinear Factorization}
\label{sec:tda_prediction}

The TDA collinear factorization has made two specific qualitative predictions regarding backward meson electroproduction, which can be verified experimentally~\cite{lansberg11, pire15, kirill15, pire18}:
\begin{itemize}
\item The dominance of the transverse polarization of the virtual-photon results in the suppression of the $\sigma_{\rm L}$ cross section by a least ($1/Q^2$): $\sigma_{\rm L}/\sigma_{\rm T}$ $< 1/Q^2$, 
\item The characteristic $1/Q^8$-scaling behavior of the transverse cross section for fixed $x_{\rm B}$ (or at fixed $\xi_{u}$), following the quark counting rules. 
\end{itemize}
The goal of the proposed $\pi^0$ measurement is to challenge these predictions. In addition, the $-u$ dependence of the separated experimental cross section will provide insight for the extraction of the $\pi N$ transition form factor $G(\Delta^2)$ (from Eqn.~\ref{eqn:G-Delta}),

\subsection{Complementary Objective: the Hadronic Approach}
\label{regge}

\begin{figure}[htb]
\centering
\includegraphics[width=0.425\textwidth]{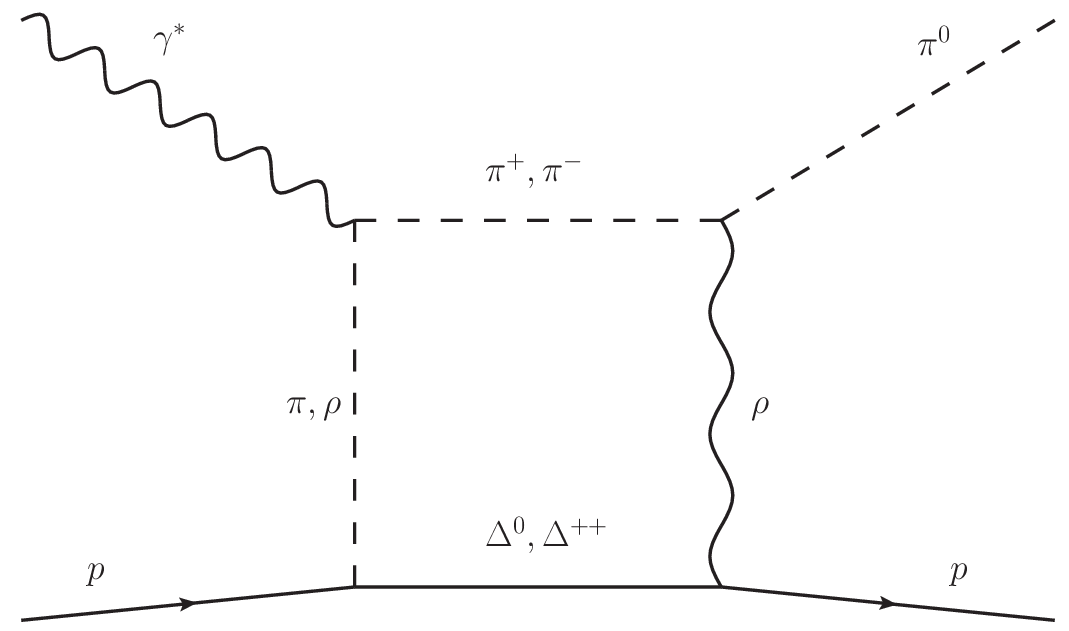}
\includegraphics[width=0.425\textwidth]{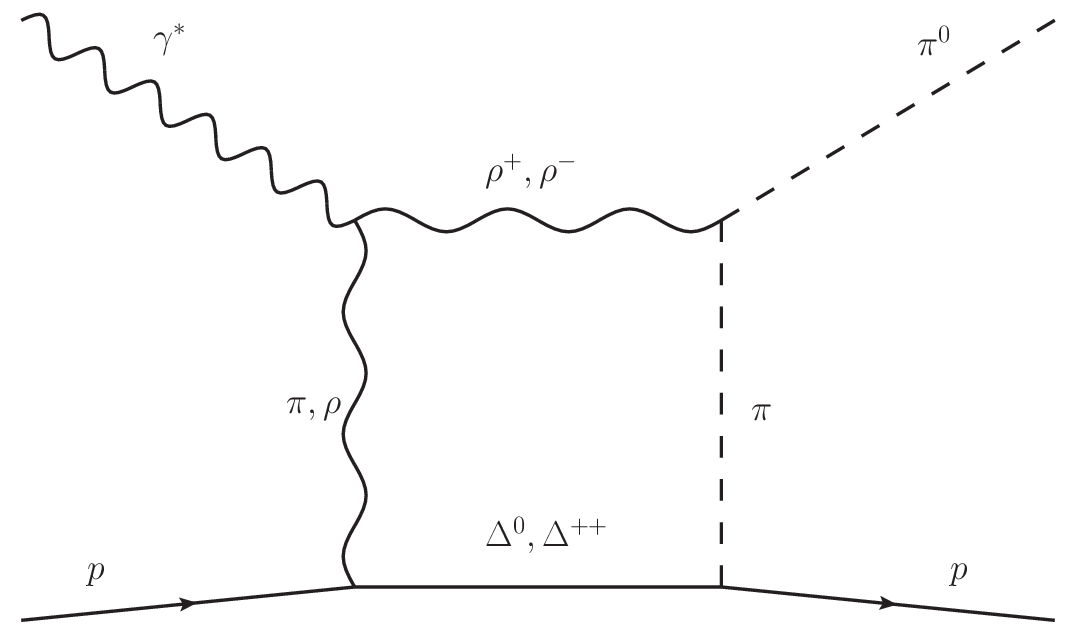}
\caption{Examples of meson exchange diagrams which contribute to forward-angle $\pi^0$ production. The left plot is an example of charged $\pi$ rescattering~\cite{laget11}; the right plot is an example vector meson contribution. These plots were created based on the original ones published in Ref.~\cite{laget18}.}
\label{fig:JML_front}
~\\[4mm]
\includegraphics[width=0.425\textwidth]{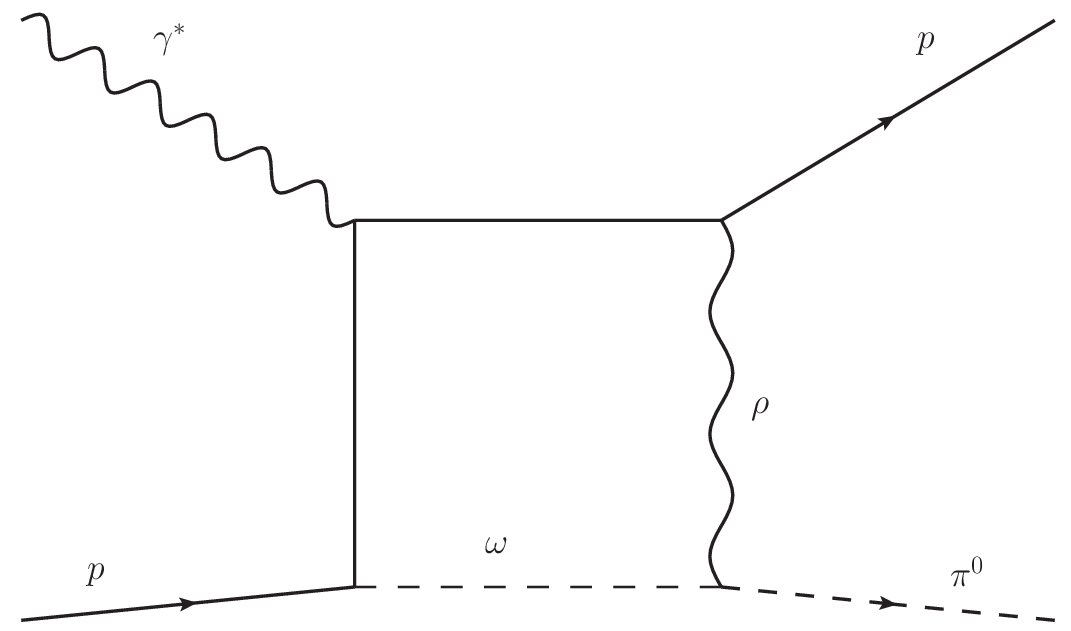}
\caption{Example of a possible meson exchange diagram which contributes to backward-angle $\pi^0$ production. This plot was created based on the original one published in Ref.~\cite{laget18}.}
\label{fig:backward_pi}
\end{figure}

The development of Regge-trajectory-based models has created a useful linkage between physics kinematic quantities and experimental observables.  Experimental observables in the JLab physics regime are often parameterized in terms of $W$, $x_{\rm B}$, $Q^2$ and $t$. By varying a particular parameter while fixing others, one can perform high precision studies to investigate the isolated dependence of the varied parameter for a given interaction.

In the Regge models, the exchange of high-spin, high-mass particles is normally taken into account by replacing the pole-like Feynman
propagator of a single particle (i.e. $\frac{1}{t-M^2}$ with the Regge (trajectory) propagator). Meanwhile, the exchange process involves a series of particles of the same quantum number (following the same Regge trajectory $\alpha(t)$), instead of single particle exchange~\cite{regge60, chew62}.
In the forward-angle $\pi^0$ electroproduction study~\cite{laget11}, J. M. Laget linked the elastic $\pi^0$ cross section to the scattering channels of $\omega p$, $\rho^+n$, $\rho^-\Delta^{++}$, diagrams shown in Fig.~\ref{fig:JML_front}. This treatment significantly improved the predictive power of the hadronic Regge-based model and led to a good agreement with the data~\cite{laget11}.

\begin{figure}[htb]
\centering
\includegraphics[width=0.75\textwidth]{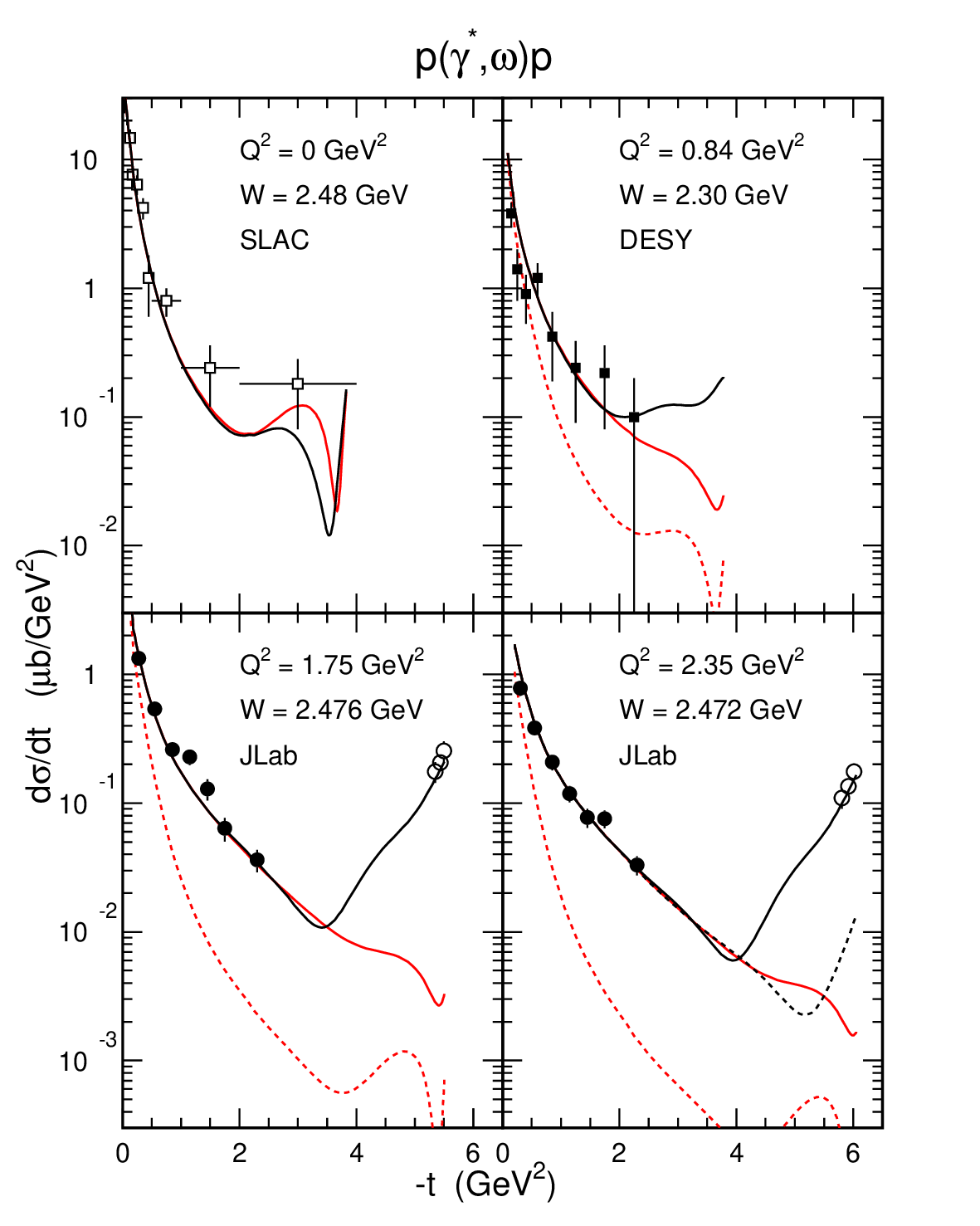}
\caption{The cross section evolution for exclusive $\omega$ meson production as a function of $-t$. Note that the bottom panels show $\omega$ electroproduction data from CLAS (dots) and Hall C (circles), the $Q^2 = 1.75$ GeV$^2$ on the left and $Q^2 = 2.35$ GeV$^2$ on the right. The dashed red curves are the predictions of the basic model when a constant cutoff mass is used in the meson electric and magnetic form factors~\cite{laget04}. The full red curves are the predictions when a $t$-dependent cutoff mass was used~\cite{laget04}. The black dashed line is the prediction of the nucleon degenerated pole only. The full line curves take into account the interference between nucleon exchange and the $\omega$ produced via nucleon exchange re-scattering on the nucleon. Plot provided by J. M. Laget through private communication~\cite{laget18}.} 
\label{laget} 
\end{figure}

Recently~\cite{laget18}, J. M. Laget indicated that the hadronic Regge-based model is capable of describing the data trend of the backward $\omega$ cross section (shown in Figs.~\ref{fig:omega},\ref{laget}), at $Q^2=1.75$ and $2.35$~GeV$^2$. The preliminary conclusion from this study was that the nucleon pole contribution (baryon exchange) alone is not enough to account for the measured cross section~\cite{laget11}.  Backward $\omega$ production requires $\rho^0$, $\rho^+n$, $\rho\Delta$ scattering channels, in addition to the nucleon pole amplitude~\cite{laget18}. Note, this approach is very similar to that used for the $\pi^0$ forward-angle study. For reference purpose, a possible $u$-channel baryon trajectory exchange diagram for $\pi^0$ production is shown in Fig.~\ref{fig:backward_pi}, and this diagram is based on the knowledge of forward-angle $\pi^0$ production (shown in Fig.~\ref{fig:JML_front}). Currently, a publication is in preparation which will contain more findings of the backward $\omega$ cross section using the hadronic Regge-based model~\cite{laget18}.

Due to a lack of systematic studies, currently available backward-angle physics data above the resonance region (most of them are summarized in Sec.~\ref{sec:exp_summary}) have limited coverage in terms of $W$, $Q^2$ and $t$ (or $u$) and therefore cannot support a full phenomenological study. However, the $u$-channel Regge-exchange study is still a useful tool to verify the key knowledge gained from the forward-angle physics program, i.e. to map out the full $-t$ evolution and extract the backward-angle slope for a given meson production process, such as the example shown in Fig.~\ref{fig:omega}. Note that the chosen kinematic setting in the proposal is made based on the existing and proposed forward-angle $\pi^0$ measurements~\cite{E12-13-010,defurne16}, i.e. $Q^2=2.0$, 3.0 and $4.0$ GeV$^2$ at fixed $x_{\rm B}=$0.36.

\section{Studying TDA through VCS and DEMP from JLab 12 GeV to EIC}
\label{sec:TDA_study_program}

With caution in mind, the $1/Q^8$ scaling behavior observed in the $\pi^+$ production (Sec.~\ref{sec:clas}) and the separated cross section ratio have shown an indication that $\sigma_T \gg \sigma_L$ for the exclusive $\omega$ electroproduction channel (Sec:~\ref{sec:omega}), which could be considered as initial evidence needed to demonstrate the validity of the TDA factorization approach in backward-angle kinematics.

\begin{table}[bt]
\centering

\setlength{\tabcolsep}{0.9em}
\caption{Status table showing the progress of TDA validation in Jefferson Lab 12 GeV. $\bigcirc$: this proposal; $\bigtriangleup$: in the early planning stage; \checkmark: parasitic data may be available to perform study; \checkmark\checkmark: confirmed by existing data.}
\label{tab:status}
\begin{tabular}{lcc}
\hline
                & $\sigma_{\rm T} > \sigma_{\rm L}$  & $1/Q^8$ Scaling  \\ \hline
$\pi^{0}$       & $\bigcirc$                         &  $\bigcirc$		\\ 
$\pi^{+}$       &                                    & \checkmark \checkmark            		\\ 
$\pi^{-}$       &                                    &            		\\ 
$K^{0}$         &                                    &            		\\ 
$K^{\pm}$       &                                    &            		\\ 
$\eta$          &  \checkmark                        &  \checkmark		\\
$\rho$          &                                    &  				\\
$\omega$        & \checkmark \checkmark              &  \checkmark	    \\
$\eta^{\prime}$ & \checkmark                         & \checkmark 		\\
$\phi$          & \checkmark                         & \checkmark  		\\ 
VCS             & $\bigtriangleup$                   & $\bigtriangleup$ \\ \hline
\end{tabular}
\end{table}

\begin{figure}[htb]
\centering
\includegraphics[width=0.85\textwidth]{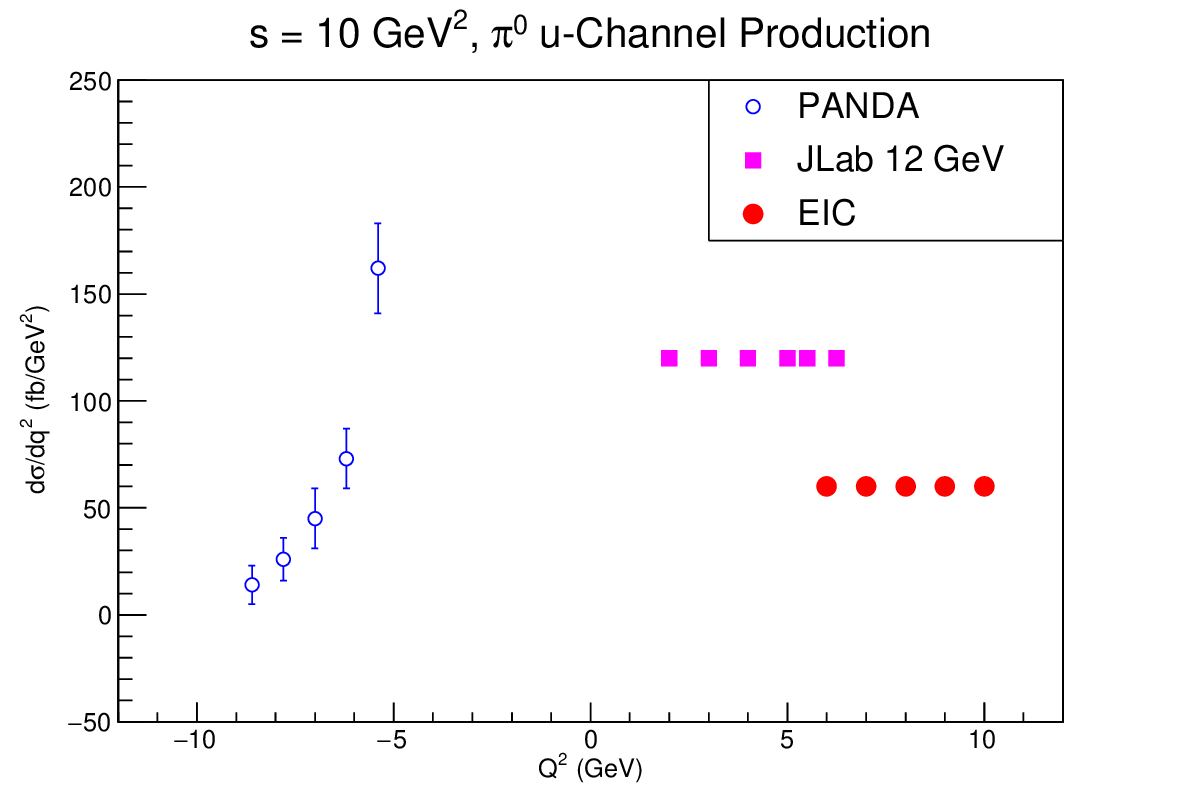}
\caption{Projected $Q^2$ evolution of $u$-channel $\pi^0$ electroproduction measurement combining coverage from projected data from Panda (blue circle), JLab 12 GeV from this proposal (magenta square) and future EIC (red circle). Note that the projected PANDA measurements is given in $d\sigma/dq^2$ at $u^\prime \sim 0$ GeV$^2$ with the assumption of $\sigma_T \gg \sigma_L$. The cross sections from this proposal (JLab 12 GeV) and EIC will be provided in $d\sigma/dt$, which can be easily converted into $d\sigma/dq^2$.  In addition, one also would have to apply a $W$ and $x_B$ correction before forming the combined plot.}
\label{fig:panda_jlab_eic}
\end{figure}
These initial successes of the TDA framework raise important and urgent questions: 
\begin{enumerate}

\item{Would the dominance of $\sigma_T$ be observed in other $u$-channel exclusive meson production channels (e.g. $\pi^0$,  $\rho$, $\eta$, $\eta^\prime$ and $u$-channel $\phi$), and  VCS?}
\item{What is the applicability region (in $Q^2$ and $W$) for a perturbative QCD description in the backward-angle region?}
\end{enumerate}
These important questions need to be answered by 12 GeV measurements. At the current stage, we are in a process of establishing a coherent and comprehensive program (with significant theory insights) that  will prioritize the backward-angle observable and lay down a path to continue studying $u$-channel physics in the future EIC. Note that an important part of this program is to further support the development of the TDA. It is our pleasure to inform the PAC that the 2020 JSA postdoc prize was awarded to one of the authors (Wenliang Li) of this proposal to develop the described $u$-channel physics program.

Tentatively, we envision a program to study TDAs systematically that consists of three stages:
\begin{description}

\item[Stage 0:] Continuing to demonstrate the existence of $u$-channel signals and studying the ``soft-hard'' transition with increasing $Q^2$.

\item[Stage 1:] Measuring the $1/Q^n$ scaling trend of $\sigma_T$ and attempting to extract $\sigma_T\gg\sigma_L$ for all single-meson production channels and VCS.

\item[Stage 2:] Extraction of TDAs by probing the single and double spin asymmetries for backward-angle meson productions. This step implies the determination of the $\Delta^2$ dependence and the $\pi N$ transition form factor $G(\Delta^2)$ defined in Eqn.~\ref{eqn:G-Delta}.

\end{description}

Stage 0 has been underway for some time, but as shown in Table~\ref{tab:status} there are still unexplored single-meson channels.  We are in the early period of the Stage 1, which includes this proposal and will likely continue through the entire 12 GeV era. Only after most of the tasks in Table \ref{tab:status} are completed with corresponding measurements can we provide answers to the important questions raised earlier.  

\subsection*{Special Role of $\pi^0$ Electroproduction}

In comparison to the $u$-channel $\omega$ or $\eta$ electroproduction processes, the reconstructed missing mass distribution for $\pi^0$ has little physics background underneath its narrow peak. This significantly reduces the complication associated with the background removal during the analysis. In addition, $\pi^0$ production has been a popular candidate for theoretical studies~\cite{laget04, lansberg07}.  All these features make it a prime choice to initiate backward-angle studies in the JLab 12~GeV era. In addition, backward $\pi^0$ production has received significant interest beyond the JLab physics program and will be studied by the $\overline{\rm P}$ANDA experiment at FAIR~\cite{panda15} through the complementary process $\overline{p}+p\rightarrow\gamma^*+\pi^0$. See Fig.~\ref{fig:panda_jlab_eic} for a $Q^2$ evolution of the $\pi^0$ cross section after combining projected data coverage from $\overline{\rm P}$ANDA, JLab 12 GeV (this proposal) and future EIC measurements. This combined $-10<Q^2<10$ GeV$^2$ range would offer a unique opportunity to challenge the universality of the TDA. 

In a recent publication, L. Szymanowski, B. Pire and K. Semonov-Tian-Shansky laid out the path for TDA from JLab 12 GeV to the EIC~\cite{szymanowski19}. Through private communication, experts unanimously agreed the electroproduction of $\pi^0$ is an ideal candidate to initiate a 12 GeV to EIC transition study.
In parallel to this proposal, the feasibility study of probing the $u$-channel $\pi^0$ process at kinematics shown in Fig.~\ref{fig:panda_jlab_eic}, has begun. The preliminary result of the study will be included in the upcoming EIC Yellow Report (YR) as one of the benchmark observables. Recently, one of the authors from this proposal (Wenliang Li) was awarded with the EIC fellowship, which will significantly accelerate the completion of the $\pi^0$ feasibility study and corresponding section of the YR.

In a broader scope, $u$-channel electroproduction is only one aspect of probing nucleon structure through backward-angle observables.  The diversified experimental programs and equipment from JLab 12 GeV offer other exciting $u$-channel physics opportunities, such as the backward-angle vector meson production and hyperon production at GlueX. The authors of this proposal are excited to inform the PAC that the first backward-angle physics focused workshop is taking place at JLab in September, 2020. One major objective is to offer a platform to connect scattered experimental and theoretical efforts together, thus, potentially forming small backward-angle physics working groups. See Appx.~\ref{app:workshop} for the full
objectives of the workshop, topics of discussion and participants list.  

Additionally, the inclusion of $u$-channel exclusive reactions in the July 14 mini-workshop on ``Physics Opportunities for Large Angle Production with CLAS'' \cite{stoler2020} indicates the growing interest in the physics opportunities available in this regime.
 
\section{Experiment Kinematics and Configuration}
\label{sec:exp_kin}

\begin{figure}[htb]
\centering
\includegraphics[width=0.6\textwidth]{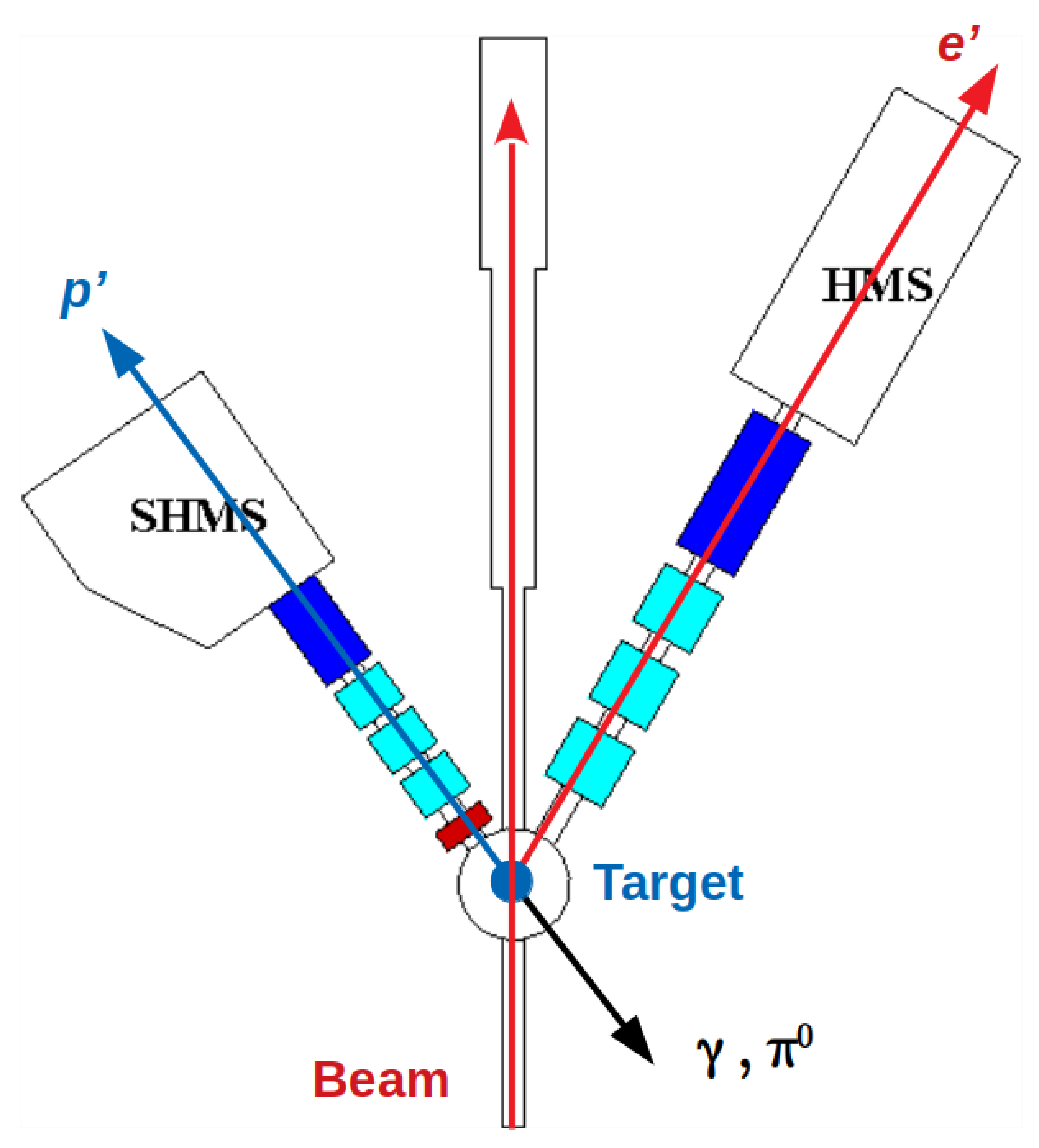}
\caption{Experimental configuration for $^1$H$(e^\prime, ep)\pi^0$ with the standard Hall C equipment. SHMS and HMS are located on the left and right side of the beam line, respectively.}
\label{fig:setup}
\end{figure}

The exclusive backward-angle $\pi^0$ electroproduction measurement is proposed to use the standard Hall~C equipment: SHMS and HMS in coincidence mode, the standard-gradient electron beam and the liquid hydrogen (LH$_2$) target. For most of settings, SHMS will be used to detect the forward going (fast) proton and HMS will be used to detect the scattered electron. The $\pi^0$ events will be selected by using the missing mass reconstruction technique. A schematic diagram of the experimental configuration for the $^1$H$(e, e^{\prime}p)\pi^0$ is shown in Fig.~\ref{fig:setup}.

\subsection{L/T/LT/TT Separation}

In the one-photon-exchange approximation, the $^1$H$(e, e^{\prime}p)X$ cross section of the $\pi^0$ and other meson production interactions ($X=\pi^\pm$, $\rho^{0}$, $\omega$, $\phi$, 2$\pi$, $\eta$ and $\eta^{\prime}$) can be written as the contraction of a lepton tensor $L_{\mu \nu}$ and a hadron tensor $W_{\mu \nu}$:
\begin{equation}
\frac{d^6\sigma}{d\Omega_{e^{\prime}}\,dE_{e^{\prime}}\,d\Omega_p\,dE_p} = |p_p| \, E_{p} \, \frac{\alpha^2}{Q^4} \frac{E_{e^{\prime}}}{E_e} \, L_{\mu\nu}  \, W^{\mu\nu}\,,
\label{Xsection_6f}
\end{equation}
where the $L_{\mu\nu}$ can be calculated exactly in QED, and the explicit structure of the $W^{\mu\nu}$ is yet to be determined. Since the final states are over constrained (either detected or can be reconstructed), as in the case of the $^1$H$(e, e^{\prime}p)\pi^0$ reaction, the cross section can be reduced further to a five-fold differential form:
\begin{equation}
\frac{d^5\sigma}{dE^{\prime} d \Omega_{e^{\prime}} d \Omega_{p}^{*} } = \Gamma_v \, \frac{d^2\sigma}{d\Omega_{p}^*}\,,
\label{Xsection_5f}
\end{equation}
where the asterisks denote quantities in the center-of-mass frame of the virtual-photon-nucleon system; $\Gamma_V$ is the virtual-photon flux factor:
$$
\Gamma_v = \frac{\alpha}{2\pi^2} \frac{E_{e^{\prime}}}{E_e} \frac{q_{L}}{Q^2} \frac{1}{(1-\epsilon)}\,,
$$
where $\alpha$ is the fine structure constant, the factor $q_L = (W^2 - m_p^2)/(2M_p)$ is the equivalent real-photon energy, which is the laboratory energy a real photon would need to produce a system with invariant mass $W$; and $\epsilon$ is the polarization of the virtual-photon which is defined as
$$
\epsilon = \left( 1 + \frac{2 |{\rm q}|^2 }{Q^2} \tan ^ 2 \frac{\theta_e}{2} \right)^{-1}\,.
$$
The two-fold differential cross section (Eqn.~\ref{Xsection_5f}) can be written in terms of an invariant cross section:
\begin{equation}
\frac{d^2\sigma}{d\Omega^{*}_\omega} = \frac{d^2\sigma}{dt~d\phi} \cdot \frac{dt}{d\cos\theta^*},
\end{equation}
where $$\frac{dt}{d\cos\theta^*} = 2|p^*||q^*|$$ is the Jacobian factor, and $p^*$ and $q^*$ are the three momentum of the proton and the virtual-photon in the CM frame.

The general form of two-fold differential cross section can be expressed in terms of the structure functions as: 
\begin{equation}
2 \pi \frac{d^2 \sigma}{dt ~ d\phi} = \frac{d \sigma_{\rm T}}{dt} + 
\epsilon ~ \frac{d \sigma_{\rm L}}{dt} + \sqrt{2\epsilon(1+\epsilon)}~ 
\frac{d\sigma_{\rm LT}}{dt} \cos \phi + \epsilon ~ 
\frac{d\sigma_{\rm TT}}{dt} \cos 2\phi \,.
\label{eqn:xsection_LT}
\end{equation}

\begin{figure}[htb]
\centering
\includegraphics[width=0.9\textwidth]{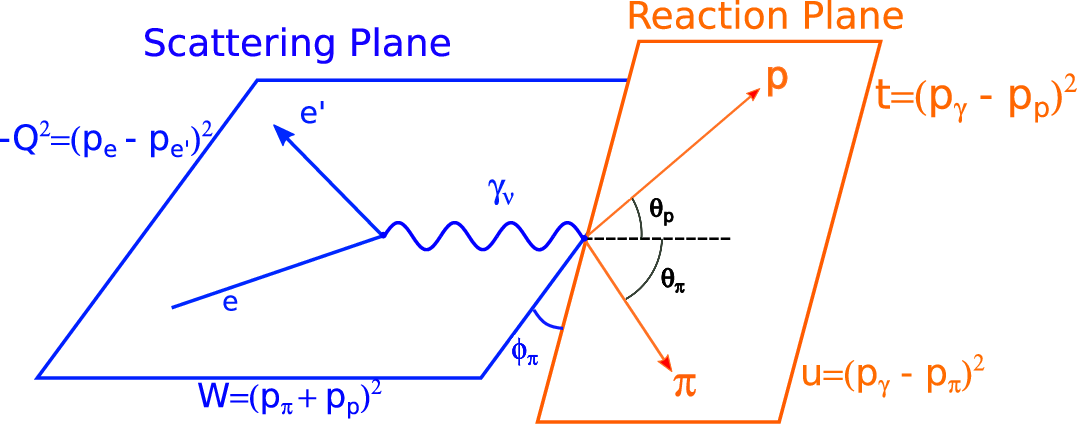}
\caption{The scattering and reaction planes for $\pi^0$ electroproduction: $^1$H$(e, e^{\prime}p)\pi^0$.  The scattering plane is shown in blue and the reaction plane is shown in orange. Note that the forward-going proton after the interaction is labelled $p$; $\gamma_\nu$ represents the exchanged virtual-photon and its direction defines the $q$-vector; $\phi_p$ ($ \phi_p = \phi_{\pi} + 180^\circ$) is defined as the angle between the scattering and reaction planes (the azimuthal angle around the $q$-vector); $\theta_p$ and $\theta_{\pi}$ denote the scattering angles of the $p$ and $\pi$ with respect to the $q$-vector, respectively. The definition of the Lorentz invariant variables such as $W$, $Q^2$, $t$ and $u$ are also shown.}
\label{fig:planes}
\end{figure}

A schematic diagram of the exclusive $\pi^0$ electroproduction reaction, $^1$H$(e, e^{\prime}p)\pi^0$, giving the definition of the kinematic variables in Eqn.~\ref{eqn:xsection_LT} is shown in Fig.~\ref{fig:planes}. The three-momentum vectors of the incoming and the scattered electrons are denoted as $\vec{p}_e$ and $\vec{p}_{e^{\prime}}$, respectively. Together they define the scattering plane, which is shown as a blue box. The corresponding four momenta are p$_e$ and p$_e^\prime$. The electron scattering angle in the lab frame is labelled as $\theta_e$. The transferred four-momentum vector $q$($\nu,\vec{q}$) is defined as (p$_e -$p$_{e^\prime}$).  The three-momentum vectors of the recoil proton target ($\vec{p}_p$ ) and produced $\pi^0$ ($\vec{p}_{\pi}$) define the reaction plane, is shown as the orange box. The azimuthal angle between the scattering plane and the reaction plane is denoted by the recoil proton angle $\phi_p$. From the perspective of standing at the Hall C beam entrance and looking downstream of the spectrometer, the forward going proton angle $\phi_p=0$ points to horizontal left of the $q$-vector, and it follows a counterclockwise rotation. The lab frame scattering angles between $\vec{p}_p$ (or $\vec{p}_\pi$) and $\vec{q}$ are labeled $\theta_p$ (or $\theta_\pi$). Unless otherwise specified, the symbols $\theta$ and $\phi$ without subscript are equivalent to $\theta_p$ and $\phi_p$, since the recoil protons will be detected during the experiment. The parallel and antiparallel kinematics are unique circumstances, and occur at $\theta = 0^\circ$ and $\theta = 180^\circ$, respectively.

The Rosenbluth separation, also known as the longitudinal/transverse (L/T) separation, is a unique method of isolating the longitudinal component of the differential cross section from the transverse component. The method requires at least two separate measurements with different experimental configurations, such as the spectrometer angles and electron beam energy, while fixing the Lorentz invariant kinematic parameters such as $x_{\rm B}$ and $Q^2$. The only physical parameter that is different between the two measurements is $\epsilon=\left(1+2\frac{|\vec{q}|^2}{Q^2}\tan^2\frac{\theta}{2}\right)^{-1}$, which is directly dependent upon the incoming electron beam energy ($E_e$) and the scattering angle of the outgoing electron.

Even though the SHMS setting at $\theta=0$ (or $\theta_{pq}=0$ for clarity) is centered with respect to the $q$-vector, corresponding to the parallel scenario for the proton (anti-parallel for $\pi$), the spectrometer acceptance of the SHMS (proton arm) is not wide enough to provide uniform coverage in $\phi$ (black events in Fig.~\ref{fig:bullseye}). A complete $\phi$ coverage over a full $u$ range is critical for the extraction of the interference terms (LT and TT) during the L/T separation procedure. To ensure an optimal $\phi$ coverage, additional measurements are required at the $\theta=\pm3^\circ$ SHMS angles (blue and red events). Constrained by the minimum SHMS angle from the beam line of $\theta_{\rm SHMS} = 5.5^\circ$, the lower $\epsilon$ measurement is only possible at two angles at some $Q^2$. However, this can be compensated by the full $\phi$ coverage at the higher $\epsilon$ measurement and the simulated distribution, thus determining the interference components (LT and TT) of the differential cross section. 

The last step of the L/T separation is to fit the experimental cross section versus $\phi$ for a given $u$ bin. The lower and higher epsilon data will be fitted simultaneously using Eqn.~\ref{eqn:xsection_LT} to ensure successful extraction of the $\sigma_{\rm T, L, LT, TT}$. The common offset between and difference between the lower and higher $\epsilon$ data set give raise to the $\sigma_{\rm T}$ and $\sigma_{\rm L}$; whereas the $\phi$ dependence signifies the $\sigma_{\rm LT}$ and $\sigma_{\rm TT}$ contribution.

\subsection{Choice of Kinematics}

The $^1$H$(e, e^{\prime}p)\pi^0$ experimental yield will be measured at $Q^2=2.0$, $3.0$, $4.0$, $5.0$ and $6.25$ GeV$^2$, at common Bjorken $x_B=0.36$. We intend to perform L/T/LT/TT separations for all except the $Q^2=6.25$ GeV$^2$ setting. One additional L/T separation study at $Q^2=2$~GeV$^2$, $W=3$ GeV will provide $W$ scaling information needed to achieve the projected experimental objective. See a summary table that includes relevant kinematics variables and spectrometer settings in Table \ref{tab:kinematics}.

\begin{figure}[htb]
\centering
\includegraphics[width=0.8\textwidth]{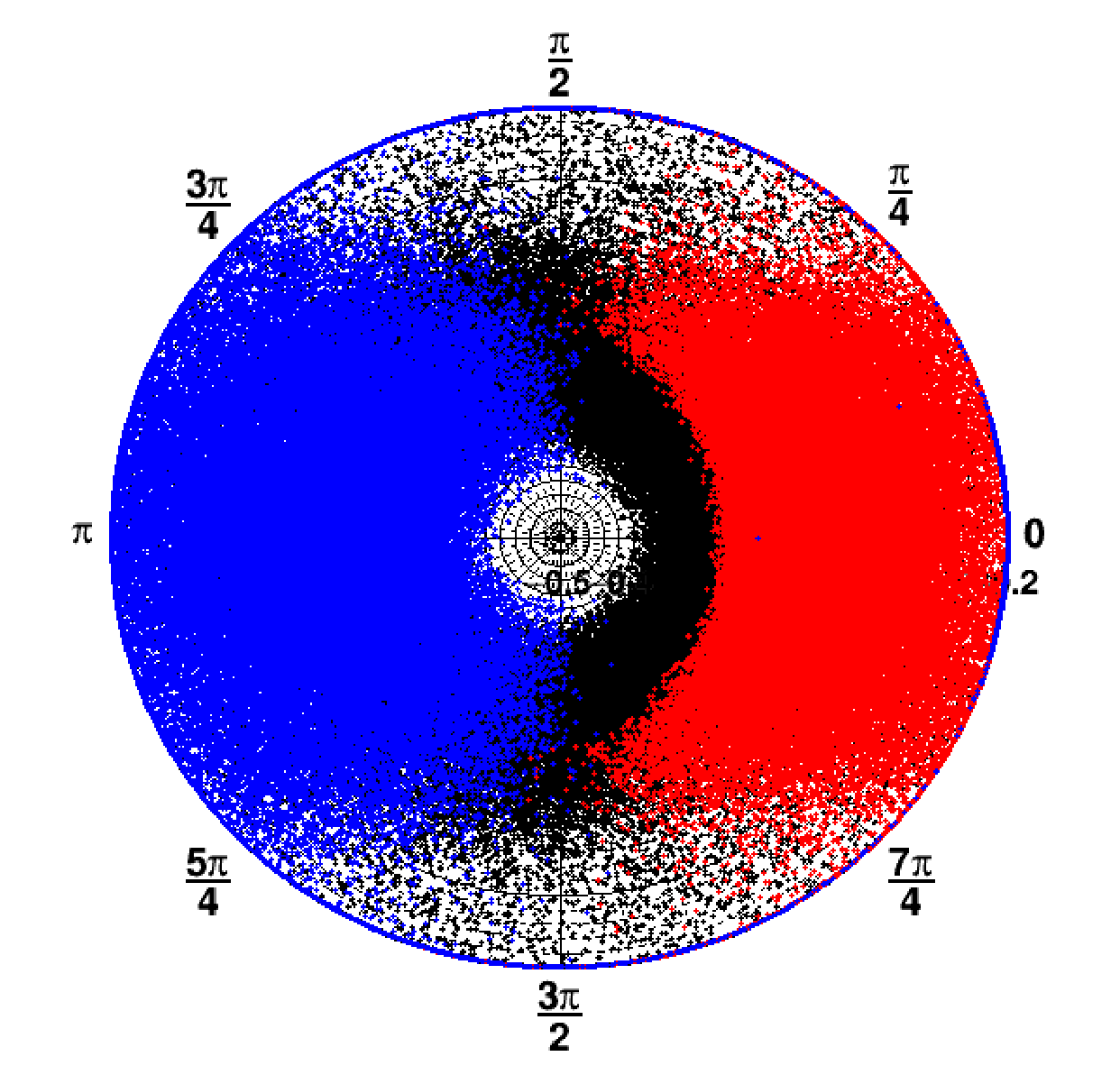}
\caption{$u'$-$\phi$ polar distributions at $Q^2 = 2$ GeV$^2$ and $\epsilon=0.52$. $(-u+0.5)$ is plotted as the radial variable and $\phi$ as the angular variable. The blue points represent data at $\theta_{pq}=+3^\circ$, black points represent data at $\theta_{pq} = 0^\circ$, and red data points represent data at $\theta_{pq} = -3^\circ$. The center of the plot represents $u=+0.5$~GeV$^2$ and the outer circle is at $u=-0.2$~GeV$^2$.}
\label{fig:bullseye}
\end{figure}

\begin{table}[htb]
\centering
%\setlength{\abovetopsep}{1ex}
%\small
\setlength{\tabcolsep}{0.35em}
\caption{Proposed kinematics for the $^1$H$(e, e^\prime p)\pi^0$ measurement. Note that the $W$ and $Q^2$ are the same for the $^1$H$(e, e^\prime p\gamma)$ reaction. For most of the settings, HMS will detect the scattered electron and the SHMS will detect the recoiled proton. For $Q^2=2$ GeV (indicated by $^*$), the SHMS will detect the electron and the HMS will detect the proton because the scattered electron momentum and angle at high $\epsilon$ are too high and too far forward for the HMS. For all settings, $u^\prime=0$ GeV$^2$. Note that at $Q^2=3.0$, and 4.0~GeV$^2$, E12-13-010 will provide the $x_B$=0.36 L/T separated cross section at $t^{\prime} \sim 0$~\cite{E12-13-010}, while $Q^2=2.0$ GeV$^2$ $x_B=$0.36 data are from Ref.~\cite{defurne16}. }
\label{tab:kinematics}

\begin{tabular}{cccccccccccc}
\hline
$Q^2$   & $W$  & $x_{\rm B}$  &  $E_{\rm Beam}$ & $\epsilon$ & $\theta_{\rm HMS}$  & $P_{\rm HMS}$  & $\theta_{\rm SHMS}$  & $P_{\rm SHMS}$ & $\theta_{pq}$  & $-t$      \\
(GeV$^2)$ & (GeV)  &      &  (GeV)        &            & (deg)                     & (GeV/c)        & (deg)                & (GeV/c)        & (deg)          & (GeV$^2$) \\ 
\hline
%2      & 2.11 & 0.36 & 4.4          & 0.52       & 32.60     & $-$1.44   &
%$-$13.71  %     & 3.51       & 0.0    &  0.0 & 5.05    \\
2.0   & 2.11  & 0.36 & 4.4$^*$  & 0.52$^*$ & 13.71$^*$ & 3.51$^*$   & $-$32.60$^*$ & $-$1.44$^*$  & $-$3.0, 0, $+$3.0   & 5.05$^*$ \\
      &       &      & 10.9$^*$ & 0.94$^*$ & 21.54$^*$ & 3.51$^*$   & $-$8.72$^*$  & $-$7.94$^*$  & $-$3.0, 0, $+$3.0   & 5.05$^*$ \\ \hline
2.0   & 3.00  & 0.20 &  6.60    & 0.32     & 29.01     & $-$1.21    & $-$6.03      &  5.90        & $-$3.0, 0 & 9.45     \\
      &       &      & 10.90    & 0.79     & 10.47     & $-$5.51    & $-$10.34     &  5.90        & $-$2.84, 0, $+$3.0   & 9.45     \\ \hline
3.0   & 2.49  & 0.36 &  6.60    & 0.54     & 26.50     & $-$2.17    & $-$11.70     &  5.00        & $-$3.0, 0, $+$3.0   & 7.79     \\
      &       &      & 10.90    & 0.86     & 11.80     & $-$4.37    & $-$16.20     &  5.00        & $-$3.0, 0, $+$3.0   & 7.79     \\ \hline
4.0   & 2.83  & 0.36 &  8.80    & 0.56     & 22.89     & $-$2.89    & $-$10.35     &  6.50   	  & $-$3.0, 0, $+$3.0   & 10.56    \\
      &       &      &  10.90   & 0.73     & 15.59     & $-$4.99    & $-$12.39     &  6.50   	  & $-$3.0, 0, $+$3.0   & 10.56    \\ \hline
5.0   & 3.13  & 0.36 &  8.80    & 0.26     & 37.36     & $-$1.38    & $-$6.23      &  8.00   	  & $-$3.0, 0 & 13.37    \\
      &       &      &  10.90   & 0.55     & 20.90     & $-$3.48    & $-$9.24      &  8.00   	  & $-$3.0, 0, $+$3.0  & 13.37    \\ \hline
6.25  & 3.46  & 0.36 &  10.90   & 0.27     & 34.18     & $-$1.66    & $-$5.59      &  9.84   	  & 0             & 16.78    \\
\hline
%\midrule
%6.25   & 3.46  & 0.36 &  10.90        & 0.27       & 34.18    &  $-$1.66   & $-$5.59    &  9.84  & 0.0  &  0.0 & 16.78       \\
%5.5  & 3.26  & 0.36 &  8.80      & 0.11       & 57.50    &  $-$2.78   & $-$3.86 %  &  8.72   & 0.0 &  0.0 & 14.69       \\
%5.5   & 3.26  & 0.36 &  10.90   & 0.45     & 24.62     &  $-$2.78   & $-$7.86      &  8.72   	  & 0.0     &  0.0     & 14.69    \\
\toprule
\end{tabular}

\end{table}

Using the backward $^1$H$(e, e^\prime p)\pi^0$ and $^1$H$(e, e^\prime p)\gamma$ physics models (see Appx.~\ref{sec:pi0_MC} for a detailed description), the estimated event rates and times for collecting the required event samples are presented in Table~\ref{tab:time}.  In order to ensure the maximum $\phi$ coverage, each ($Q^2,\epsilon$) point requires three proton spectrometer (SHMS in most cases) angle settings: left ($\theta_{pq}=-3^\circ$), center ($\theta_{pq}=0^\circ$) and right ($\theta_{pq}=+3^\circ$) with respect to the $q$-vector, as shown in Fig.~\ref{fig:bullseye}. 

$W$ versus $Q^2$ distributions (the `diamond' distributions) for all settings are shown in Fig.~\ref{fig:diamond}. The L/T separated cross sections are planned at $Q^2=2.0$, $3.0$, $4.0$ and $5.0$~GeV$^2$. These measurements will provide the $-u$ dependence for $\sigma_{\rm L}$ and $\sigma_{\rm T}$ at nearly constant $Q^2$ and $W$, in addition to the behavior of $\sigma_{\rm L}$/$\sigma_{\rm T}$ ratio as function of $Q^2$. Note there are two $\epsilon$ measurements at each $Q^2$ setting: the red diamonds indicate low $\epsilon$ measurements and black for the high $\epsilon$ measurements. The $Q^2=6.25$~GeV$^2$ setting is chosen to test the $Q^2$ scaling nature of the unseparated cross section, but only one $\epsilon$ setting is available due to limitations on the accessible spectrometer angles. 

These proposed measurements will provide the following insights into the $Q^2$ dependence of the TDA formalism:

\paragraph*{\bf{L/T Separation at $\mathbf Q^2$=2~GeV$^2$}}

The experimental insights from Figs.~\ref{fig:park18}, \ref{fig:omega_sep} and \ref{fig:ratio} reveal TDA's ability of capturing the general trend of cross section and $\sigma_L/\sigma_T$ ratio behaviour as a function of $Q^2$. However, it is interesting to note that, at $Q^2 \sim$1.6~GeV$^2$, the TDA completely mis-predicted the cross section and the T/L ratio for $\pi^+$ and $\omega$ production.
Based on these observations, we would draw a tentative conclusion: in the region of $1.6<Q^2<2.6$~GeV$^2$, the nucleon wavefunction undergoes a transition; $Q^2$=2.5~GeV$^2$ is a boundary point where the TDA factorization starts to become valid. Also, we expect the $\sigma_T$ begins to become dominant (due to the fast drop of $\sigma_L$ in this transition region), and the predicted $\sigma_T/\sigma_L$ ratio ($R$) in the range of $2< R < 4$.

\paragraph*{\bf L/T Separation at $\mathbf Q^2$=3, 4, 5 GeV$^2$}
The L/T separated cross sections at these kinematic points yield the core data of the proposed experiment. In this kinematics region, if the TDA collinear factorization hypothesis is valid, we expect to clearly observe: $\sigma_T > \sigma_L$.  Based on our parameterization of the Defurne, et al. Hall A data \cite{defurne16}, our parameterization of a GPD calculation by Goloskokov \& Kroll \cite{gk11}, and the observed forward-backward peak ratios in $\omega$ electroproduction \cite{li19}, we estimate ratios of: $R>5$ at $Q^2=3$ GeV$^2$ and $R\sim100$ at $Q^2=5$ GeV$^2$. However, the expected extremely low $\sigma_L$ contribution to the cross section ($\sigma_L \sim 0$), will make the accurate determination of $\sigma_L$ impossible within proposed running time. 
For these extremely large values of $R$, our goal is to set a lower bound on $R$, as accurate very large ratio values within the proposed running time are not feasible.

\paragraph*{\bf Cross Section at $\mathbf Q^2$=6.25 GeV$^2$}
As it will be shown in the time estimation in Sec.~\ref{sec:kin}, the measurement at $Q^2=6.25$ GeV$^2$ requires less time than $Q^2=5.0$, since no L/T separation is intended due to the spectrometer angle limitations. At this kinematics point, we assume the complete domination of $\sigma_T$ over $\sigma_L$ and $\sigma_L\sim0$ (based on the results from the lower $Q^2$ points). In this case, the measured cross section would only consist of contributions from $\sigma_T$ and the interference term $\sigma_{TT}$. With the cross section model extracted from the $Q^2=2$, $3$, $4$ and $5$ GeV$^2$, we will be able to complete the projected $\sigma_T \propto 1/Q^8$ test from 2 to 6.2 GeV$^2$. This spread of coverage $q^2$ ($\Delta q^2\sim 4$~GeV$^2$) is similar to the spread of $q^2$ coverage of the $\overline{\rm P}$ANDA TDA study.

\subsection{``Soft-hard'' Transition through $u$-channel Phenomenology Study}
\label{sec:u_dep}

\begin{figure}[htb]
\centering
\includegraphics[width=0.49\textwidth, trim={0cm 0cm 1.5cm 0cm},clip]{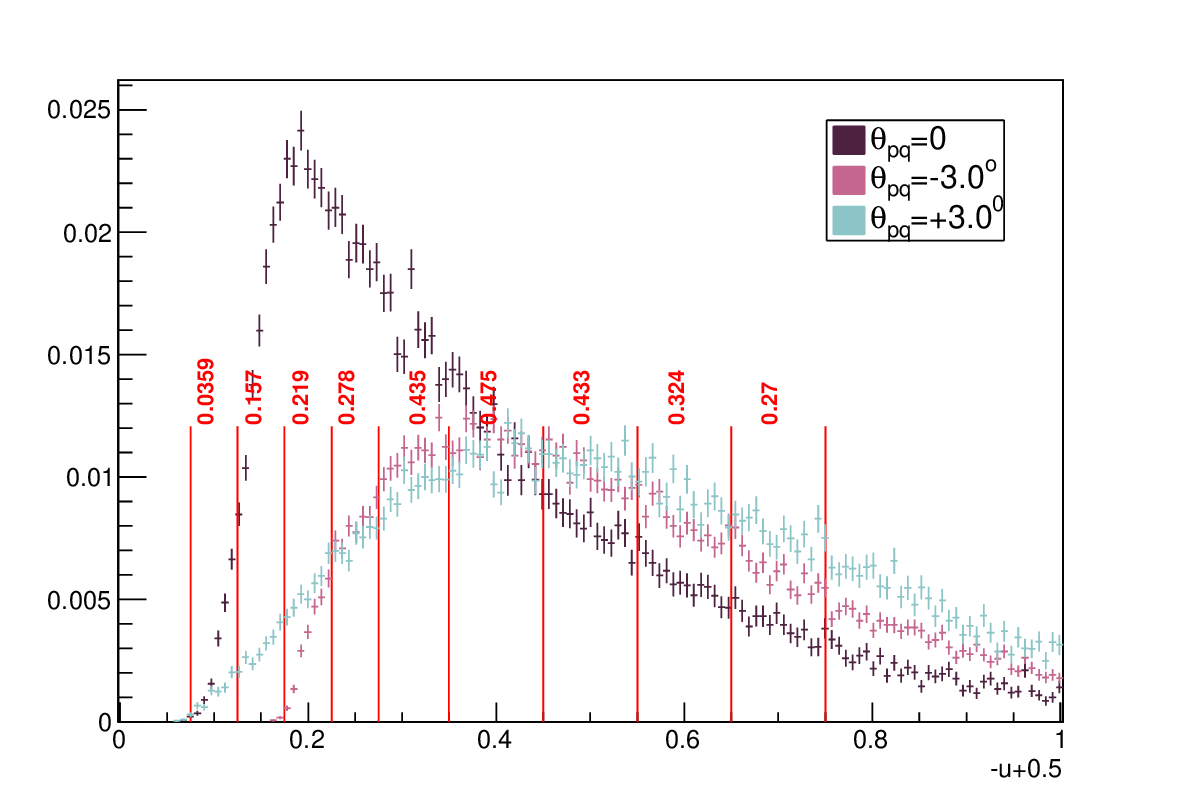}
\includegraphics[width=0.49\textwidth, trim={0cm 0cm 1.5cm 0cm},clip]{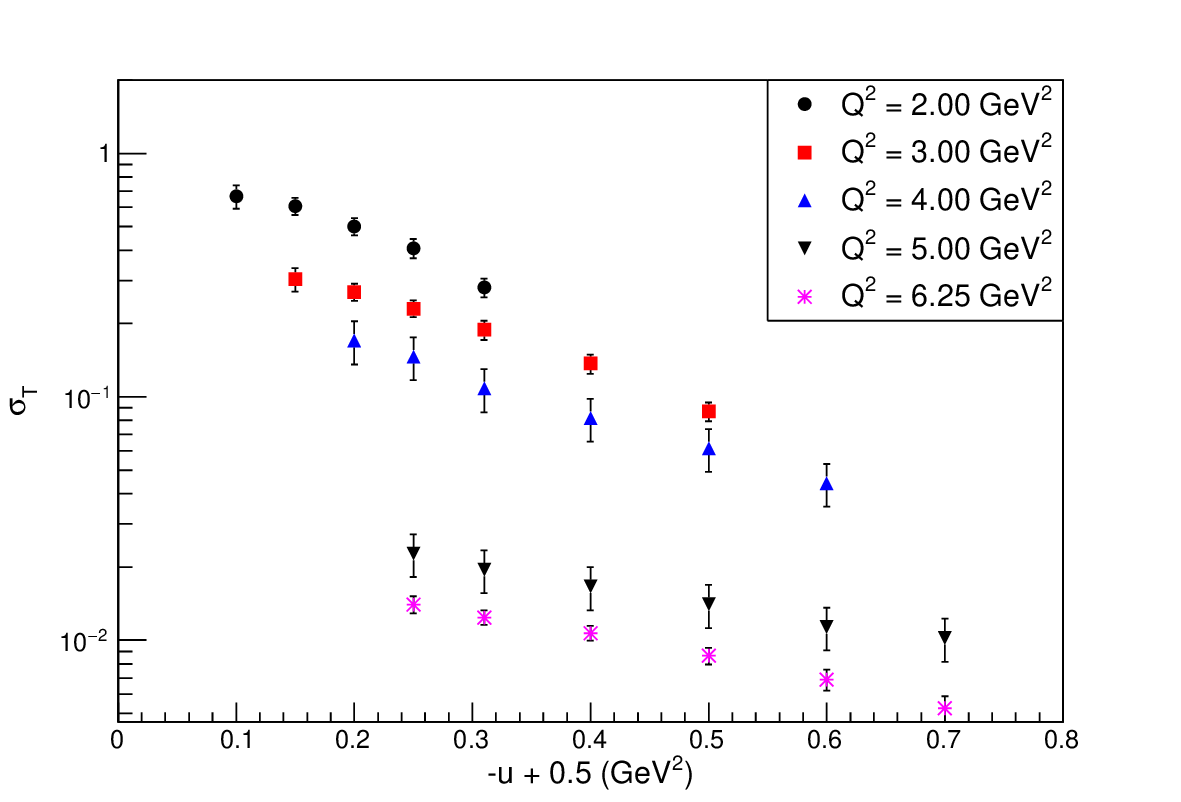}
\caption{(a) shows the simulated $-u$ distributions at $Q^2=4$~GeV$^2$, $\epsilon=0.73$ generated by the Hall C Monte Carlo (SIMC). Distributions from all three angle settings: 0$^\circ$ (black), -3$^\circ$ (red) and +3$^{\circ}$ (cyan) are overlapped. The spectrometer acceptance and the diamond (Fig.~\ref{fig:diamond}) cuts are applied. (b) demonstrates $-u$ coverage for $\sigma_T$ as a function of $-u$. In the presence of backward-angle peaks, the total (or separated) cross sections are expected to fall as $-u$ increases (as described by Eqn.~\ref{eqn:xi_u}). Here, in order the maintain the continuity of the $u$ dependence, an offset of $+0.5$ GeV$^2$ to the $u$ coverage is introduced for all settings. The $\sigma_T$ is in arbitrary unit. Note that the purpose of the plot is to demonstrate the $-u$ coverage and falling behaviour of cross sections in all $Q^2$ settings. The shown slope values do not represent any accurate prediction. }
\label{fig:u_dep}
\end{figure}

As described in the introduction, one of the main objectives of the proposed measurement is to better understand the mechanism of a ``soft-hard transition'' in $u$-channel physics, which remains an important open question.  To accomplish the stated objective, we propose to utilize the phenomenology tools developed at HERA, by examining the $t$ (or $u$ in this case) dependence of the cross section at different $Q^2$ settings, then comparing the extracted transverse size of interaction (using the fitted slope) to the hadronic size. For a ``soft'' Regge-exchange picture, the transverse size of the interaction is on the order of the hadronic size; for a ``hard'' QCD regime, the transverse size of interaction $\ll$ hadronic size~\cite{abramowicz95}. Note that the proposed measurement is the first attempt to apply such a methodology in the $u$-channel kinematics.

In the presence of a backward-angle peak, the events ($\theta_{pq}=0^\circ$, $\pm3^\circ$) will be binned in five $u$ bins, as shown in Fig.~\ref{fig:u_dep}(a). All $u$ bins will have equal statistics. Also see the projected $-t$ coverage for all $Q^2$ settings in Fig.~\ref{fig:u_dep}(b).

The standard formula with the exponential $-t$ dependence is replaced by the $-u$ dependence to address the rising cross section in the u-channel kinematics, given by
\begin{equation}
\frac{d\sigma_{L,T}}{du} = A\cdot e^{-b\cdot|u|},
\label{eqn:trans}
\end{equation}
where $A$ and $b$ are free fitted parameters. The parameter, $b$, in the above equation can be rigorously linked to the transverse size of the $\gamma^*p$ interaction region, as given below,
\begin{equation}
r_{int} = \sqrt{|b|} \, \hbar c,
\end{equation}
where $\hbar c =0.197$ GeV$\cdot$fm. In some terminology, $r_{int}$ is also referred to as the interaction radius. The same approach for extracting $r_{int}$ was successfully applied to extract the $-t$ dependence in the forward-wide angle $\pi^+$ exclusive electroproduction in Hall C (using the 6 GeV data)~\cite{basnet19}.

At the same time, one must point out the two foreseeable challenges when performing the described phenomenological study:
\begin{itemize}

\item The proposed measurement only offers $\sim0.5$ GeV$^2$ coverage in $-u$, which provides a smaller lever arm than other similar studies in meson production channels. See $-u$ coverage at different $Q^2$ settings in Fig.~\ref{fig:u_dep}(b).

\item The expected suppression and large uncertainty in $\sigma_L$ will make the extraction of its $-u$ dependence less conclusive in that case.  See Table~\ref{tab:kinematics} for reference.

\end{itemize} 
With these in mind, we are optimistic that the proposed measurement will determine the $-u$ dependence of $\sigma_T$ at the proposed $Q^2$ settings where L/T separation separation data is available. The $Q^2$ dependence of the measured slopes, and corresponding size of interaction region, will provide new insights into the ``soft-hard'' transition in this unique kinematic regime

\subsection{Beam Spin Asymmetry Measurement}
\label{sec:pi0_BSA}
As mention in the introduction section, the $\pi^0$ BSA will come for free with the planned data of this proposal. The study is similar to the one descried in Sec.~\ref{sec:BSA}.

The $A^{\sin \phi}_{LU}$ will be obtained at fixed $x_B=0.36$ and $Q^2=2$, $3$, $4$ and $5$ GeV$^2$. At each setting, BSA will be obtained in two separate methodologies: 
\begin{enumerate}
    \item Reconstruct $A^{\sin \phi}_{LU}$ based on the separated cross sections: $\sigma_T$, $\sigma_L$, and $\sigma_{LT}$ using Eqn.~\ref{eqn:bsa}.
    \item Utilize the readily available electron beam polarization and extract $A^{\sin \phi}_{LU}$ directly from the data as described by Eqn.~\ref{Def_BSA}.
\end{enumerate}
Here, $\sigma_{LT}$ and $\sigma_{LT^\prime}$ are identical in terms of helicity amplitude product terms. The only difference between the two is that $sigma_{LT}$ is a real part of the products and $\sigma_{LT^prime}$ is an imaginary part of the products. The determinations of  $A^{\sin \phi}_{LU}$ with $\sigma_{LT}$ and $\sigma_{LT^\prime}$ would provide information sensitive to the phase between the two.  

The interference ($\sigma_{LT}$ and $\sigma_{TT}$) contributions at the parallel kinematics: $u_{min} < u < 0.5$ GeV$^2$, are expected to be very small (in this proposal), which makes the accurate extraction of the BSA a difficult task (with significant uncertainty comparable to $-t \sim 6.8$ GeV$^2$  in Fig.~\ref{Fig_BSA_t_dependence}). However, we will be able to offer two separate verification (as stated above) to validate the phenomenological observation from CLAS: sign change of BSA in $u$-channel kinematics like in Fig.~\ref{Fig_BSA_t_dependence} and \ref{Fig_BSA}. In order to accomplish this we would request to utilize the existing beam polarimeter at Hall C. 

We aim to obtain the beam polarization to a precision of $dP/P=2$-$3$\%. This will require semi-regular polarization measurements with the existing Moller polarimeter. These measurements will be made after every beam pass change and each measurement lasts one shift.

\subsection{$W$ Scaling Correction and $\mathbf Q^2=2$ GeV$^2$, $\mathbf W=3$ GeV Setting}

\begin{figure}[htb]
\centering
\includegraphics[width=0.76\textwidth]{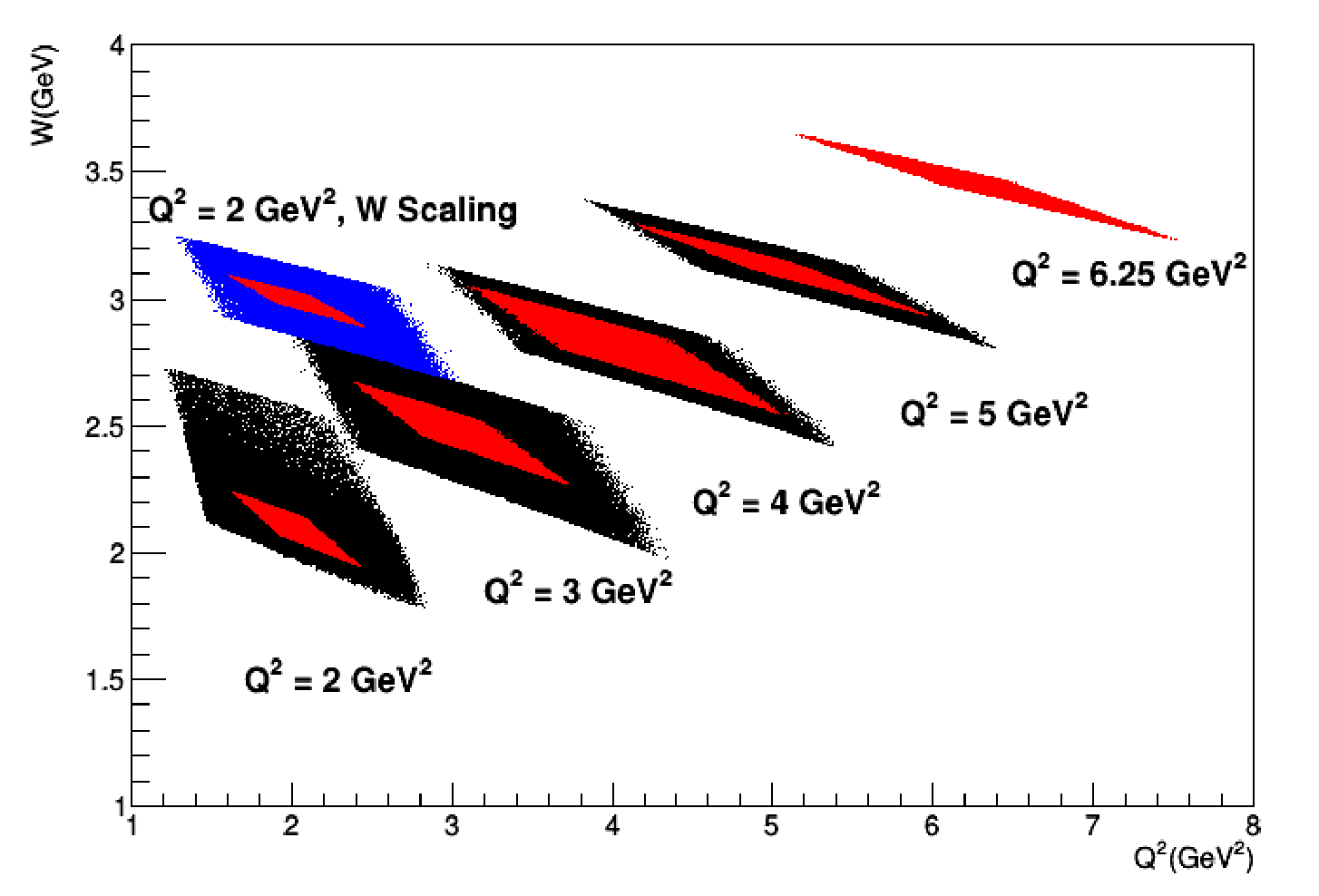}
\caption{$W$ vs $Q^2$ diamonds for the $Q^2=2.0$, $3.0$, $4.0$, $5.0$ and $6.25$ GeV$^2$ settings. The black diamonds are for the higher $\epsilon$ settings and the red diamonds are for the lower $\epsilon$ settings. The diamonds for the $W$ scaling setting are shown separately (in blue and red). Note that there is only one $\epsilon$ setting for $Q^2=6.25$ GeV$^2$. The overlap between the black and red diamond is critical for the L/T separation at each setting. The boundary of the low $\epsilon$ (red) data coverage will become a cut for the high $\epsilon$ data.}
%Note that the $Q^2=5.5$ GeV$^2$ setting has measurement at only one $\epsilon$ value. u}
\label{fig:diamond}
\end{figure}

In the ideal case, one would select kinematics regions where $W$ and $x_B$ are fixed among measurements at different $Q^2$ settings. However, the narrow spectrometer acceptance and correlated nature of $Q^2$-$W$ (within the coincidence acceptance) make measurements at a fixed $W$ and $x_B$ an impossible task, as demonstrated in Fig.~\ref{fig:diamond}. 

When performing a study of L/T-separated cross sections versus $u$ or $t$, one must use a scaling procedure to correct for the small $W$ dependence in the measured cross section, namely, scaling different $W$ values to a common $W_{norm}$. The standard $W$ scaling formalism is as follows: 
\begin{equation}
\frac{W-M_p}{W_{norm}-M_p},
\label{eqn:w_scaling}
\end{equation}
where $M_p$ is the proton mass. This procedure is our best estimate for the $W$ correction based on the $t$-channel meson production data. The necessary data to validate this relationship for backward-angle meson production have never been acquired, so this relationship was assumed to apply in the backward $\omega$ case ($\Delta W \sim 0.1$ GeV)~\cite{wenliang17, li19}.

In this proposal, keeping $x_B$ fixed at 0.36 means that $\Delta W \sim 1$ GeV as $Q^2$ is varied from 2 to 5 GeV$^2$.  This raises an important question, could Eqn.~\ref{eqn:w_scaling} correct the $W$ dependence? The $W$ dependence across the diamond for a single L/T separation is not that large (Fig.~\ref{fig:diamond}). However, it will play a role when comparing data-sets from the different $Q^2$ with each other.  An inaccurate $W$ dependence measurement
would increase the uncertainty, or even bias the result, when determining the exponent factor ($n$) of the $\sigma_T \propto 1/Q^n$ scaling test.

To verify the $W$ scaling procedure with experimental data, we propose performing a $u$-channel L/T separated cross section measurement at $Q^2=2.0$ GeV$^2$, $W=3$ GeV and $x_B=0.20$. Along with planned separated cross section at $Q^2=2.0$ GeV$^2$, $W=2.11$, one can get a clean $W$ correction for $\sigma_L$ and $\sigma_T$ independently.

\subsection{Singles Rate Estimation}
\label{sec:singles}

\begin{figure}[htb]
\centering
\includegraphics[width=0.80\textwidth]{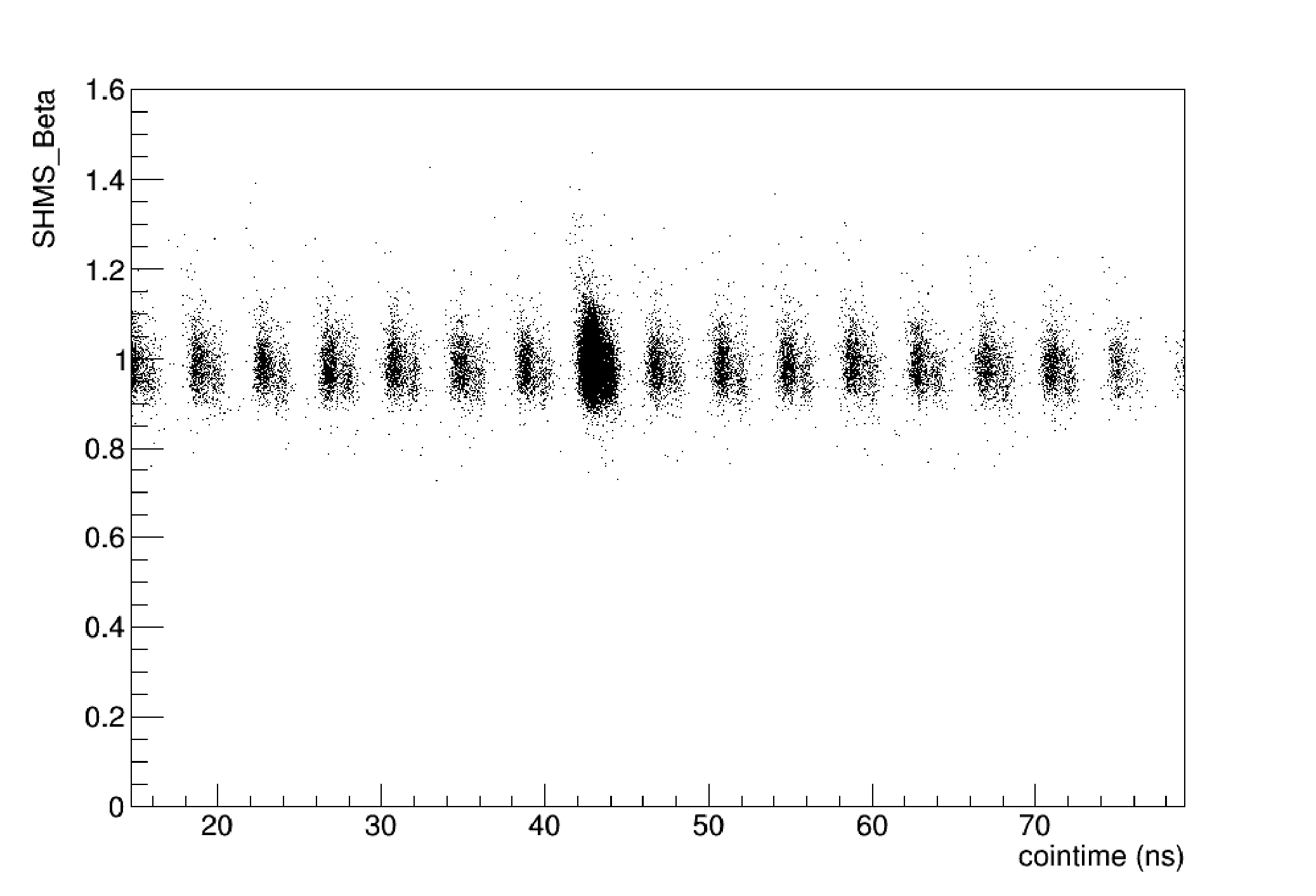}
\caption{Example SHMS $\beta$ versus HMS-SHMS coincidence time distribution from the recent 12 GeV Kaon-LT experiment (E12-09-011). No acceptance or PID cuts are applied.  The 4~ns individual beam bunch spacing from the CEBAF accelerator is easily resolved.  Within each bunch, a doublet is observed, with the left darker band due primarily to $p(e,e'\pi^+)n$ and $p(e,e'K^+)\Lambda$ coincidences, and the right fainter band due to $p(e,e'p)X$ coincidences, the offset being due to the longer time of flight of protons in the hadron arm.}
\label{fig:coin}
\end{figure}

All singles rate estimates assume the use of a 70~$\mu$A beam on a 8~cm LH$_2$ cryogenic target and the detection efficiencies listed in Table \ref{tab:eff}.

For the purpose of calculating online random coincidence rates, the hadron arm trigger rate was taken as equal to the raw trigger rate, i.e. no distinguishing between pions, kaons and protons in the hadron trigger. Assuming the online ELREAL trigger is set up to suppress the combined trigger contribution from $\pi^-$ and $K^-$ by 5:1, the electron arm trigger rate was taken to be electrons plus $(\pi^- + K^- + \bar{p})/5$. The electron arm rate is calculated as $$
{\rm E~Rate} =e^- + (\pi^- + K^- + \bar{p})/5 \,,$$ and hadron arm rate as: $$ {\rm H~Rate} =e^+ + \pi^+ + K^- + p\,.$$ Here, we also assume the trigger contribution from $e^+$ is small compared to other particles, therefore is neglected. The random coincidence rate is then given by $$({\rm E~Rate})\cdot({\rm H~Rate})\cdot\Delta t\,,$$ where the coincidence resolving time was taken to be $\Delta t=70$~ns. 

Since the roles of SHMS and HMS differ, the singles rate estimations for the $Q^2=$3 to 6.25 GeV$^2$ and $W$-scaling settings are listed in Table~\ref{tab:singles_1}; the singles rate estimation for $Q^2=$2 GeV$^2$ is in Table~\ref{tab:singles_2}. In all cases, the resulting online real$+$random rates are well below the expected capability of the HMS+SHMS data acquisition system. The real coincidence rate is based on our estimated SIMC yield without any cuts applied.

It is important to note that the random and real coincidence rates listed in Table~\ref{tab:singles_1} and Table~\ref{tab:singles_2} are for the online DAQ rates, which only take into account the $\pi^-$ rejection ratio on the electron arm (assuming 70 ns resolving time).
Fig.~\ref{fig:coin} shows an example of SHMS $\beta$ vs. the HMS-SHMS coincidence time spectrum from a recent 12 GeV Hall C experiment.  It shows we typically resolve the prompt coincidence peak to less than 2~ns, so a cut on this automatically would reduce the random contributions by a factor of $\sim$35 (i.e. 70~ns/2~ns).  Random coincidence subtraction and particle ID cuts will reduce this further, to a few percent of the reals physics events in the offline analysis.

\begin{table}[ht]
\centering
%\setlength{\abovetopsep}{1ex}
%\small
\setlength{\tabcolsep}{0.38em}
\caption{Calculated singles rates for $Q^2=$ 3, 4, 5 and 6.25~GeV$^2$ and the $W$-scaling settings. In these settings, the SHMS is on the positive and the HMS is on the negative polarity. Note that these calculated random and real rates are for the online DAQ only, and do not correspond to the actual event and background offline analysis rates.}
\label{tab:singles_1}
\begin{tabular}{c|ccc|cccc|cc}
\hline
           & \multicolumn{3}{|c|}{SHMS}   & \multicolumn{4}{|c|}{HMS}                &              &             \\     
$\epsilon$ & $\pi^+$ & $K^+$ & $p$   & $e$   & $\pi^-$  & $K^{-}$      & $\bar{p}$   & Random Coin. &  Real Coin. \\
           & (kHz)   & (kHz) & (kHz) &(kHz)  & (kHz)    & (kHz)        & (kHz)       &  (Hz)        &  (Hz)       \\ \hline
             \multicolumn{10}{c}{$Q^2=$2.0 GeV$^2$, $W=3$ GeV, $x_B=0.20$}                                                             \\ \hline  
0.32       &  169    &  62   &  51   &  2.3  & 221      &  7.2          & 0.08       & 950          & 0.01        \\
0.79       &   94    &  35   &  38   & 113   & 55       &  5.3         &  0.60       & 1500         & 0.68        \\ \hline
             \multicolumn{10}{c}{$Q^2=$3.0 GeV$^2$, $W=2.49$ GeV, $x_B=0.36$}       
                                                                \\ \hline
0.54       &  29     &  11	 &  18	 &   4   & 18       &  0.7         &  0.01       & 31           & 0.06        \\
0.86       &  10     &  4    &  8    &   52	 & 126      &  12          &  2          & 123          & 0.30        \\ \hline
             \multicolumn{10}{c}{$Q^2=$4.0 GeV$^2$, $W=2.83$ GeV, $x_B=0.36$}                                                               \\ \hline
0.55       &  18     &  9    &  10   &   4   & 11       &  0.7         &  0.04       & 16           & 0.05        \\
0.73       &  10     &  4    &  5    &   15  & 5        &  0.5         &  0.04       & 21           & 0.22        \\ \hline
             \multicolumn{10}{c}{$Q^2=$5.0 GeV$^2$, $W=3.13$ GeV, $x_B=0.36$}                                                               \\ \hline    
0.26       &  46     &  25   &  17   &   0.4 & 33       &  1           &  0.02       & 44           & 0.005       \\
0.55       &  13     &  7    &  6    &   4	 & 8        &  0.7         &  0.05       & 10           & 0.040       \\ \hline
             \multicolumn{10}{c}{$Q^2=$6.25 GeV$^2$, $W=3.46$ GeV, $x_B=0.36$}                                                               \\ \hline     
0.27       &  37     &  23	 &  13   &   0.3 & 25       &  1           &  0.04       & 28           & 0.005       \\
\hline
\end{tabular}
\vspace{5mm}
\caption{Calculated singles rates at $Q^2=$2~GeV$^2$. In these settings, the HMS is on the positive and the SHMS is on the negative polarity.}
\label{tab:singles_2}
\begin{tabular}{c|ccc|cccc|cc}
\hline
           & \multicolumn{3}{|c|}{HMS}   & \multicolumn{4}{|c|}{SHMS}                &              &             \\     
$\epsilon$ & $\pi^+$ & $K$   & $p$   & $e$  & $\pi^-$  & $K^{-}$       & $\bar{p}$   & Random Coin. &  Real Coin. \\
           & (kHz)   & (kHz) & (kHz) &(kHz) & (kHz)    & (kHz)         & (kHz)       &  (Hz)        &  (Hz)       \\ \hline
             \multicolumn{10}{c}{$Q^2=$2.0 GeV$^2$, $W=2.21$ GeV, $x_B=0.36$}                                                               \\ \hline     
0.52       &  18     &   6   &  18   &  4   &   56     &   1.0         &  0.01       & 45           &  0.09       \\
0.94       &  7      &   2   &  9    &  5   &   16     &   1.3         &  0.1        & 10           &  4.70       \\ \hline

\end{tabular}

\end{table}

\subsection{Kinematic Checks and Normalization with Elastic Scattering}

The elastic $^1$H$(e, e^\prime)p$ and $^1$H$(e, e^\prime p)$ measurements are extremely useful tests for determining the systematic uncertainties in single arm and coincidence measurements. The fixed position of the elastic peak ($W^2$ for the single arm case and along missing energy/missing momentum in the coincidence) allows one to verify the spectrometer central angle momentum and determine potential offsets. In addition, the well known $e$-$p$ elastic cross section provides verification of the normalization and establishes acceptance boundaries.

The use of elastic scattering for calibration has been performed extensively and has been a standard procedure in Hall C measurements in both the 6~GeV and 12~GeV eras. Due to the strict systematic requirements of the L/T separation procedure, we will perform additional checks at kinematics close to the planned measurements. In this section, we briefly discuss our kinematics choices and beam time requirements for the elastic checks.  

\subsubsection{Single Arm Elastic Checks}

As described above, the single arm elastic scan measures the hydrogen elastic peak. There are three unknown parameters to be determined from these measurements: beam energy, spectrometer central angle, and spectrometer central momentum.  A series of high quality measurements will provide precise constraints to these parameters.

For the HMS, we can rely on the rigid connection to the target station pivot to ensure that variations in the pointing angle are relatively small when rotating the spectrometer to various angles. Because of this excellent pointing reproducibility, one can assume that, by and large, any offset to the spectrometer central angle is a fixed value, with minimal variation (on the order of 0.2 mrad) as the spectrometer is rotated. Similarly, one can assume that the deviation of the spectrometer central momentum is a fixed value due to the very linear response of the HMS dipole. Hence, by measuring the position of the reconstructed proton mass peak ($W = M_p$) over a range of angles and at several beam energies, one has several constraints on the spectrometer kinematic offsets. In certain $Q^2$ measurement settings, the HMS spectrometer momentum is considerably higher than during the standard 6 GeV operation. We will take additional optics data at $P_{HMS}=5.51$ GeV/c (our highest HMS momentum setting) to ensure that any saturation corrections that may be needed are properly understood and applied.

A similar study will be carried out with the SHMS. In this case, the smaller angles and higher energies accessible provide a very large lever-arm for constraining the central scattering angle and momentum. The elastic peak position is expected to shift from $\sim -7$~MeV/mrad to $\sim -17$~MeV/mrad as one rotates from 5.5 degrees to 18 degrees. At fixed beam energy, the dependence on the central spectrometer momentum is relatively flat, but by making measurements at several beam energies, one can extract the angle and momentum offsets.

A study of this nature can be carried out in concert with data taking during the $\pi^0$ measurement. At each setting, one need acquire 10,000 elastic events each. Note that this is not an issue since all $\pi^0$ measurements require much longer running times.

\subsubsection{Elastic Coincidence Checks}

\begin{table}[htb]
\centering
%\setlength{\abovetopsep}{1ex}
%\small
\setlength{\tabcolsep}{0.5em}
\caption{Elastic coincidence $^1$H$(e, e^\prime p)$ kinematics. Settings indicated by $^*$ will have HMS detecting proton and SHMS detecting electron; Setting indicated by $^+$ will have HMS detecting electron and SHMS detecting proton. Assumed 70 $\mu$A beam current. 10000 events which corresponds to 1\% statistical error.}

\label{tab:heep_kin}
\begin{tabular}{cccccccc}
\hline
$E_{beam}$ & $Q^2$      & $\theta^{\prime}_e$ & $p^{\prime}_e$ & $\theta_p$  & $p_p$      & Coincidence Rate & Time    \\
(GeV)      & (GeV$^2$)  & (deg)               & (GeV)          & (deg)       & (GeV)      & (Hz)            & (Hours) \\ \hline
%6.6        &   4.13     &  21.75              &  4.40          &  32.94      &  3.00      &                 \\ 

%/*--------------------------------------------------*/
% Not possible due to the SHMS rotating angle
%4.4$^+$    &   5.56$^*$ &  55.85$^*$          &  1.44$^*$      &  18.34$^*$  &  3.78$^*$      &                 \\ 
%4.4$^+$    &   1.47$^*$ &  17.45$^*$          &  3.62$^*$      &  48.87$^*$  &  1.44$^*$      &                 \\ 

4.4$^+$    &   2.34$^+$    &  23.70$^+$        & 3.15$^+$       &  39.95$^+$  &  1.97$^+$      & 371   & 1  \\ %5305
4.4$^*$    &   2.68$^*$    &  25.15$^*$       &  2.97$^*$       &  37.12$^*$  &  2.17$^*$      & 251   & 1  \\ %3582

% %/*--------------------------------------------------*/
% % +/- symmetry test
% 6.6$^*$    &   4.137$^*$ &  38.2$^*$          & 2.195 $^*$       &  26.91$^*$  &  3.0$^*$  &                 \\ 

6.6$^+$    &   4.18$^+$    &  21.95$^+$       &  4.37$^+$       &  32.69$^+$    &  3.03$^+$    & 30    & 1 \\ %432
6.6$^{+}$  &   3.00$^{+}$  &  17.35$^{+}$     &  5.00$^{+}$     &  39.21$^{+}$  &  2.36$^{+}$ & 170   & 1 \\ %2424
6.6$^{*}$  &   3.00$^{*}$  &  17.35$^{*}$     &  5.00$^{*}$     &  39.21$^{*}$  &  2.36$^{*}$ & 323   & 1 \\ %4613
6.6$^*$    &   1.32$^{*}$  &  1.55 $^{*}$     &  5.90$^*$       &   53.43$^{*}$  &  1.345$^{*}$     & 4500  & 1 \\ %46750
% 10.550     1.316     5.899   -53.433     1.345    0.165E-03    0.685E-05a

8.8$^*$    &   1.61$^*$    &  8.70$^*$        &  7.94$^*$       &  51.71$^*$    &  1.53$^*$    & 3272  & 1 \\ %46750

%/*--------------------------------------------------*/
% Polarity swap
8.8$^*$    &   4.32$^*$ &  15.80$^*$           &  6.50$^*$      &  34.77$^*$    &  3.10$^*$  & 0.8       & 4 \\ %11.87
%8.8$^*$    &   1.49$^*$ &  8.35$^*$            &  8.00$^*$      &  52.85$^*$    &  1.46$^*$  &       \\ 
10.9$^*$   &   1.99$^*$ &  7.80$^*$            &  9.84$^*$      &  49.30$^*$    &  1.76$^*$  & 167       & 1 \\ %2381
%10.9$^*$   &   4.09$^*$ &  11.90$^*$           &  8.72$^*$      &  37.25$^*$    &  2.97$^*$  & 1.1       & 3 \\ %15.90
\hline
\end{tabular}

\end{table}

In addition to constraining the kinematic offsets as described above, the elastic data taken will enable us to check the normalization of the single arm and coincidence acceptance. In particular, examining the elastic yield across the spectrometer momentum acceptance has provided rigorous checks of our knowledge of the spectrometer response.  In the $^1$H($e, e^\prime p$) reaction, the scattered electron is detected in one arm and recoil proton in another, and these signals are in a coincidence mode. The missing energy: $E_m$ and three components of missing momentum: $p^{\perp}_m$, $p^{\parallel}_m$, $p^{oop}_m$.

Ideally, one would choose a coincidence elastic data point at each beam energy used by the experiment. It is also of benefit to choose the kinematics to sample a similar angle and momentum range used in the experiment.  Table \ref{tab:heep_kin} shows the selected $^1$H($e, e^\prime p$) kinematics. One would like to use the elastic coincidence data to totally overlap the kinematics used in the experiment, however, this is not possible for all settings. We chose these kinematics in strong favor of SHMS, due to its shorter operation history. The electron and proton arm assignments are described in detail in the Table~\ref{tab:heep_kin} caption. In order to utilize the available spectrometer angle and momentum ranges, while minimizing the run time (run at lowest possible $Q^2$ value), the spectrometer magnets are at opposite polarities to the corresponding $\pi^0$ measurement. This is based on the assumption that reversing the magnet polarity will not affect the spectrometer response. Note that there are two elastic settings at $E_{beam} = 6.6$ GeV, $Q^2=3.0$~GeV$^2$, where each of the SHMS and HMS will perform both the $e$ and $p$ arm role. This will serve as a good check to see any potential offsets and discrepancies in the spectrometer response when the polarity reversed. Additionally, one can also check for potential effects from hadrons (protons in this case) punching through the collimator and investigate hadron absorption effects in the detector stack.

The expectation for the length of the elastic runs is to collect 10,000 coincidence events at 70~$\mu A$ of beam current.  Due to magnet polarity reversal and other experimental overhead, the minimum time listed for each setting in Table~\ref{tab:heep_kin} is 1~hr.

\subsection{Particle Identification}

\begin{figure}[htb]
\centering
\includegraphics[width=0.523\textwidth]{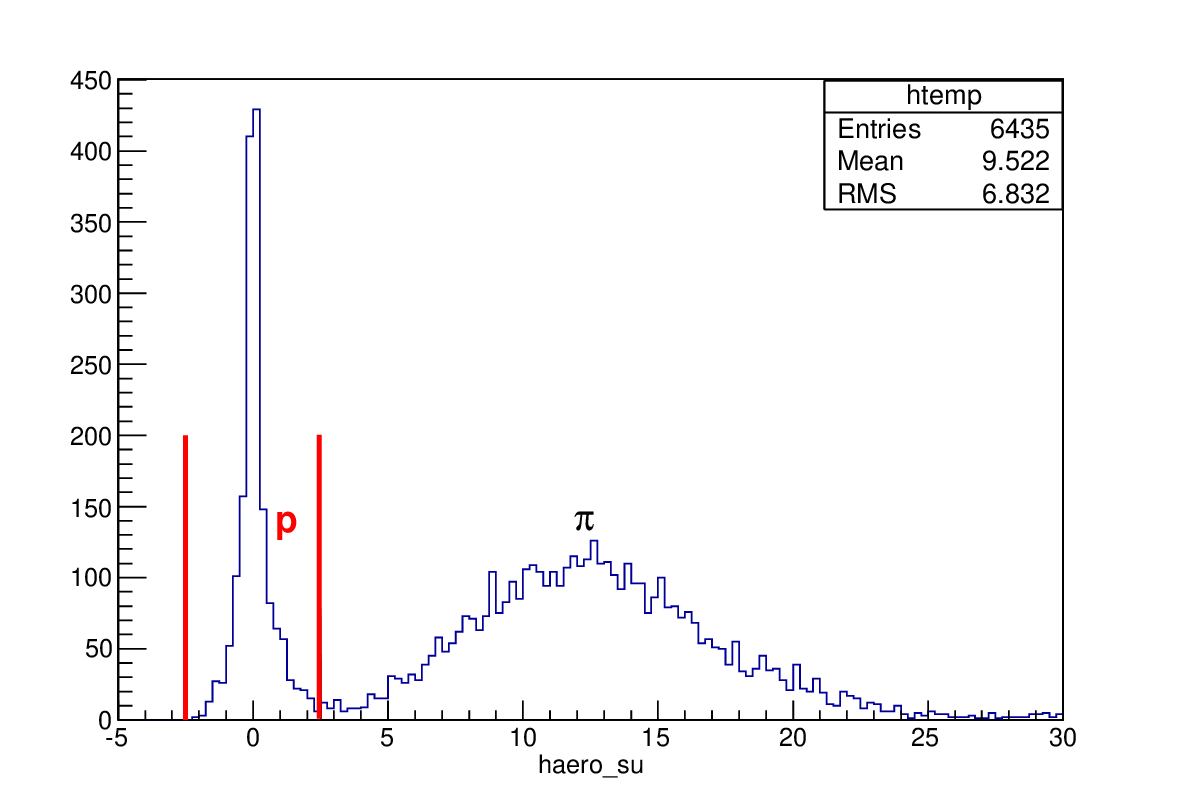}
\includegraphics[width=0.47\textwidth]{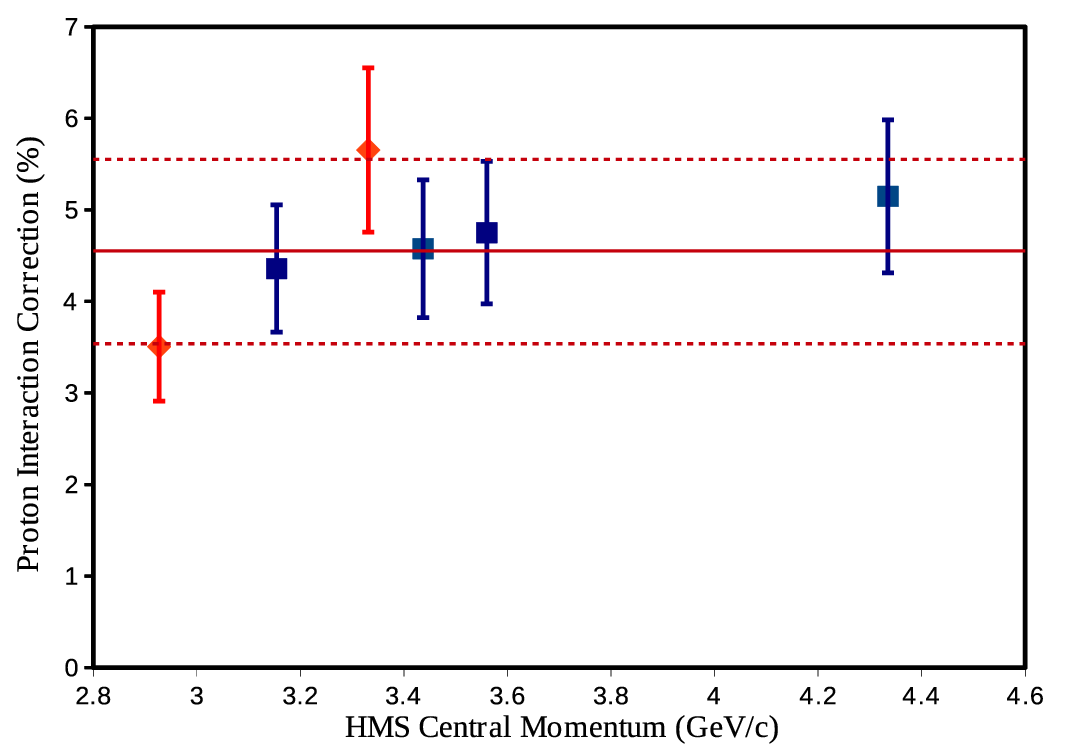}
\caption{(a) HMS aerogel Cherenkov cuts that discriminate protons from pions at $P_{HMS}=2.93$ GeV/c; (b) HMS proton loss percentage due to scattering in the HMS as function of central momentum. Red dots are for the $\omega$ data; where black dots are for the elastic scattering data. Both results are from 6 GeV studies in the F$_\pi$-2 experiment.}
\label{fig:efficiency}
\end{figure}

\begin{table}[htb]
\centering
\caption{Anticipated HMS and SHMS detection efficiency based on past operation experience.}
\label{tab:eff}
\setlength{\tabcolsep}{3.0em}
\setlength{\abovetopsep}{2em}
\renewcommand{\arraystretch}{1.4}
\begin{tabular}{lc}
\hline                                                                              
HMS Tracking                                           &  0.95      \\
SHMS Tracking                                          &  0.95      \\
HMS Aerogel for proton PID                             &  0.95      \\
SHMS Aerogel for proton PID                            &  0.95      \\
HMS proton scattering loss                             &  0.95      \\
SHMS proton scattering loss                            &  0.95      \\
HMS:5.9 msr Acceptance for $\delta=-10\%$ to $+10\%$   &  0.90      \\
SHMS: 3.5msr Acceptance  for $\delta=-15\%$ to $+20\%$ &  0.90      \\
\end{tabular}	
\end{table}

For the most $\pi^0$ settings, the SHMS will detect the forward going protons and HMS will detect the scattered electrons. Both spectrometers will be equipped with full Cherenkov detector packages (gas and aerogel Cherenkov detectors), and the signals from these detectors can be used to reject the $\pi$ and $K$ events. 

For the $Q^2=2$~GeV$^2$ settings, the SHMS will be used as the electron arm and it requires the installation of the Noble Gas Cherenkov detector to reject $\pi^-$ events. A variety of gases with different refractive indices are available to perform $e$/$\pi$ separation at different particle momenta.  We propose at $P_{SHMS}=-1.44$~GeV/c, the NGC is filled and circulated with mixture of Argon \& Nitrogen gas at 1~atm; and at $P_{SHMS}=-7.94$~GeV/c, the NGC is filled and circulated with Neon gas at 1~atm. Note that at $P_{SHMS}\sim -1.5$~GeV/c, the $\beta$ vs. coincidence beam bunch structure will be capable of providing a clean $e/\pi$ separation.  

In the SHMS proton detection case, the Noble Gas Cherenkov detector will remain installed despite higher proton multiple scattering probability. The reason is to benefit the $e$-$p$ elastic scattering measurement, where in some settings, the SHMS will detect the scattered electron in the elastic runs. At $P_{SHMS}=$3, 5, 6.5 and 9.5~GeV/c, the NGC will be filled and circulated with Argon-Nitrogen, Hydrogen, Neon and Helium gas respectively. The Heavy Gas Cherenkov (HGC) detector will be filled with C$_4$F$_{10}$ at 1~atm. Under normal operation, the HGC gas pressure is reduced at momenta higher than 7~GeV/c to ensure good $K$/$\pi$ separation. However, this step is not necessary for the proton identification (to save time). The primary methodology for the proton identification during the $\pi^0$ measurement involves: 1) Examining the $\beta$ vs. coincidence time distribution; 2) Placing a threshold cut on the Aerogel Cherenkov Detector (ACD). These points are further elaborated below.

At 8~GeV/c SHMS momentum, the coincidence time information provided by the RF reference and hodoscope triggers (from both spectrometers) cannot provide clean $\pi$/$K$ separation. However, proton coincidence triggers arrive significantly later ($\sim$10~ns) than those of $\pi$/$K$, therefore a clean proton separation is expected in the $\beta$ vs. cointime distribution for all kinematics settings (less separation at the highest momentum setting).

As complementary to examining the coincidence timing, the method of placing ACD cuts to exclude events beyond the applied threshold further cleans the proton event sample. An example of ACD distribution from the 6~GeV analysis is shown in Fig.~\ref{fig:efficiency}(a). For the $\pi^0$ measurement, we propose to utilize the SHMS aerogel tray with $n=1.011$ throughout the run. This corresponds to a Cherenkov threshold momentum of 3.315~GeV/c for $K$ and 6.307~GeV/c for proton. The ACD threshold cut will be applied at $3.5$ photoelectrons. It is worth noting that the HGC (at 1~atm) has a threshold momentum of 9~GeV/c for kaon and ($P_{th}\ll$ 11~GeV/c) for proton, and it can be used to help $K/P$ separation at the high SHMS momentum settings ($P_{SHMS}=9.84$ GeV/c). Note that $\pi$ will generate HGC and ACD signals in all measurement settings. 

In the HMS detecting proton case ($P_{SHMS}=3.51$ GeV/c), we propose the HMS aerogel tray with $n=1.030$ throughout the run. This corresponds to a Cherenkov threshold momentum of 2.0~GeV/c for $K$ and 3.8~GeV/c for proton. The proton PID efficiency has been studied during the 6 GeV operation (see efficiency in Table ~\ref{tab:eff}) at a similar momentum setting, the ACD PID (negligible $\pi$ contamination) and tracking efficiency of 95\% were determined with the elastic data.  We are confident that the combined $\beta$ vs coincidence information and cuts from the Cherenkov package provide clean separation of proton events from pion and kaon events at the highest momentum setting. 
   
The dominant reaction for the recoil protons inside of the spectrometers is inelastic scattering (mainly pion production), elastic and (quasi) elastic scattering (with heavier elements than hydrogen).  In the case of pion production and (quasi) elastic scattering, a secondary pion, proton or neutron is emitted along the path of the recoil proton momentum, and therefore has a probability to generate a valid trigger. The $pp$ and $pn$ total cross sections are dependent on the proton momentum, and are estimated to be 43~mb at 3.5~GeV/c, where the elastic cross section is 1/3 of the total cross section. Extensive studies during the 6~GeV era, as an example, revealed the proton loss is around $5\%$ for the HMS in the momentum range of 2 to 4 GeV/c. For the SHMS, we assume similar 5\% loss (based on similar material in radiation length to the HMS) for all momentum range, and detailed studies will be performed with elastic data at the $\pi^0$ measurement momentum range.

\subsubsection{Critical Hardware Replacement}

As described in the previous section, the HMS ACD plays an important role in the proton PID when used as the proton arm at $Q^2=2$ GeV/c$^2$, $W=2.11$ GeV. Based on recent operation experience, the performance of the HMS ACD PMT tubes were identified as degraded. We would like to urge for the refurbishment or replacement of the damaged HMS ACD PMTs before this proposed measurement, so that one could achieve the stated systematic uncertainties in Table~\ref{tab:sys_err_tab}.

\subsection{Physics Background Contribution}
\label{sec:background}

\begin{figure}[htb]
\centering
\includegraphics[width=0.4\textwidth, trim={0cm 0cm 0cm 3.8cm},clip, angle=90]{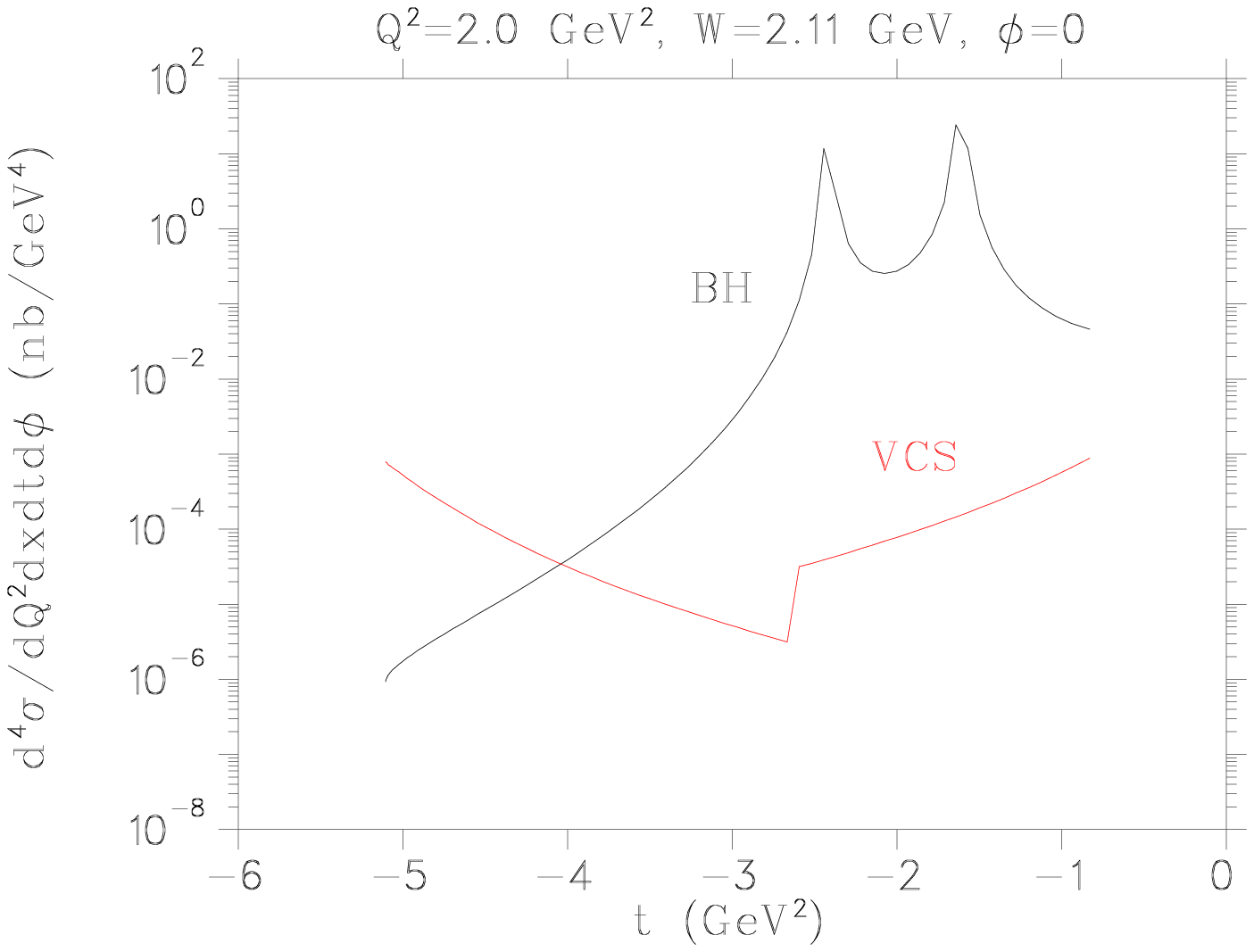}
\includegraphics[width=0.4\textwidth, trim={0cm 0cm 0cm 3.8cm},clip, angle=90]{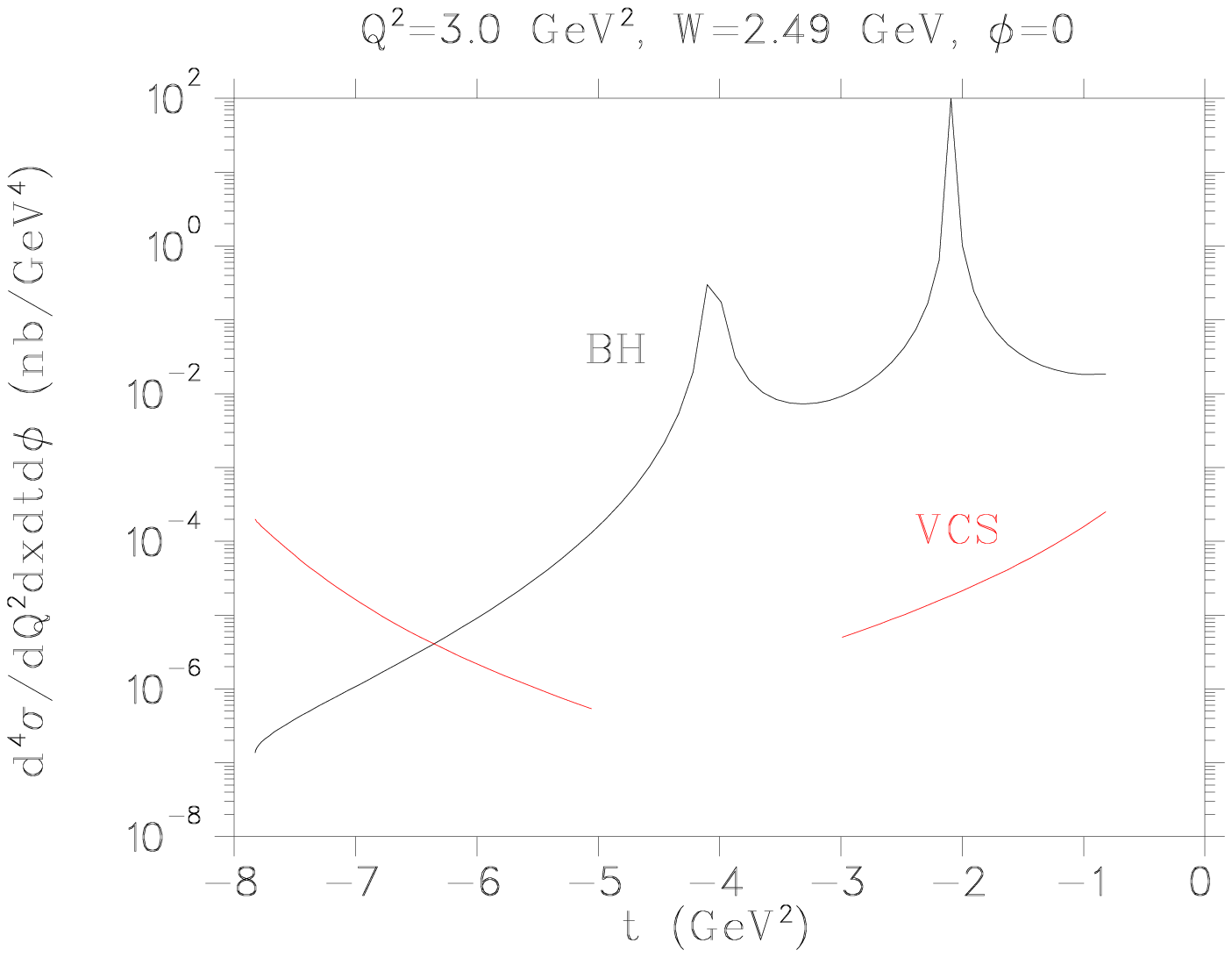}\\
\includegraphics[width=0.4\textwidth, trim={0cm 0cm 0cm 3.8cm},clip, angle=90]{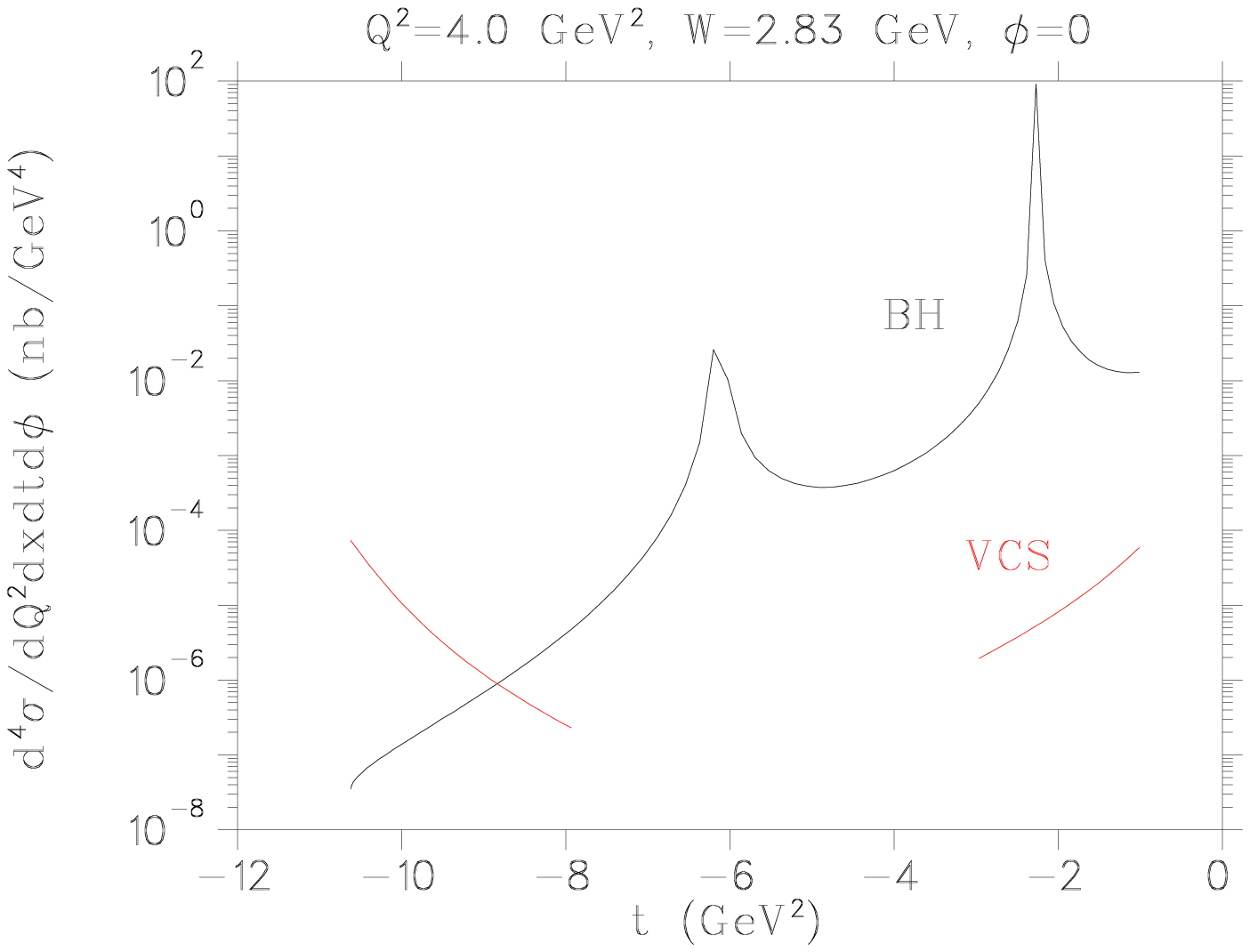}
\includegraphics[width=0.4\textwidth, trim={0cm 0cm 0cm 3.8cm},clip, angle=90]{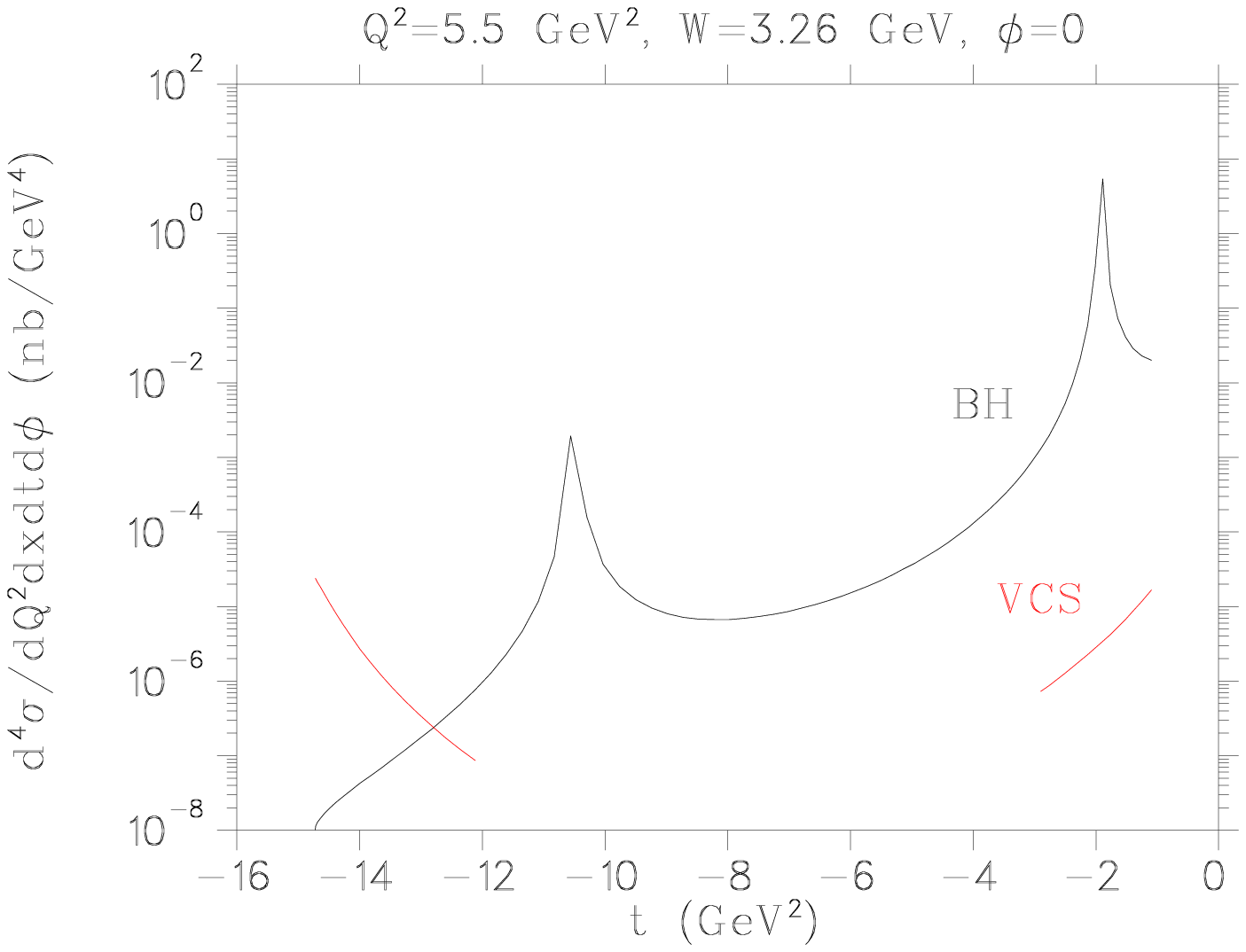}
\caption{$t$ evolution of the VCS model and BH at $Q^2=$2, 3, 4 and 5.5~GeV$^2$ (red lines). The BH contribution is plotted in black. The objective of these plots is to demonstrate the general trend of the projected VCS in relation to the classic BH contribution at $t=t_{min}$ and $t=t_{max}$.  The most important observation here is: BH dominates at small $-t$, and is suppressed at extremely large $-t$.  The discontinuation (in the middle $-t$ range) is due to the fact that the backward-angle VCS model is parameterized separately from the forward-angle model. We have not focused on parameterizing the model that handles the forward-backward transition region, since this is far outside the kinematics of this proposed measurement. For further details, see Appx.~\ref{sec:gamma_MC}.}
\label{fig:DVCS}
\end{figure}

\begin{figure}[htb]
\centering
\subfloat[$Q^2$=2.0 GeV, $\epsilon=$0.52]{\includegraphics[clip,trim=1mm 1mm 10mm 10mm,width=0.4\textwidth]{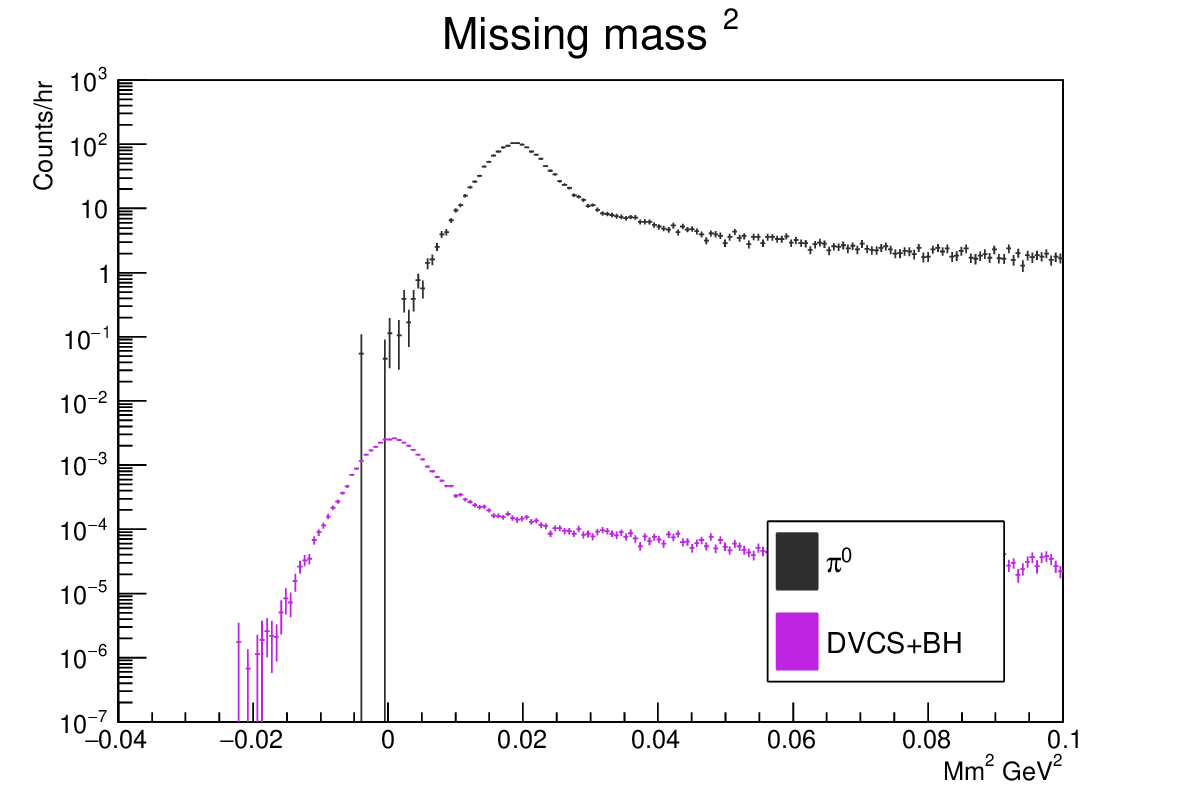}}
\subfloat[$Q^2$=2.0 GeV, $\epsilon=$0.94]{\includegraphics[clip,trim=1mm 1mm 10mm 10mm,width=0.4\textwidth]{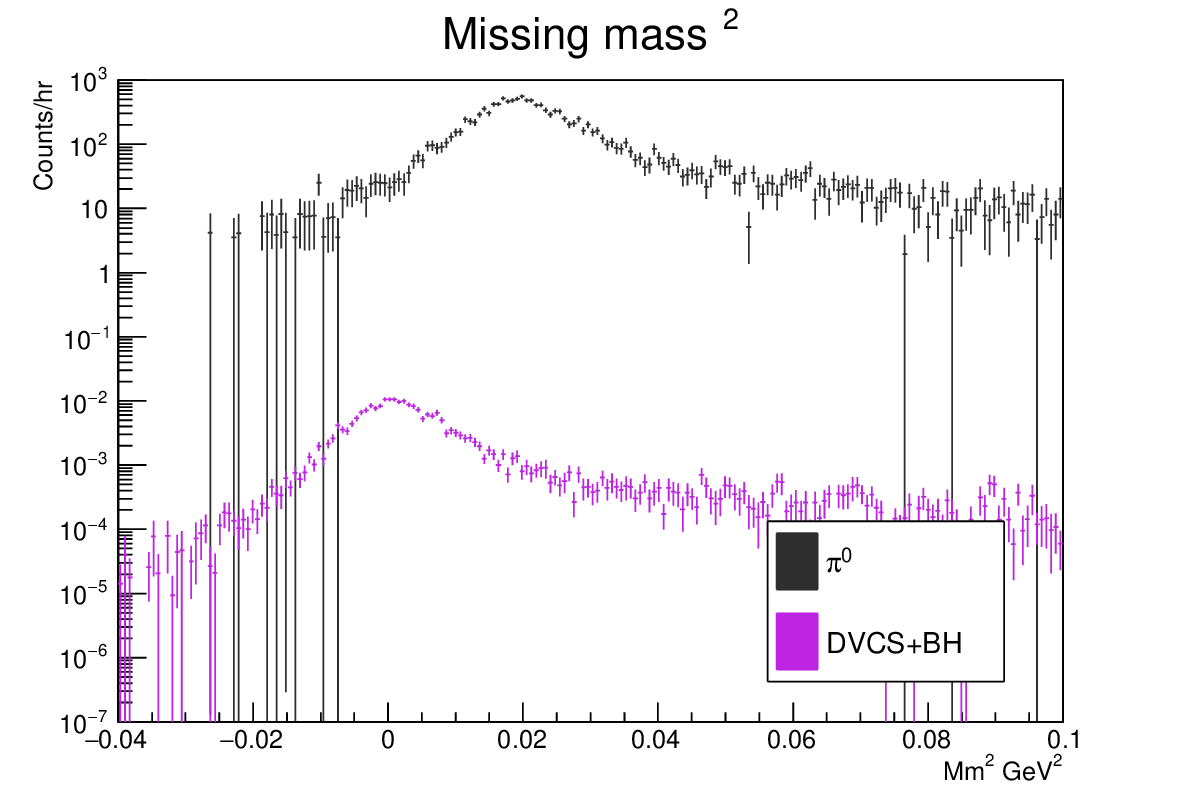}} \\
\subfloat[$Q^2$=3.0 GeV, $\epsilon=$0.54]{\includegraphics[clip,trim=1mm 1mm 10mm 10mm,width=0.4\textwidth]{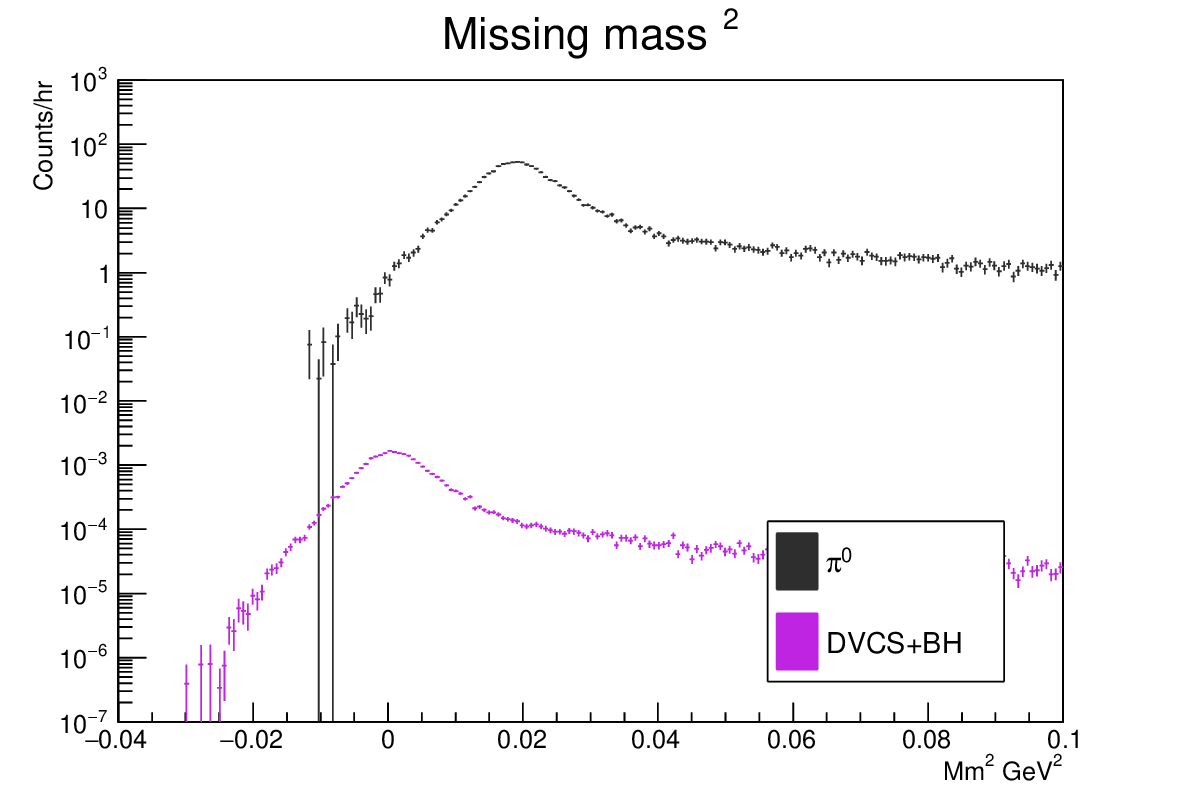}} 
\subfloat[$Q^2$=3.0 GeV, $\epsilon=$0.86]{\includegraphics[clip,trim=1mm 1mm 10mm 10mm,width=0.4\textwidth]{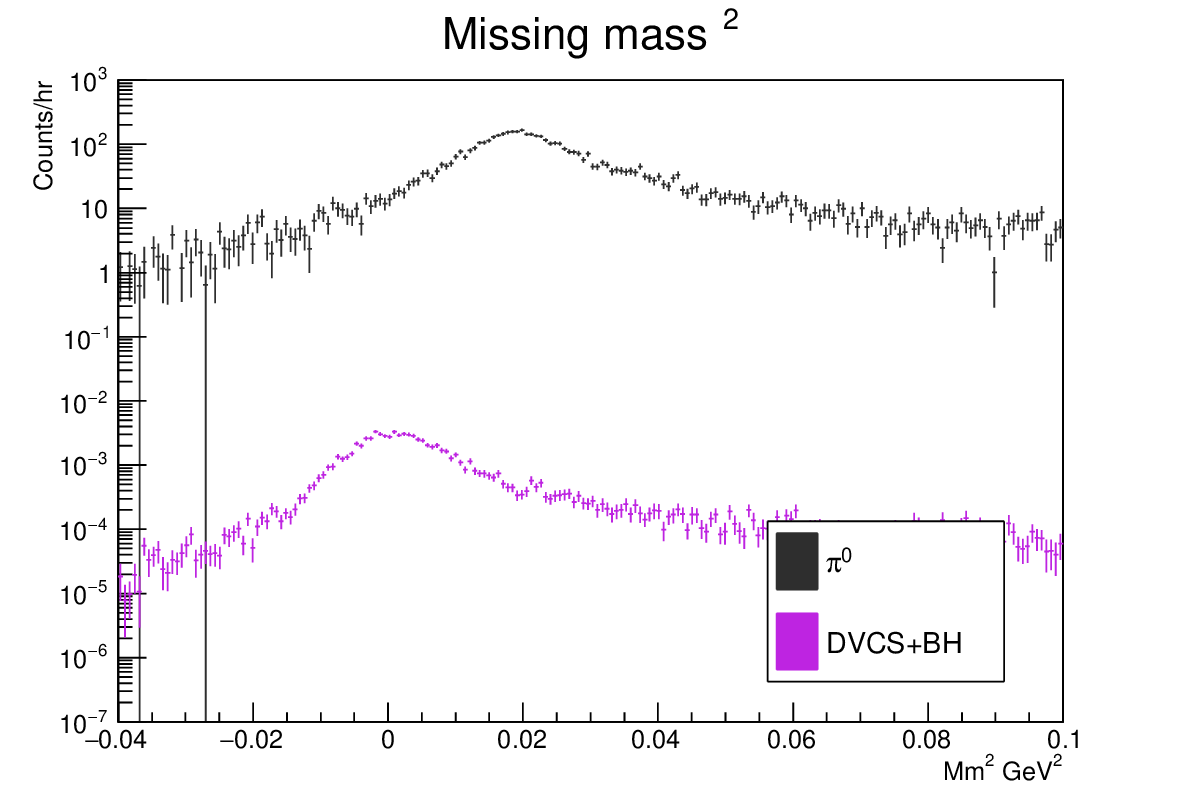}} \\
\subfloat[$Q^2$=4.0 GeV, $\epsilon=$0.56]{\includegraphics[clip,trim=1mm 1mm 10mm 10mm,width=0.4\textwidth]{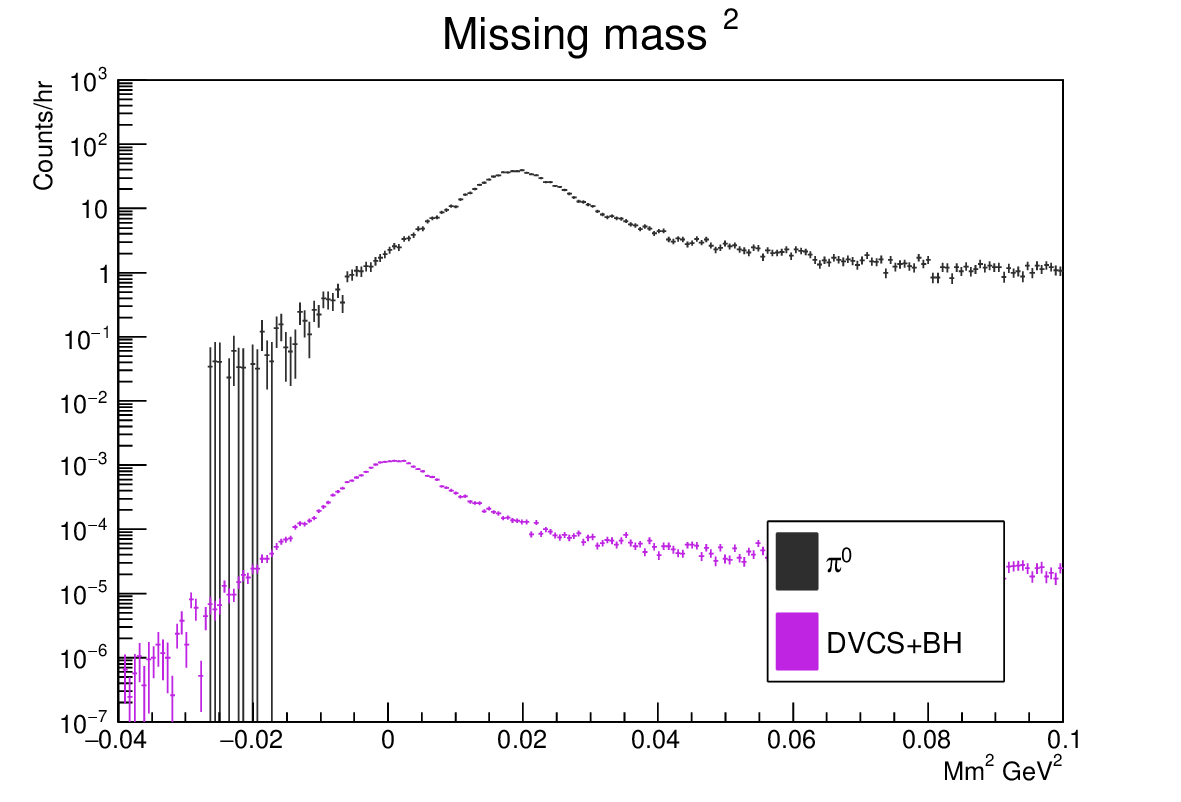}} 
\subfloat[$Q^2$=4.0 GeV, $\epsilon=$0.73]{\includegraphics[clip,trim=1mm 1mm 10mm 10mm,width=0.4\textwidth]{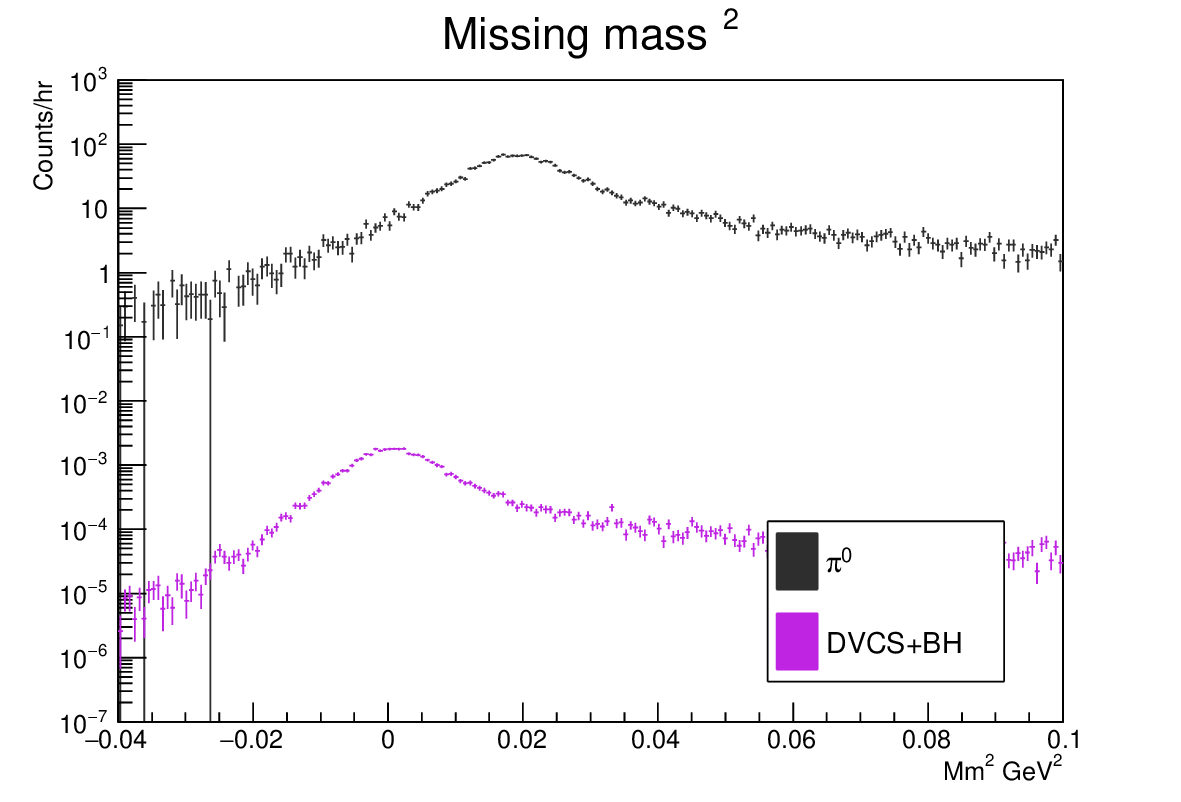}} \\
%\subfloat[$Q^2$=5.5 GeV, $\epsilon=$0.44]{\includegraphics[clip,trim=1mm 1mm 10mm 10mm,width=0.4\textwidth]{plots_eps_files/Mm2_550_326_44_+0000.eps}}
\caption{Simulated missing-mass-squared (M$_m^2$) distribution of $^1$H$(e, e^\prime p)X$ process at $Q^2=$2, 3 and 4 GeV. The $\pi^0$ distributions are in black; the backward VCS is in magenta. A cut at $M_m^2=$0.1 GeV$^2$ will be applied to minimize the VCS background. Simulation will be used to examine/subtract the small level of VCS contamination directly underneath the $\pi^0$ peak. } 
\label{fig:mm2}
\end{figure}

In comparison to backward-angle $\omega$ electroproduction~\cite{wenliang17, li19}, $\pi^0$ production has much less physics background from other mesons (such as $\eta$ and $\rho$).  A contributing physics background under the coincidence missing mass peak comes from the VCS process, whose missing mass peak is near $m_x=0$~GeV. An improved $u$-channel VCS model (based on the results from E93-050 and forward-angle DVCS model) and standard Bethe-Heitler (BH) formulation are used, see their implementation in Fig.~\ref{fig:DVCS}). The simulated $m_x^2$ distributions for the backward $\pi^0$ and $\gamma$ at $Q^2=$2, 3 and 4~GeV$^2$ are shown in Fig.\ref{fig:mm2}. The black distributions are for the $\pi^0$ events and the magenta distributions are for single $\gamma$ events, both distributions are normalized to 1~$\mu$C of beam charge. The $\pi^0:\gamma$ production ratio is $\sim$1000:1 in the simple simulation models used (the physics cross section models are described in Appx.~\ref{sec:pi0_MC} and \ref{sec:gamma_MC}).
A $m^2_x$ cut of 90~MeV$^2$ should exclude most of the single $\gamma$ events.  After events are binned in the $u$ and $\phi$, the shape and width of the $m_x$ peak will change slightly due to differences in the kinematics coverage ($Q^2$ and
$W$). Given the reconstructed resolution, the standard missing mass cut will not completely separate the two event distributions. Monte Carlo simulation will be needed to estimate the single $\gamma$ contamination for background subtraction purposes and place sensible cuts. This contamination is expected to be much less than 1\%. 
Furthermore, the single $\gamma$ physics background is unlikely to come from the Bethe-Heitler (BH) contribution. In Fig.~\ref{fig:DVCS}, the Virtual Compton Scattering (VCS) model used to estimate the $\gamma$ background is plotted on the same $-t$ axis as the classic BH formalism, and in the backward kinematics regime ($-t$ reaches the maximum value), the BH contribution is suppressed by a factor of 1000 compared to VCS. 

The lowest possible limit for the two pion production phase-space is likely to start playing a role at $M_m \sim 270$ MeV, which corresponds to $M_m^2= 0.06$ GeV$^2$, it will continue into the $\eta$ and $\rho$ mass region. We will place a cut at $M_m^2=0.05$ GeV$^2$ to reject potential 2$\pi$ contamination due to resolution effects. We expect  the background contribution from 2$\pi$ production (after cuts) to be very small and it will be included as part of the analysis simulation.

\subsection{Non-Physics Background}

Once a combination of online hardware and offline software cuts had determined that there is a coincidence between an electron in the HMS and a proton in the SHMS, there remain several backgrounds of the incoherent ‘non-physics’ variety: random coincidences and events from the end caps of the target cell. The online coincidence rates are presented and described in Sec.~\ref{sec:singles}, and the online electronic coincidence resolving window will be roughly 70~ns. 

Offline, our excellent coincidence time resolution enables us to reduce the relevant resolving time to 2~ns with negligible inefficiency. This is the first level of suppression of random coincidences. A cut on the missing mass variable reduces the final random coincidence contamination to the few percent level. The missing (or undetected residual) mass is reconstructed from the final electron and detected hadron 4-momenta:
\begin{equation}
M^2=(P_e - P_{e^\prime} + P_{tar} - P_h)^2\,.
\end{equation}

The missing mass cut does a lot more than random coincidence reduction. To the extent that particle identification is flawless, real coincidences with larger inelasticity than $p(e, e^\prime p)\pi^0$ are completely removed. Finally, the model dependence of the experimentally determined cross sections due to radiative effects are reduced as well.

Both spectrometers will detect the aluminum target end windows in all configurations, so window background subtractions are necessary. Because the aluminum windows are each 4~mil thick, the ratio of protons in the windows to protons in the liquid hydrogen is about 10\%. However, based on the previous operational experience such as the F$_\pi$-2 experiment, the surviving window background: $p(e, e^\prime\pi^+)n$ and $p(e, e^\prime p)\omega$, after cuts was found to be only 1\%~\cite{tanja15, wenliang17}. The Hall C ``empty'' target consists of two 40~mil thick aluminum windows separated by 8~cm, which can tolerate up to 30~$\mu$A beam current. Thus, our ``empty'' data come in 3 times $=$ (40 mil $\times$ 30~$\mu$A)/(4 mil $\times$ 90~$\mu$A) faster than window events on the real target. The empty target measurement overhead will be about 10\% of total data taking.

\subsection{Systematic Errors}
 
\begin{table}[h!]

\caption{Estimated systematic uncertainties for the proposed $\pi^0$ measurement. The systematic uncertainties in each column are added quadratically to obtain the total systematic uncertainty shown in the last row. The systematic uncertainty from the F$_{\pi}$-2-$\omega$ analysis \cite{wenliang17} is also listed for comparison purpose. The $\pi^0$ has smaller systematic uncertainties since it does not require multiple background fitting and subtraction from other neighboring mesons (such as the $\rho^0$ underneath the $\omega peak)$. }

\label{tab:sys_err_tab}
\setlength{\tabcolsep}{1 em}

\begin{tabular}{lccc}

\hline
Correction        	          & Uncorrelated &  $\epsilon$ Uncorrelated & Correlated  \\
                  	          & (Pt-to-Pt)   &  $u$ Correlated          &  (scale)    \\
               	              & (\%)         &  (\%)               & (\%)        \\ \hline
SHMS+HMS Tracking             &              &  0.6                &  1.2        \\
SHMS+HMS Triggers             &              &  0.1                &             \\
SHMS/HMS Detectors            &              &                     &  0.2        \\
Target Thickness              &              &  0.2                &  0.8        \\
CPU Live Time                 &              &  0.2                &             \\       
Electronic Live Time          &              &  0.2                &             \\       
Coincidence Blocking          &              &                     &  0.2        \\ \hline
Beam charge                   &              &  0.5                &  0.5        \\
PID                           &              &  0.2                &             \\
Acceptance                    &  0.6         &  0.6                &  1.0        \\
Proton Interaction            &              &                     &  1.0        \\  
Radiative Corrections         &              &  0.3                &  1.5        \\ 
Kinematics Offset             &  0.4         &  1.0                &             \\ 
Model Dependence              &  0.7         &                     &             \\ \hline 
$\pi^0$ Total                 &  1.0         &  1.4                &  2.5        \\ \hline 
F$_\pi$-2-$\omega$ Total      &  2.9         &  1.9                &  2.7        \\
\hline
\end{tabular}

\vspace{1cm}

\setlength{\tabcolsep}{0.5em}
\caption{Statistical error projection for each $Q^2$-$\epsilon$ setting. These estimates are based on the cross section model (presented in Table~\ref{tab:time}); and the assumed cross section ratio ($\sigma_T/\sigma_L$) based on the previous $u$-channel $\omega$ analysis \cite{wenliang17}.}
\begin{tabular}{ccccccccccc}
\hline
$Q^2$     & $W$    & $\epsilon_{1}$ & $\epsilon_{2}$ & $\Delta\epsilon$  & $\delta\sigma_{1}$  & $\delta\sigma_{2}$ & $\delta\sigma_L$  &  $\delta\sigma_T$   &  $\sigma_T/\sigma_L$ &  $\delta(\sigma_T/\sigma_L)$ \\
(GeV$^2)$ & (GeV)  &        &       &       & (\%)  &  (\%)  & (\%)       &  (\%)    &        &  (\%)  \\ \hline 
2.0       & 2.11   &  0.52  & 0.94  & 0.42  & 3.0   &  2.5   &  25        &  12      &  2     &  20    \\
2.0       & 3.00   &  0.32  & 0.79  & 0.47  & 5.6   &  5.5   &  46        &  12      &  2     &  25    \\
3.0       & 2.49   &  0.54  & 0.86  & 0.32  & 3.0   &  2.6   &  70        &  12      &  5     &  20    \\
4.0       & 2.83   &  0.55  & 0.73  & 0.18  & 3.5   &  3.1   &  800       &  20      &  30    &  40    \\
5.0       & 3.13   &  0.27  & 0.55  & 0.28  & 8.6   &  7.8   &  8300      &  20      &  200   &  50    \\
6.25      & 3.46   &  -     & 0.36  & -     & -     &  8.5   &  -         &  -       &  -     &  -     \\
\hline
% 5.5       & 3.26   &  -     & 0.44  & -     & -     &  4.5   &  -         &  -       &  -     &  -     \\
\end{tabular}
\end{table}

For all of the measurements proposed here, we have chosen the target length to be 10~cm. This is longer than the 6~cm used in L/T separation experiments in Hall C during the 6~GeV era. Using a longer target is possible because of the larger SHMS $y$-target acceptance compared to the Short Orbit Spectrometer (SOS). Even at very small angles, the extended target presents no problem for SHMS to project the background events from the target cell wall. The HMS $y$-target acceptance could potentially be problematic since it will be used at rather large angles (up to $\sim$37 degrees). We anticipate minimal extra uncertainty due to the use of the longer target.

The resulting anticipated systematic uncertainties are listed in Table~\ref{tab:sys_err_tab}. Our estimates are based on the proven experience with the HMS+SOS during the 6~GeV era and HMS+SHMS operation since the 12~GeV commissioning. 

\subsection{Projected Error Bars, Rates and Time Estimation}
\label{sec:kin}

The unseparated cross sections at low and high $\epsilon$ values: $\epsilon_1$ and $\epsilon_2$, can be expressed in terms of the separated cross sections $\sigma_L$ and $\sigma_T$,
\begin{align}
\sigma_{1} &= \sigma_T + \epsilon_{1}\,\sigma_L = \sigma_T \, (1 + \frac{\epsilon_{1}}{R}) \\
\sigma_{2} &= \sigma_T + \epsilon_{2}\,\sigma_L = \sigma_T \, (1 + \frac{\epsilon_{2}}{R})  
\end{align}
where $\sigma_1$ and $\sigma_2$ represent the unseparated cross sections at $\epsilon_1$ and $\epsilon_2$, respectively; $R$ is the transverse-longitudinal (T-L) ratio defined as 
\begin{equation}
R=\frac{\sigma_T}{\sigma_L}\,.
\end{equation}

Through substitution and manipulation of equations above, $\sigma_T$ and $\sigma_L$ can be expressed in terms of $\sigma_1$ and $\sigma_2$:
\begin{align}
\sigma_L &= \frac{\sigma_1-\sigma_2}{\epsilon_1 - \epsilon_2} \\
\sigma_T &= \frac{\sigma_2 \epsilon_1- \sigma_1 \epsilon_2}{\epsilon_1 - \epsilon_2}
\end{align}

By differentiating $\sigma_L$ and $\sigma_T$, the percentage errors can be expressed as, 
\begin{align}
\frac{\delta \sigma_T}{\sigma_T} (\%) & = \frac{1}{\epsilon_1 - \epsilon_2} \sqrt{\epsilon^2_1 \left(\frac{\delta\sigma_1}{\sigma_1} \right)^2 \left(1+ \frac{\epsilon_2}{R}\right)^2 + \epsilon^2_2 \left(\frac{\delta\sigma_2}{\sigma_2}\right)^2 \left(1+\frac{\epsilon_1}{R} \right)^2} \\
\frac{\delta \sigma_L}{\sigma_L} (\%) & = \frac{1}{\epsilon_1 - \epsilon_2} \sqrt{\epsilon^2_1 \left(\frac{\delta\sigma_1}{\sigma_1}\right)^2 (R + \epsilon_1)^2 + \epsilon^2_2 \left(\frac{\delta\sigma_2}{\sigma_2}\right)^2 (R+\epsilon_2)^2}
\end{align}
where $\delta\sigma_1$ and $\delta\sigma_2$ are the total statistical uncertainties of $\sigma_1$ and $\sigma_2$, respectively. The error magnification factor is $1/(\epsilon_1-\epsilon_2)$.

\begin{figure}[h]
\centering
\includegraphics[width=0.66\textwidth]{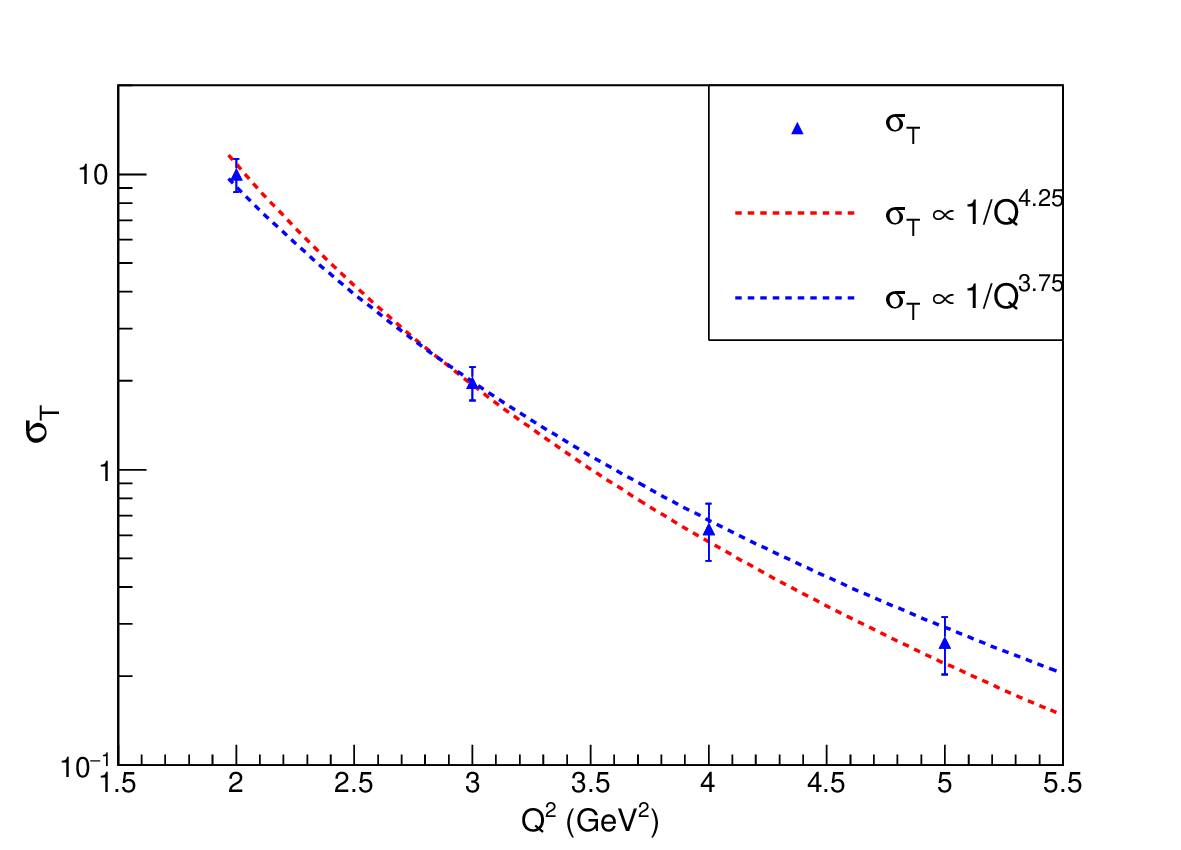}
\includegraphics[width=0.33\textwidth]{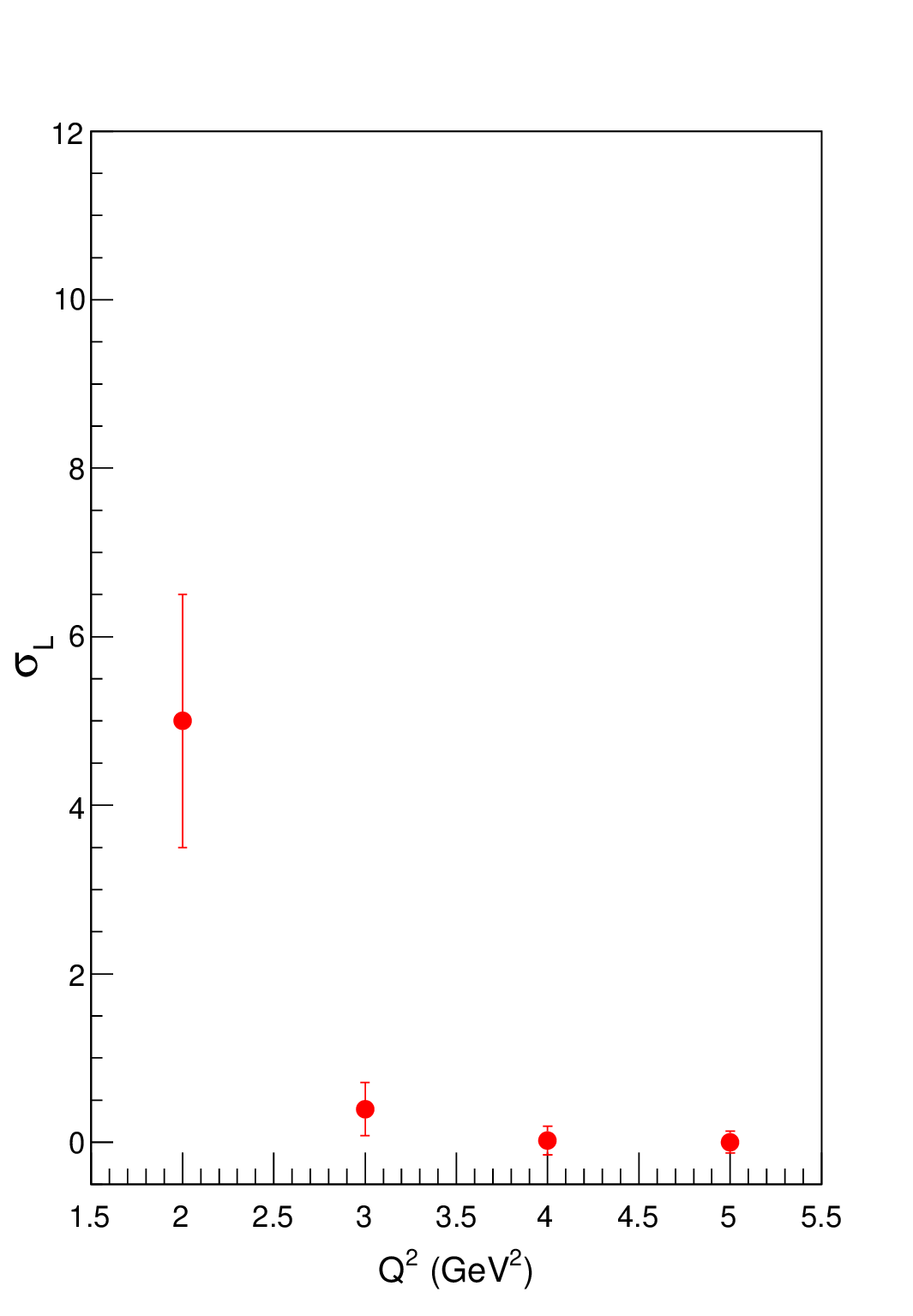}
\caption{$\sigma_T$ as function of $Q^2$. This is the figure of merit that demonstrates the second TDA postulation: $\sigma_T \propto 1/Q^8$. Our estimate of $\sigma_L$ as function of $Q^2$.}
\label{fig:cross_section}
\end{figure}

\begin{figure}[h]
\centering
\includegraphics[width=0.8\textwidth]{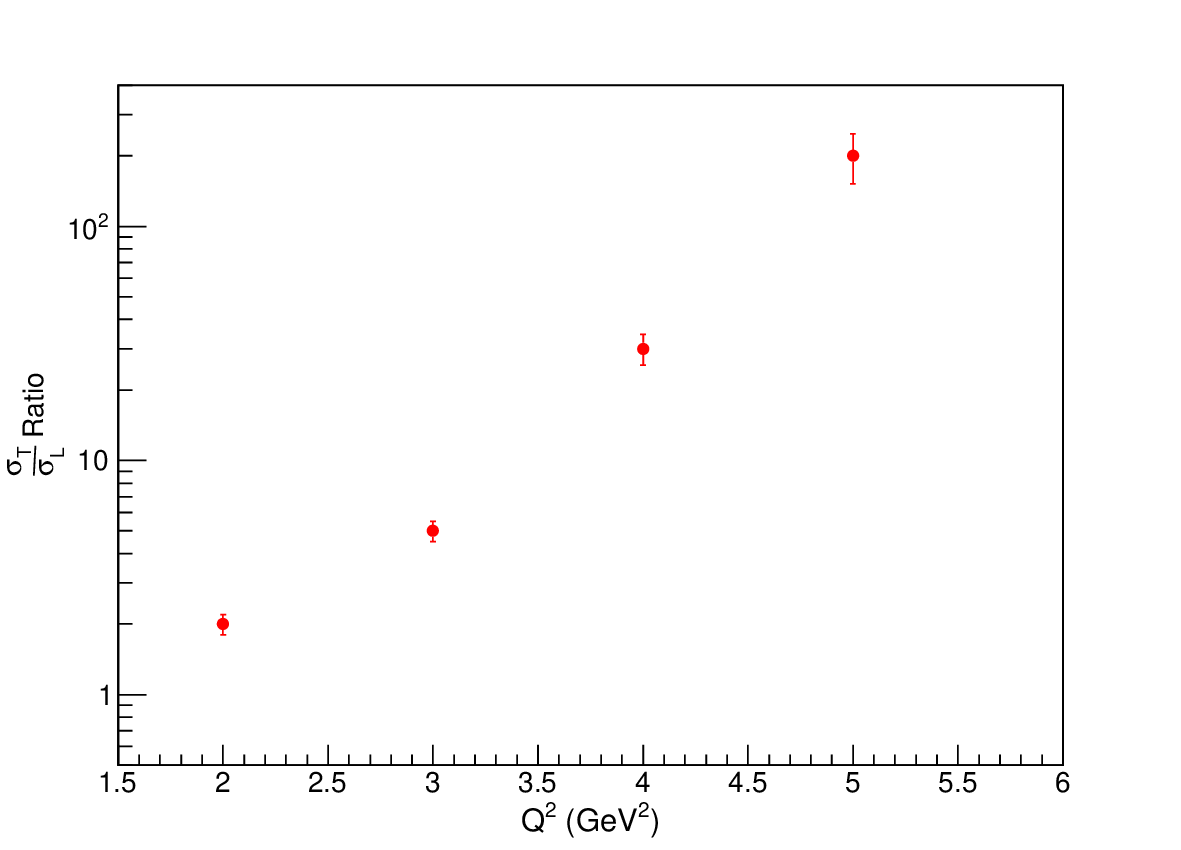}
\caption{$\sigma_T/\sigma_L$ ratio as function of $Q^2$. This is the figure of merit that demonstrates the first TDA postulation: $\sigma_T\ll \sigma_L$ or $\sigma_T/\sigma_L < 1/Q^2$.}
\label{fig:target_ratio}
\end{figure}

The determination of the running time at each $Q^2$-$\epsilon$ setting is dictated by the observable, shown in Figs.~\ref{fig:cross_section} (a) and \ref{fig:target_ratio}, for testing the TDA hypothesis. The logic behind our decisions is as follows: 
\begin{enumerate}

\item For the $\sigma_T$ scaling shown in Fig.~\ref{fig:cross_section} (a), our goal is to demonstrate that $\sigma_T (u^\prime=0) \propto 1/(Q^2)^n$ and with an uncertainty for $n=4\pm0.25$, assuming the $1/(Q^8)$ scaling hypothesis.  

\item In order to demonstrate the dominance of $\sigma_T\gg\sigma_L$ separated cross section, i.e. the ratio $R=\sigma_T/\sigma_L$, the uncertainty $\delta(\sigma_L/\sigma_T)$ should be kept less than $<50\%$. 

\item For a fixed $Q^2$, the statistical uncertainty balance between different epsilon settings, $\delta\sigma_1$ and $\delta\sigma_2$ ,after the acceptance and diamond cuts must be maintained. Note that all data are divided in five different $u$ bins, where $\delta\sigma_1$ and $\delta\sigma_2$ are the average value between the uncertainty at the lowest $-u$ and the five-bin-average for low and high $\epsilon$, respectively.

\end{enumerate}

The total time required for each $Q^2$-$\epsilon$ setting is listed in Table~\ref{tab:time}, this time will be shared equally by the hadron arm angle settings. Times have been increased by 10\% to account for data taking from the aluminum ``dummy'' target, needed to subtract contributions from the target cell walls. The time required to complete the $\pi^0$ measurement is 706 (PAC) hours.

\begin{table}[ht]
\centering
\caption{Estimated event rate (per hour) and total requested PAC time for the proposed measurement. These estimations take into account cuts such as the spectrometer acceptance cut, missing mass cut and the diamond cut (see Fig.~\ref{fig:diamond}). 70~$\mu$A beam current is assumed. Note that the estimated time presented in the table includes ($\pi^0$+Heep) dummy target running time (10\%). Heep time is scaled up be additional 1 hours per setting inelastic optics study (verification purpose) on a carbon target. Four $E_{beam}$ polarization measurements are planned, one measurement at each beam energy. All planed measurements assume the standard accelerator gradient at the time of running (here we assume 2.2~GeV/pass). }
\label{tab:time}

\setlength{\tabcolsep}{0.2em}

\begin{tabular}{cccccccc}
%\toprule 
%\multicolumn{5}{c}{$\pi^{0}$ Rates for Central Angles} \\
%\toprule 
\hline                                                                              
$Q^{2}$                 & $W$   &  $\epsilon$ & $E_{\rm Beam}$ [Pass] & Physics Rate & Background Rate & PAC Time & PAC Time \\
(GeV$^2$)               & (GeV) &             & (GeV)          & (per Hour)   & (per Hour)      & (Hours)  & (Days)   \\ \hline
2.0                     & 2.11  &  0.52       &  4.4 [2]       &  140         & 0.01            &  33      &  1.4     \\
                        &       &  0.94       &  10.9 [5]      &  500         & 0.05            &  10      &  0.4     \\ \hline 
2.0                     & 3.00  &  0.32       &  6.6 [3]       &  14          & $<$0.01         &  66      &  2.8     \\
                        &       &  0.79       &  10.9 [5]      &  73          & $<$0.01         &  27      &  1.1     \\ \hline
3.0                     & 2.49  &  0.54       &  6.6 [3]       &  60          & $<$0.01         &  60      &  2.5     \\
                        &       &  0.86       &  10.9 [5]      &  140         & 0.01            &  27      &  1.1     \\ \hline
4.0                     & 2.83  &  0.56       &  8.8 [4]       &  40          & $<$0.01         &  60      &  2.5     \\
                        &       &  0.73       &  10.9 [5]      &  80          & $<$0.01         &  40      &  1.7     \\ \hline 
5.0                     & 3.31  &  0.26       &  8.8 [4]       &  4           & $<$0.01         &  132     &  5.5     \\
                        &       &  0.55       &  10.9 [5]      &  11          & $<$0.01         &  47      &  2.0     \\ \hline 
6.25                    & 3.46  &  0.36       &  10.9 [5]      &  2.63        & $<$0.01         &  88      &  3.7     \\ \hline 
Subtotal                &       &             &                &              &                 &  590     &  24.6    \\ \hline  
$^1$H$(e, e^{\prime}p)$ &       &             &                &              &                 &  28      &  1.2     \\ \hline  
$E_{Beam}$ change       &       &             &                &              &                 &  52      &  2.2     \\ \hline  
Optics study            &       &             &                &              &                 &  4       &  0.2     \\ \hline
$E_{Beam}$ Polar. &       &             &                &              &                 &  32       &  1.3     \\ \hline  
Total Time          &       &             &                &              &                 &  706     &  29.4    \\

\end{tabular}	

\end{table}

\section{Closing Remarks}

In a short summary, the proposed exclusive electroproduction $^1$H$(e, e'p)\pi^0$ measurement is a standard Hall C L/T experiment which utilizes the HMS+SHMS configuration, unpolarized LH$_2$ target and standard accelerator gradient (at the time of running). The total run time is 706 PAC hours (29.4 PAC days). Technically, the experiment is ``straightforward''.

We aim to perform L/T separation over a $Q^2$ range of 2-5~GeV$^2$ and two additional settings: $Q^2$=2~GeV$^2$ ($W=3$~GeV) and $Q^2$=6.25~GeV$^2$ to check the $W$ and $Q^2$ scaling.

\begin{enumerate}
\item At each measurement setting, data are binned in five $-u$ bins, with the lowest valued $-u$ bin corresponding to $-t=-t_{max}$. The observed $-t$ dependence will reveal the anticipated $u$-channel $\pi^0$ peak, which will facilitate the comparison to the forward-angle exclusive data from E12-13-010.

\item  Once the backward-angle peaks are confirmed, the separated cross sections: $\sigma_T$ and $\sigma_L$ at $Q^2=$2, 3, 4 and 5 GeV$^2$ will be extracted. The transverse size of interaction that is determined from the $u$-dependence of the separated cross sections at each setting, which provides a good phenomenological handle to study the ``soft-hard'' transition in $u$-channel physics.

\item The anticipated $\sigma_T/\sigma_L$ ratios, will test the TDA predicted dominance of $\sigma_T$ over $\sigma_L$. Quantitatively, the criteria for $\sigma_T\gg\sigma_L$: $\sigma_T/\sigma_L$ increases as a function of $Q^2$ and reaches $\sigma_T/\sigma_L > 10$ at $Q^2=5$~GeV. 

\item The $\sigma_T\propto 1/Q^n$ scaling test for $2< Q^2< 6.25$ GeV$^2$, especially, at $Q^2=6.25$ GeV$^2$ is dependent on the $\sigma_T\gg\sigma_L$ from the above item. The scaling result will further validate the TDA factorization scheme and we expect to extract the exponent factor to an accuracy of $\delta n=\pm0.25$,

\end{enumerate}
With these results, one would conclude the backward-angle ($u$-channel) factorization scheme offered by TDA is valid for the exclusive $\pi^0$ electroprodcution channel.

In our opinion, the proposed measurement is an example of how we can utilize the existing experimental apparatus and refined techniques, but slightly tweak our perspective to explore a largely unknown kinematics territory. After this proposal, we anticipate a wave of $u$-channel meson electroproduction and photoproduction measurements to emerge ($\eta$, $\rho$, $\omega$, $\phi$, even J/$\psi$) and perhaps $u$-channel VCS. 

\clearpage

\appendix

\section{Monte Carlo model of Deep Exclusive $\pi^0$ Production in $u$-channel}

\label{sec:pi0_MC}

The Monte Carlo studies needed for this proposal require a reaction model for an experimentally unexplored region of kinematics.  This appendix describes the model and the constraints used. The differential cross section for exclusive $\pi$ production from the nucleon can be written as
\begin{equation}
  \frac{d^{5} \sigma}{dE' d\Omega_{e'} d\Omega_{\pi}} = \Gamma_{V} \frac{d{^2}
  \sigma}{d\Omega_{\pi}}.
\end{equation}
The virtual-photon flux factor $\Gamma_{V}$ is defined as
\begin{equation}
  \Gamma_v=\frac{\alpha}{2\pi^2} \frac{E'}{E} \frac{K}{Q^2}\frac{1}{1-\epsilon},
\end{equation}
where $\alpha$ is the fine structure constant, $K$ is the energy of real photon equal to the photon energy required to create a system with invariant mass equal to $W$ and $\epsilon$ is the polarization of the virtual-photon.
\begin{equation}
  K=(W^2-M_p^2)/(2 M_p)
\end{equation}
\begin{equation}
  \epsilon=\left(1+\frac{2 |\mathbf{q}|^2}{Q^2} \tan^2\frac{\theta_{e}}{2}
  \right)^{-1},
\end{equation}
where $\theta_{e}$ is the scattering angle of scattered electron.

The two-fold differential cross section $\frac{d{^2} \sigma}{d\Omega_{\pi}}$ in the lab frame can be expressed in terms of the invariant cross section in center of mass frame of the photon and nucleon,
\begin{equation}
  \frac{d^2 \sigma}{d\Omega_\pi}= J \frac{d^2 \sigma}{dt d\phi},
\end{equation}
where $J$ is the Jacobian of transformation of coordinates from lab $\Omega_{\pi}$ to $t$ and $\phi$ (CM). 

In the one-photon exchange approximation, the unpolarized nucleon cross section for $n(e,e^{\prime}\pi^{-})p$
can be expressed in four terms. Two terms correspond to the polarization states of the virtual-photon (L and T) and two states correspond to the interference of polarization states (LT and TT),
\begin{equation}
  d\sigma_{UU} =  \epsilon  \frac{d\sigma_{\mathrm{L}}}{dt}
  + \frac{d\sigma_{\mathrm{T}}}{dt} + 
  \sqrt{2\epsilon (\epsilon +1)} \frac{d\sigma_{\mathrm{LT}}}{dt} \cos{\phi}
  + \epsilon  \frac{d\sigma_{\mathrm{TT}}}{dt} \cos{2 \phi},
  \label{eqn:cross-2}
\end{equation}
where $\phi$ is the angle between lepton plane and hadron plane (Fig.~\ref{fig:planes}). The first two terms of Eqn.~\ref{eqn:cross-2} correspond to the polarization states of the virtual-photon (L and T) and last two terms correspond to the interference of polarization states (LT and TT). 

The following data and calculations were used as constraints on the parameterizations used in this model:
\begin{itemize}
\item
From Hall A, $L/T/LT/TT$ separated experimental data of exclusive electroproduction of $\pi^0$ on $^1$H are available at $x_{\rm B}=$0.36 and three different $Q^2$ values ranging from 1.5 to 2 GeV$^2$ \cite{defurne16}. Of these three, we use only the data set at $Q^2$=1.75~GeV$^2$, as it spans the widest $t$-range, $0.184<-t<0.284$~GeV$^2$~\cite{defurne16}.
\item
A GPD-based handbag-approach calculation by Goloskokov and Kroll \cite{gk11} for the E12-13-010 proposal \cite{E12-13-010} at $x_{\rm B}=$0.36, $Q^2$=3.0, 4.0, 5.5~GeV$^2$~\cite{gk11}.

\end{itemize}

Since both of these data and calculations are for forward-angle kinematics, we used the following prescription to obtain a crude model for the unique backward-angle kinematics proposed here.  
\begin{itemize}
\item
The $t$-dependence of the T/LT/TT structure functions at each $Q^2$ were fitted with functions of the form $a+b/(-t)$, which gave good fits over the range $-t_{min}<-t<0.8$~GeV$^2$ with a minimum of fit parameters. $\sigma_L$ displayed very little $t$-dependence over the region for which there were data, so it was simply taken as a small constant value with $t$ (about 1~nb/GeV$^2$, but with magnitude dropping as $Q^2$ increases).
\item
Since the electroproduction data in Fig.~\ref{fig:omega} display a forward to backward-angle peak ratio of about 10:1, we estimate the magnitude of the backward angle cross sections by switching the $u$-slope for $t$-slope in the above equations, and divide by ten.
\item
Linear interpolation was performed between the parameterized values at fixed $Q^2$=1.75, 3.0, 4.0, 5.5~GeV$^2$ to obtain the L/T/LT/TT cross sections for the exact $Q^2$ needed for each event in the SIMC Hall C Monte Carlo
simulation.
\item
After the parameterization of $\sigma_{L,T,LT,TT}$ for $-u$ and $Q^2$, we assume the same $W$ dependence as used in \cite{blok08} for exclusive $\pi^+$ electroproduction at similar $x_{\rm B}$, which is $(W^2-M^2)^{-2}$ where $M$ is the proton mass.
\end{itemize}

Clearly, this model can only be described as a `best guess' of the actual DEMP $\pi^0$ cross sections in this unexplored regime.  It is anticipated that some parasitic $\pi^0$ backward angle data will be acquired in the Hall C DEMP experiments, E12-09-011, E12-19-006, which can be used to improve the crude model used here.

\section{Monte Carlo model of $\gamma$ Production in $u$-channel}
\label{sec:gamma_MC}

The backward-angle VCS+BH model is based on the C++ code by C. Munoz Camacho and H. Moutarde (CEA-Saclay, IRFU/SPhN), which itself is derived from a Mathematica package written by P. Guichon and M. Vanderhaeghen.

The Bethe-Heitler (BH) formulas are exact and valid over the full kinematic range of the experiment. As shown in Fig.~\ref{fig:DVCS}, the BH process does not exhibit a backward-angle peak, and has minimal contribution in the large $-t$ probed by this experiment.

For the Virtual Compton Scattering (VCS)  process, the code makes use of a file of Compton Form Factors which were computed by Guichon and Vanderhaeghen in a grid over the range: $0.2<x<0.9$, $1<Q^2<14~\text{GeV}^2$, $0<-t<3.0~\text{GeV}^2$.  To extend the model to the backward-angle regime, we applied the same ansatz as assumed in the $\pi^0$ model, i.e. we estimate the magnitude of the backward angle cross sections by switching the $u$-slope for $t$-slope, and divide by ten.  The discontinuity in the red VCS curves in Fig.~\ref{fig:DVCS} is caused by this assumption, with the gap between the backward-angle and forward-angle curves due to the lack of Compton Form Factors at wide angles (far outside the acceptance of the experiment in any case).

\section{Differences and Connections to Other Approved $\pi^0$ Measurements}
\label{sec:other_mea}

In this section, we attempt to connect the proposed $\pi^0$ measurement with other approved $\pi^0$ measurements. It is interesting that the combined kinematics coverage from different experiments offer a complete $-t$ evolution at certain kinematics settings. Also note that these measurements utilize direct detection method (of the decayed photons) to reconstruct the $\pi^0$ events, in contrast to the missing mass technique utilized in this proposal.  

There were two approved $\pi^0$ production oriented measurements at Hall C, see their brief description as follows:
\begin{description}

\item[E12-13-010 and E12-06-114]  Hall C E12-13-010 proposed by the Neutral Particle Spectrometer (NPS) Collaboration in PAC 41, is a dedicated measurement that will map out the DVCS and exclusive $\pi^0$ electroproduction cross sections at various bins of $x_B$, $W$ and $Q^2$. Its main physics objective involves studying GPDs and establish the applicable range of the forward-angle factorization scheme. At the same kinematic settings as this proposal, $x_B=0.36$, $Q^2=3$ and $4$ GeV$^2$, the NPS will offer L/T separated cross sections at low $-t$, in particular, $-t=-t_{min}$ or $-t^\prime=0$ GeV$^2$. Note that in this kinematics region $-u=-u_{max}$. This is sometimes referred to as the ``low $-t$ region''. Hall A E12-06-114 experiment has the same physics goal and similar kinematics coverage as E12-13-010, which will offer complementary data set. The full L/T separation for all settings depends on the completion of both programs.

\item[E12-14-005] is a companion program to the Wide Angle Compton scattering experiment (WACS), and it focuses on the detection of $\pi^0$ photoproduction at a range of large scattering angles. In the range of $s$: $2.8<s<10.1$ GeV$^2$, the NPS will be used to detect $\pi^0$ produced in $50^\circ$ to $110^\circ$ in CM angle. In this region, $u \approx t$. This is sometimes referred to as the ``high $-t\approx -u$ transition region''.

\item[Large Angle $\pi^0$ Measurement at CLAS 12] With its large acceptance, the default CLAS 12 detector package is capable of extracting $\pi^0$ cross sections up to a large angle ($0<\theta<130^\circ$ in the lab frame) at the same kinematics setting as the proposed measurement. These results would serve as an even better complementary data set which allows one to study with high precision the hard $\rightarrow$ soft $\rightarrow$ hard structure transition, which corresponds to forward-angle-peak $\rightarrow$ wide-angle-plateau $\rightarrow$ backward-angle-peak in the measured cross section. Interestingly, the measurement of BSA and cross section ($\phi$) modulation sensitive to $\sigma_{LT}$ has already generated interesting physics insight, see Sec.~\ref{sec:BSA}.  

\end{description}

As described in the main section of the proposal, this proposal $\pi^0$ has unique $-t = -t_{max}$ or $-u=-u_{min}$ kinematics (a low $u$ region). No other physics measurement acquires such kinematics at JLab. The combined $-t$ dependence (two mentioned $\pi^0$ measurements along with this proposal) will yield a complete $-t$ evolution (after appropriate kinematics correction). Such example is shown in Fig.~\ref{fig:omega}: the forward-peak, the wide-angle-plateau and the backward-peak will be projected on the same $-t$ axis. These combined features will significantly reduce the Regge-model uncertainty in any description of proposed data, specially at the overlapped $Q^2=3$ and $4$ GeV$^2$ settings.

\section{Further Details on $u$ Channel Workshop in September 2020}

The first backward-angle ($u$-channel) physics workshop is set to take place in Jefferson Lab in September 21-22, 2020. The scientific program is expected to comprise roughly 15 talks.   If an in-person meeting is not possible, the workshop will be held virtually.

The objectives of the workshop are as follows:
\begin{itemize}
\item Offer a platform to connect scattered experiment and theory efforts together, thus, potentially forming small backward-angle physics working groups.
\item Generate discussions on the implications the backward-angle physics and probe the physics case for a systematic backward-angle physics research program.
\item Inspire future backward-angle physics data mining or dedicated studies, including the JLab 12 GeV program, and $rm\overline{P}$ANDA/FAIR.
\item Discuss the feasibility of including backward-angle physics in the EIC scientific program.
\end{itemize}
JLab event page: \url{https://www.jlab.org/conference/BACKANGLE}

\subsection*{$u$ Channel Workshop Agenda and Speakers List}

\label{app:workshop}

Tentative list of discussion topics and speakers are listed below. A up-to-date agenda can also be found on the JLab indico page: \url{https://www.jlab.org/indico/event/375/}. 
\begin{itemize}

\item Meson-Nucleon Transition Distribution Amplitude, Lech Szymanowski (NCNR, Poland).

\item Studying Vector Meson Electroproduction at Large Momentum Transfer, J-M Laget (Jefferson Lab, USA). (Not confirmed)

\item Regge Phenomenology through $u$-channel Processes , Christian Weiss (Jefferson Lab, USA).

\item Backward Charged $\pi^+$ Electroproduction from CLAS 6,  Kijun Park (Hampton University Proton Therapy Institute, USA).

\item Backward Exclusive $\omega$ Electroproduction from JLab 6 GeV Hall C, Garth Huber (University of Regina, Canada).

\item Backward Meson Electroproduction from JLab 12 GeV Hall C Kaon LT Experiment, Stephen Kay (University of Regina, Canada).

\item Backward Meson Electroproduction Opportunities at EIC, Wenliang (Bill) Li (William and Mary, USA).

\item Backward Opportunities at JLab 12 GeV Hall A with BigBite and Super BigBite Spectrometers, Carlos Ayerbe (Mississippi State, USA).

\item Large Angle $\pi^0$ Production at CLAS 12, Stefan Diehl (University of Connecticut, USA).

\item Studying TDA with $p\overline{p}\rightarrow e^ + e^- \pi^0$ at the $\rm\overline{P}$ANDA Experiment, Stefan Diehl (University of Connecticut, USA).

\item Wide Compton Scattering at Hall C, Bogdan Wojtsekhowski (Jefferson Lab, USA)

\item $\omega$ Photoproduction off Proton Target at Backward Angles, B.-G. Yu (Korea Aerospace University, Korea).

\item Backward-angle $\omega$ Photoproduction at $E_{\gamma} = 8.8$ GeV with GlueX Experiment, Justin Stevens, (William and Mary, USA).

\item Backward Baryon-Antibaryon Photoproduction at GlueX Experiment, Hao Li (Carnegie Mellon, USA).

\item $\Sigma$-$K^+$ Photoproduction at Backward Angle at GlueX Experiment, Nilanga Wickramaarachchi (Old Dominion, USA).

\item High-energy collisions in UPCs and at an EIC, Spencer Klein (Lawrence Berkeley National Laboratory, USA).

\end{itemize}

\end{document}